\documentclass[fleqn,usenatbib]{mnras}

\usepackage{newtxtext,newtxmath,ulem,hhline}

\usepackage[T1]{fontenc}

\DeclareRobustCommand{\VAN}[3]{#2}
\let\VANthebibliography\thebibliography
\def\thebibliography{\DeclareRobustCommand{\VAN}[3]{##3}\VANthebibliography}

\usepackage{graphicx}	
\usepackage{amsmath}	
\usepackage[dvipsnames]{xcolor}
\usepackage{tikz}
\usepackage{tabularx}

\newcommand{\equ}[1]{eq.~(\ref{eq:#1})}
\newcommand{\equs}[1]{eqs.~(\ref{eq:#1})}
\newcommand{\equm}[1]{(\ref{eq:#1})}

\newcommand{\equnp}[1]{eq.~\ref{eq:#1}}

\newcommand{\se}[1]{Section~\ref{sec:#1}}

\newcommand{\fig}[1]{Fig.~\ref{fig:#1}}
\newcommand{\figs}[1]{Figs.~\ref{fig:#1}}
\newcommand{\figss}[1]{\ref{fig:#1}}
\newcommand{\Fig}[1]{Figure~\ref{fig:#1}}

\newcommand{\tab}[1]{Table~\ref{tab:#1}}
\newcommand{\be}{\begin{equation}}
\newcommand{\ee}{\end{equation}}
\newcommand{\bea}{\begin{eqnarray}}
\newcommand{\eea}{\end{eqnarray}}

\newcommand{\ltsima}{$\; \buildrel < \over \sim \;$}
\newcommand{\lsim}{\lower.5ex\hbox{\ltsima}}
\newcommand{\gtsima}{$\; \buildrel > \over \sim \;$}
\newcommand{\gsim}{\lower.5ex\hbox{\gtsima}}

\newcommand{\const}{\rm const.}

\def\la{\langle}

\newcommand{\msun}{\,{\rm M}_\odot}

\newcommand{\ifm}[1]{\relax\ifmmode#1\else$\mathsurround=0pt #1$\fi}
\newcommand{\kms}{\ifmmode\,{\rm km}\,{\rm s}^{-1}\else km$\,$s$^{-1}$\fi}

\newcommand{\kpc}{\,{\rm kpc}}
\newcommand{\pc}{\,{\rm pc}}

\newcommand{\K}{\,{\rm K}}

\def\cms{\,{\rm cm}^{-2}}

\def\dmax{d_{\rm max}}

\def\tsc{\,{t_{\rm sc}}}
\def\rsi{\,{r_{\rm s,i}}}
\def\rsf{\,{r_{\rm s,f}}}
\def\chif{{\chi_{\rm f}}}

\def\M*{M_{\rm *}}
\def\Mv{M_{\rm v}}
\def\Rv{R_{\rm v}}

\def\Kdegree{{\rm K}}

\def\Pi{\varpi_{_{\rm I}}}

\title[Geometrical and metallicity effects]{Effects of Cloud Geometry and Metallicity on Shattering and Coagulation of Cold Gas, and Implications for Cold Streams Penetrating Virial Shocks}

\author[Yao et al.]{
Zhiyuan Yao$^{1}$\thanks{E-mail: zhiyuan.yao@mail.huji.ac.il},
Nir Mandelker$^{1}$\thanks{E-mail: nir.mandelker@mail.huji.ac.il},
S. Peng Oh$^{2}$,
Han Aung$^{1}$,
Avishai Dekel$^{1,3}$
\\
$^{1}$Racah Institute of Physics, The Hebrew University, Jerusalem 91904, Israel\\
$^{2}$Department of Physics, University of California, Santa Barbara, CA 93106, USA\\
$^{3}$SCIPP, University of California, Santa Cruz, CA 95064, USA
}

\date{Accepted XXX. Received YYY; in original form ZZZ}

\pubyear{2024}

\begin{document}
\label{firstpage}
\pagerange{\pageref{firstpage}--\pageref{lastpage}}
\maketitle

\begin{abstract}
Theory and observations reveal that the circumgalactic medium (CGM) and the cosmic web at high redshifts are multiphase, with small clouds of cold gas embedded in a hot, diffuse medium. A proposed mechanism is `shattering' of large, thermally unstable, pressure-confined clouds into tiny cloudlets of size $\ell_{\rm shatter}\sim {\rm min}(c_{\rm s}t_{\rm cool})$, with $c_{\rm s}$ and $t_{\rm cool}$ the gas sound speed and cooling time, respectively. These cloudlets can then disperse throughout the medium like a `fog', or recoagulate to form larger clouds. We study these processes using idealized numerical simulations of thermally unstable gas clouds. We expand upon previous works by exploring the effects of cloud geometry (spheres, streams, and sheets), metallicity, and the inclusion of an ionizing UV background. We find that `shattering' is triggered by clouds losing sonic contact and rapidly imploding, leading to a reflected shock which causes the cloud to re-expand and induces Richtmyer-Meshkov instabilities (RMI) at its interface. In all cases, the expansion velocity of the cold gas is of order the cold gas sound speed, $c_{\rm s,c}$, slightly smaller in sheets than in streams and spheres due to geometrical effects. After fragmentation the cloudlets experience a drag force from the surrounding hot gas, which due to their low relative velocity, $\Delta v\sim c_{\rm s,c}$, is always dominated by condensation of hot gas onto the clouds rather than ram pressure. This eventually causes the cloudlets to recoagulate into one or several large clouds. We distinguish between `fast' and `slow' coagulation regimes depending on how much the cold clouds have dispersed prior to decelerating and co-moving with the hot gas. We show analytically, and confirm using simulations, that sheets are always in the `fast' coagulation regime while streams and spheres have a maximum overdensity for rapid coagulation. This depends on both the cloud radius and its temperature relative to the cooling floor when it loses sonic contact, contrary to previous models which found a dependence on cloud size only. The critical overdensity for spheres is 
smaller than for streams, such that 
the coagulation efficiency increases from spheres to streams to sheets. After shattering, spheres and streams develop a lognormal distribution of clump masses that peaks at the resolution scale, even if this is below $\ell_{\rm shatter}$, but is also well fit by $N(>m)\propto m^{-1}$ over a wide range of masses. 
This suggests that a minimal cloud size should be set by other mechanisms such as thermal conduction or external turbulence. We apply our results to the case of cold streams feeding massive ($\Mv\gsim 10^{12}\msun$) high-$z$ ($z\gsim 2$) galaxies from the cosmic web, finding that streams are likely to shatter upon entering the CGM through the virial shock. This offers a possible explanation for the large clumping factors and covering fractions of cold gas in the CGM around such galaxies, and may be related to galaxy quenching by providing a mechanism to prevent cold streams from reaching the central galaxy.
\end{abstract}




\section{Introduction}
\label{sec:intro}
Only a small fraction of the Universe's baryons are found in galaxies, including both stars and interstellar gas \citep[e.g.][]{Peeples.etal.14,Tumlinson.etal.2017,Wechsler.Tinker.18}. The majority of baryons, and also the majority of metals, reside in the \textit{circumgalactic medium} (CGM, gas outside galaxies but within dark matter halos), and the \textit{intergalactic medium} (IGM, gas outside dark matter halos). Besides their importance for the cosmic baryon budget, the physical properties and chemical composition of the C/IGM offer valuable insight into galaxy evolution, since they supply galaxies with fresh gas and also act as a reservoir for their ejected, enriched gas \citep[the cosmic baryon cycle, e.g.][]{Putman.etal.2012,McQuinn16,Tumlinson.etal.2017}. Moreover, the distribution of neutral hydrogen (HI) in the high-$z$ IGM can be used to constrain cosmic reionization, structure formation, and the nature of dark matter through the Lyman-$\alpha$ (Ly$\alpha$) forest 
\citep[e.g.][]{Rauch98,Viel.etal.13,Lidz.Malloy.14,McQuinn16,Eilers.etal.18}.

\smallskip
Gas in the C/IGM is highly diffuse and difficult to directly observe. It has traditionally been traced using absorption line spectroscopy along lines of sight to distant QSOs or galaxies \citep[e.g.][]{Bergeron86,Hennawi.etal.06,Steidel.etal.10,Lehner.etal.22}. In recent years, the advent of new integral field unit (IFU) spectographs such as KCWI on Keck and MUSE on the VLT have enabled emission line studies of the CGM and IGM around galaxies at $z\gsim 3$ \citep{Steidel.etal.00,Cantalupo.etal.14,Martin.etal.14a,Martin.etal.14b,Umehata.etal.19}. Both emission and absorption line studies reveal that the gas in and around galaxy halos has a complex multiphase structure \citep{Tumlinson.etal.2017}. Surprisingly, large amounts of cold ($\sim 10^4 {\rm K}$) gas have been observed in the outskirts of galaxy halos, which cannot be in hydrostatic equilibrium with the halo gravitational potential. During \textit{cosmic noon}, at $z\sim (2-6)$ near the peak of cosmic galaxy formation, this cold gas is inferred to be denser than the ambient hot gas within which it is embedded by a factor $\chi\equiv \rho_{\rm c}/\rho_{\rm h}\sim 10^2-10^3$, and to be composed of tiny clouds with sizes of $l\sim N_{\rm H}/n_{\rm H}\lsim 50\pc$ \citep{Cantalupo.etal.14, Hennawi.etal.15, Borisova.etal.16}. The cold gas has order unity area covering fractions, $f_{\rm C}\sim O(1)$, and mass fractions, $f_{\rm M}\sim O(1)$ \citep{Pezzulli.Cantalupo.19}. However, its volume filling factor is tiny, $f_{\rm V}\sim f_{\rm M}/\chi\sim 10^{-3}$, making it extremely clumpy \citep{Cantalupo.etal.19} and its apparent abundance difficult to explain \citep{FG.Oh.2023}.

\smallskip
Recent theoretical advances have shed new light on these issues.  A major new insight \citep[][]{McCourt.etal.18} is that when the cooling time of a gas cloud is much less than its sound-crossing time such that it cannot cool isobarically, it does not cool isochorically as had been presumed \citep[][]{Field.65, Burkert.Lin.00}. Rather, the cloud `shatters' into many small fragments that lose sonic contact, causing them to contract independently and subsequently disperse, similar to a terrestrial fog \citep{McCourt.etal.18,Gronke.Oh.20b}. The typical size of the resulting cloudlets is expected to be of order the \textit{minimal cooling length}, $\ell_{\rm shatter} \sim {\rm min}(c_{\rm s} t_{\rm cool})$, with $c_{\rm s}$ and $t_{\rm cool}$ the sound speed and cooling time, and the minimal value obtained at $T \sim 10^4 {\rm K}$. For typical CGM conditions at $z\sim (2-3)$, this is $\gsim 10\pc$, consistent with inferred cloud sizes. This would explain the vastly different area covering and volume filling factors, since for $N$ droplets of size $l$ dispersed throughout a system of size $R$, 
$f_{\rm C}/f_{\rm V}\sim R/l\gg 1$ \citep{FG.Oh.2023}. A `fog' can also explain a host of additional observations in the CGM, High Velocity Clouds, quasar Broad Line Regions, and the interstellar medium \citep{Gronke.etal.17,McCourt.etal.18,Stanimirovic.Zweibel.2018,FG.Oh.2023,Sameer.etal.24}.

\smallskip
Despite the many appealing features of the shattering model, numerous puzzles remain. It is unclear under what conditions a large cooling cloud will shatter, with some suggesting this depends on the final overdensity between the cold and hot gas \citep{Gronke.Oh.20b} and others that it depends on the thermal stability conditions in the initial cloud \citep{Waters.Proga.19a,Das.etal.21}. Even when clouds do shatter in 3D simulations, they do not appear to do so hierarchically as was initially proposed by \citet{McCourt.etal.18}. Rather, if the initial cloud is large ($r_{\rm cl}\gg \ell_{\rm shater}$) and highly non-linear ($\delta \rho/\rho \gg 1$) when it loses sonic contact it initially cools isochorically, then becomes strongly compressed by its surroundings until its central pressure overshoots, and finally it explodes into many small fragments \citep{Gronke.Oh.20b}. This process is sometimes referred to as `splattering' \citep{Waters.Proga.19a,Waters.Proga.23}, and seems to be due to vorticities generated by Richtmyer-Meshkov instabilities \citep[RMI;][]{Richtmyer1960TaylorII, Meshkov1969InstabilityOT, Zhou.2017a,Zhou.2017b}, which explains why it is not seen in 1D simulations \citep{Waters.Proga.19a,Das.etal.21}. Additional fragmentation mechanisms have been proposed, such as shredding by collisions with larger fragments \citep{Jennings.Li.21}, or rapid rotation of clumps \citep[`splintering';][]{Farber.Gronke.23}. We will hereafter use the term `shattering' to refer to the general process of cold gas fragmentation into numerous small clouds, but note that the process may be very different than that originally proposed by \citet{McCourt.etal.18}.

\smallskip
Even if a cloud initially shatters, the resulting cloudlets may recoagulate to form larger clouds. 
\citet{Gronke.Oh.20b} found that a cloud remained shattered if its \textit{final} overdensity 
\be
\chif\equiv\rho_\mathrm{s,f}/\rho_\mathrm{bg},
\ee
where $\rho_\mathrm{s,f}$\footnote{The subscript 's' refers to 'spheres', 'streams', or 'sheets'.} is the cloud density at the temperature floor and $\rho_\mathrm{bg}$ is the background density, was above a critical value of $\chi_\mathrm{f,crit}\sim 300$ with a weak dependence on cloud size of $(r_{\rm cl}/\ell_{\rm shatter})^{1/6}$. The origin of this threshold remains unclear. 
Several coagulation mechanisms have been discussed in the literature \citep[see summary in][]{FG.Oh.2023}. These include direct collisions, similar to dust grain growth in protoplanetary disks, and coagulation due to the advective flow generated by hot gas condensing onto a cold cloud. In the latter, the inflow velocity can be set by thermal conduction (or numerical diffusion; \citealp{Elphick.etal.91,Elphick.etal.92,Koyama.Inutsuka.04,Waters.Proga.19b}) or, more relevant for our purposes, by the so-called mixing velocities through turbulent mixing layers \citep{Gronke.Oh.23}. Of particular interest is that in the case of turbulent mixing layers, the coagulation can be modeled as an effective force between two clouds, which scales as $r^{-n}$, with $r$ the distance between the clouds and $n=2$, $1$, or $0$ for clouds in $3$, $2$, or $1$ dimensions, similar to the gravitational force \citep{Gronke.Oh.23}.

\smallskip
Studying this process in a cosmological context is extremely challenging, since even in the most advanced cosmological simulations employing novel methods to enhance the resolution in the CGM \citep{vandeVoort.etal.19, Hummels.etal.19, Suresh.etal.19, Peeples.etal.19} or the IGM \citep{Mandelker.etal.19b,Mandelker.etal.21}, cell sizes are still significantly larger than $\ell_{\rm shatter}$. As a result, the amount and extent of cold, dense, low-ionization gas in the CGM increases with resolution and is not converged. Furthermore, different simulations disagree on the magnitude of the effect of enhanced CGM refinement, at least in part due to the different subgrid models employed for galaxy formation physics, such as stellar and AGN feedback, galactic winds, and gas photoheating and photoionization. This has obscured the details of why higher resolution leads to more cold gas in the CGM, where this cold gas comes from, and what a meaningful convergence criterion for the formation of multiphase gas might be. Numerically, it seems that the formation of multiphase gas requires resolving the cooling length at $T\sim 10^5\Kdegree$ where isochoric cooling modes are stable \citep{Das.etal.21,Mandelker.etal.21}, though it is unclear how generic this is and other convergence criteria have been proposed \citep{Hummels.etal.19,Gronke.etal.22}. 

\smallskip
For these reasons, shattering and coagulation are most commonly studied using idealized simulations and numerical models. In the vast majority of cases, such models assume a spherical or quasi-spherical cloud or distribution of clouds. However, many systems in the C/IGM where these processes are important are filamentary (cylindrical) or planar in nature. Modern cosmological simulations reveal strong accretion shocks around intergalactic filaments \citep{Ramsoy.etal.21,Lu.etal.24} and sheets \citep{Mandelker.etal.19b, Mandelker.etal.21} that make up the `cosmic web' of matter on Mpc-scales, similar to virial accretion shocks around massive dark-matter halos \citep[][]{Rees.Ostriker.77, White.Rees.78, Birnboim.Dekel.03, Stern.etal.20}. The post-shock gas in the high-$z$ cosmic-web can shatter \citep{Mandelker.etal.19b,Mandelker.etal.21,Lu.etal.24}, with the resulting cold cloudlets in intergalactic sheets potentially explaining observations of extremely metal-poor Lyman-limit systems \citep[LLSs, e.g.][]{Robert.etal.19, Lehner.etal.22}. The small-scale structure of cosmic filaments and sheets have additional important consequences for a wide variety of issues, including how gas is accreted onto DM halos, interpretations of Ly$\alpha$ forest statistics, measured dispersions of FRBs, radiative transfer and the self-shielding of photoionized gas, and the overall cosmic census of baryons \citep[see discussion in][]{Mandelker.etal.21}. On smaller scales, gas accretion onto massive galaxies at high-$z$ is thought to be dominated by cold streams flowing along cosmic-web filaments, which penetrate the halo virial shock and flow freely towards the central galaxy \citep[e.g.][]{Dekel.Birnboim.06,Dekel.etal.09a}. The interaction of these cold streams with the hot CGM can lead to shattering and the formation of small-scale cold clouds \citep{Mandelker.etal.20a,Lu.etal.24}. Finally, Filamentary structures are expected around both inflowing and outflowing gas clouds in the CGM due to cloud-wind interactions \citep[e.g.,][]{Gronke.Oh.18,Gronke.Oh.20,Banda.etal.16,Banda.etal.19,Li.etal.20,Sparre.etal.20,Tan.etal.23,Tan.Fielding.24}. 

\smallskip
An additional complication arises due to the metallicity of the gas, which affects the cooling rates and therefore $\ell_{\rm shatter}$ and the resulting cloud sizes, as well as the strength of coagulation forces. While most studies in the literature assume solar metallicity gas \citep[though see][]{Das.etal.21}, gas in the high-$z$ cosmic web is expected to have much lower metallicity, $Z\sim(10^{-4}-0.1)Z_{\odot}$ \citep{Mandelker.etal.21}. Similarly, the presence of a UV background (UVB) is typically not included in studies of shattering although it too can influence cooling rates and cloud sizes. 

\smallskip
In this paper, we explore the effects of geometry and metallicity on shattering and coagulation using idealized 3D simulations of spherical clouds, cylindrical filaments, and planar sheets with metallicity values in the range $(0.03-1.0)Z_{\odot}$. In Section~\ref{sec:methods}, we introduce our numerical tools and simulation methods. In Section~\ref{sec:geometry}, we compare the evolution of shattering and coagulation among planar, cylindrical, and spherical geometries at solar metallicity. In Section~\ref{sec:metallicity}, we extend the cylindrical geometry to lower metallicity and include a UVB. In \se{sizes} we discuss the size distribution of clumps formed through shattering. We present a model for the shattering criteria in both streams and spheres in \se{shatter}, and tentatively apply this to the case of cold streams penetrating the halo virial shock in \se{application}. In \se{caveats} we address caveats casued by additional physical processes not included in our analysis, before concluding in \se{summary}. 

\section{Numerical methods}
\label{sec:methods}
In this section we describe the details of our simulation setup. 

\subsection{Simulation Code}
We use Eulerian AMR code \texttt{Ramses} \citep{2002A&A...385..337T} to perform 3D idealized numerical simulations. We adopt the multi-dimensional MonCen limiter \citep{1977JCoPh..23..276V} for the piecewise linear reconstruction, the Harten-Lax-van Leer-Contact (HLLC) approximate Riemann hydro solver \citep{1994ShWav...4...25T} for calculating fluxes at cell interfaces, and the MUSCL-Hanchock scheme \citep{doi:10.1137/0905001} for the Godunov integrator. The adiabatic index is $\gamma=5/3$, and the Courant factor is $0.6$.

\subsection{Radiative cooling}
We utilize the standard \texttt{Ramses} cooling module, which accounts for atomic and fine-structure cooling for our assumed metallicity values. When comparing the three geometries in section~\ref{sec:geometry}, we assume solar metallicity for both cold and hot phases and set a temperature floor at $T_\mathrm{floor}=1.68\times10^{4}\,$K. When focusing on cold streams in section~\ref{sec:metallicity}, we follow \citet{Mandelker.etal.20a} and assume metallicity values of $Z_\mathrm{bg}=0.1\,Z_\odot$ for the background CGM and $Z_\mathrm{s}=0.03\,Z_\odot$ for the streams, and include photoheating and photoionization from a $z=2$ \citet{Haardt.Madau.96} UVB. At our assumed densities, the equilibrium temperature between the radiative cooling and the UV heating is roughly at $T_\mathrm{floor}$ which we adopt in \se{geometry}. 
In all cases we shut off cooling above $0.8\,T_\mathrm{bg}$ to prevent the cooling of background \citep[e.g.][]{Gronke.Oh.18,Mandelker.etal.20a}.

\subsection{Clump finder}
To quantify the degree of cold gas fragmentation, we utilize the built-in \texttt{Ramses} clump finder module \texttt{PHEW} \citep{2015ComAC...2....5B}, which is a parallel segmentation algorithm for 3D AMR datasets. It detects connected regions above a certain density threshold based on a `watershed' segmentation of the computational volume into dense regions, followed by a merging of the segmented patches based on the saddle point density. Basically, each clump is centered on a local density peak (whose density is above the density threshold) and includes all surrounding gas with densities above both the density threshold and any local saddle points or density minima. Two neighbouring peaks separated by a saddle-point whose density is above the secondary saddle density threshold are then `merged' into a single clump. If the saddle density is below this secondary threshold, the two peaks represent two distinct clumps. 
We choose the clump density threshold to be the initial density of warm gas (see \se{ICs} below), while the saddle density threshold is the geometric mean of the initial warm gas density and the final cold gas density.


\subsection{Grid structure and boundary conditions}
\label{sec:BCs}

The coordinates can be generalized by $(x_1,x_2,x_3)$, which represents 
$(x,y,z)$, $(r,\phi,z)$, and $(r,\theta,\phi)$ in sheets, streams, and spheres, respectively. 
The sheets are 
aligned with the $yz$ plane while the stream axis is aligned with the $z$-axis. Sheets are initially confined to the region $|x|\le r_\mathrm{s,i}$, while streams and spheres are initially confined to $r\le r_\mathrm{s,i}$. Here $r_\mathrm{s,i}$ represents the initial sheet half-thickness, cylindrical radius, and spherical radius for the three different geometries, with cold gas always occupying the region $|x_1|\le r_\mathrm{s,i}$. 

\smallskip
The simulation region is a cubic box with size $L=32\,r_\mathrm{s,i}$. 
We use a statically refined mesh with higher resolution around the cold gas. In our fiducial setup, the maximal refinement level is 10 corresponding to a minimal cell size $\Delta=L/1024=r_{\rm s,i}/32$ valid in the region $|x_1|<3\,r_\mathrm{s,i}$. The cell size increases by a factor of 2 at $|x_1|=[3,6,9,12]r_{\rm s,i}$, reaching a maximal value of $r_{\rm s,i}/2$ at $|x_1|>12\,r_\mathrm{s,i}$. 

\smallskip
We adopt outflow conditions for all boundaries, such that the gradients of all hydrodynamic variables are set to zero. We note that while periodic boundary conditions along the stream axis and within the plane of the sheet would have been preferable to model the idealized cases of infinitely long streams and sheets, for technical reasons to do with our clump finder we were forced to adopt the same boundary conditions on all six boundaries. We opted to implement outflow boundary conditions everywhere as these are necessary to allow correct entrainment flows to develop perpendicular to the sheet plane and the stream axis, which are necessary for properly modeling coagulation. While this has no impact on our analysis of spheres, we find that streams contract along their axis and sheets within the plane due to coagulation along these axes induced by entrainment flows that develop after the initial shattering. 
To avoid any potential effects of these boundary conditions on our analysis in streams and sheets, we restrict our analysis of these geometries to a narrower box, excluding gas within $10\,r_{\rm s,i}$ of the boundaries along the stream axis $(x_3)$ and within the sheet plane $(x2, x3)$. 
We find this narrower box to be unaffected by this contraction over the run time of our simulations, $\sim (10-20)t_{\rm sc}$, where $t_{\rm sc}\equiv r_{\rm s,i}/c_{\rm s,c}$ is the sound crossing time of the initial cloud radius at the sound speed of cold gas, $c_{\rm s,c}$, with a temperature of $T_{\rm c}\sim T_{\rm floor}\gsim 10^4{\rm K}$.

\begin{table*}
   \centering
   \caption{Summary of simulations analysed throughout this work. From left to right we list the initial cloud geometry (sheet, stream, or sphere); the ratio of the equilibrium cloud density to its initial density, $\eta\equiv \rho_{\rm s,f}/\rho_{\rm s,i}$; the initial density contrast between the cloud and the background, $\chi_{\rm i}\equiv \rho_{\rm s,i}/\rho_{\rm bg}$; the final density contrast between the cloud and the background, $\chi_{\rm f}\equiv \rho_{\rm s,f}/\rho_{\rm bg}=\eta\chi_{\rm i}$; the initial cloud radius, $r_{\rm s,i}$, in $\kpc$; the cloud radius where it loses sonic contact, $r^*$, in $\kpc$; the final equilibrium cloud radius, $r_{\rm s,f}$, in $\kpc$; the shattering lengthscale at the initial cloud metallicity, $\ell_{\rm shatter}\equiv{\rm min}(c_{\rm s}t_{\rm cool})$, in $\pc$; the initial cloud metallicity, $Z_{\rm s}$, in solar units; the background metallicity, $Z_{\rm bg}$, in solar units; the number of cells per initial cloud radius, $r_{\rm s,i}/\Delta$; whether or not a \citet{Haardt.Madau.96} UVB is assumed; whether or not the end result is a shattered cloud; and the section where the simulation is discussed.}
   \begin{tabular*}{0.9\textwidth}{cccccccccccccc}
       \hline
       Geometry & $\eta$ & $\chi_{\rm i}$ & $\chi_{\rm f}$ & $r_{\rm s,i}$ & $r^{*}$  & $r_{\rm s,f}$ & $\ell_{\rm shatter}$  & $Z_{\rm s}$ & $Z_{\rm b}$ & $r_\mathrm{s,i}/\Delta$ & UVB & Shattering & Section\\
                 &                  &                 &               &   [kpc]             &   [kpc]     & [kpc]               &   [pc]                 & [$Z_{\sun}$]  & [$Z_{\sun}$]  &         &       &            &             \\
       \hline
       Sheet  &10 &  10 &  100 & 3 & 3 & 0.30 & 3.7 & 1.0 & 1.0 & 32 & Off & False & \ref{sec:geometry}\\
       Sheet  &10 &  20 &  200 & 3 & 3 & 0.30 & 3.7 & 1.0 & 1.0 & 32 & Off & False & \ref{sec:geometry}\\
       Sheet  &10 &  40 &  400 & 3 & 3 & 0.30 & 3.7 & 1.0 & 1.0 & 32 & Off & False & \ref{sec:geometry}\\
       Sheet  &10 &  60 &  600 & 3 & 3 & 0.30 & 3.7 & 1.0 & 1.0 & 32 & Off & False & \ref{sec:geometry}\\
       Sheet  &10 & 100 & 1000 & 3 & 3 & 0.30 & 3.7 & 1.0 & 1.0 & 32 & Off & False & \ref{sec:geometry}\\
       \hline
       Stream &10 &  10 &  100 & 3 & 3    & 0.95 & 3.7 & 1.0 & 1.0 & 32 & Off & False & \ref{sec:geometry}, \ref{sec:metallicity}, \ref{sec:shatter}\\
       Stream &10 &  20 &  200 & 3 & 3    & 0.95 & 3.7 & 1.0 & 1.0 & 32 & Off & True/Borderline  & \ref{sec:geometry}, \ref{sec:metallicity}, \ref{sec:shatter}\\
       Stream &10 &  40 &  400 & 3 & 3    & 0.95 & 3.7 & 1.0 & 1.0 & 32 & Off & True  & \ref{sec:geometry}, \ref{sec:metallicity}, \ref{sec:shatter}\\
       Stream &10 &  60 &  600 & 3 & 3    & 0.95 & 3.7 & 1.0 & 1.0 & 32 & Off & True  & \ref{sec:geometry}, \ref{sec:metallicity}, \ref{sec:shatter}\\
       Stream &10 & 100 & 1000 & 3 & 3    & 0.95 & 3.7 & 1.0 & 1.0 & 32 & Off & True  & \ref{sec:geometry}, \ref{sec:metallicity}, \ref{sec:shatter}\\
       Stream &5  &  80 &  400 & 3 & 3    & 1.34 & 3.7 & 1.0 & 1.0 & 32 & Off & True  & \ref{sec:shatter}\\
       Stream &30 &  10 &  300 & 3 & 2.66 & 0.55 & 3.7 & 1.0 & 1.0 & 32 & Off & True  & \ref{sec:shatter}\\
       Stream &10 &  10 &  100 & 30& 30   & 9.49 & 3.7 & 1.0 & 1.0 & 32 & Off & False  & \ref{sec:shatter}\\
       Stream &40 &  10 &  400 & 30& 30   & 4.74 & 3.7 & 1.0 & 1.0 & 32 & Off & True  & \ref{sec:shatter}\\
       \hline
       Sphere &10 &  10  &  100 & 3 & 3 & 1.39 & 3.7 & 1.0 & 1.0 & 32 & Off & False  & \ref{sec:geometry}, \ref{sec:shatter}\\
       Sphere &10 &  20  &  200 & 3 & 3 & 1.39 & 3.7 & 1.0 & 1.0 & 32 & Off & True  & \ref{sec:geometry}, \ref{sec:shatter}\\
       Sphere &10 &  40  &  400 & 3 & 3 & 1.39 & 3.7 & 1.0 & 1.0 & 32 & Off & True  & \ref{sec:geometry}, \ref{sec:shatter}\\
       Sphere &10 &  60  &  600 & 3 & 3 & 1.39 & 3.7 & 1.0 & 1.0 & 32 & Off & True  & \ref{sec:geometry}, \ref{sec:shatter}\\
       Sphere &10 &  100 & 1000 & 3 & 3 & 1.39 & 3.7 & 1.0 & 1.0 & 32 & Off & True  & \ref{sec:geometry}, \ref{sec:shatter}\\
       Sphere &30 &  6   &  180 & 3 & 3 & 0.97 & 3.7 & 1.0 & 1.0 & 32 & Off & True  & \ref{sec:shatter}\\
       Sphere &30 &  10  &  300 & 3 & 3 & 0.97 & 3.7 & 1.0 & 1.0 & 32 & Off & True  & \ref{sec:shatter}\\
       Sphere &10 &  11  &  110 & 3 & 3 & 1.39 & 3.7 & 1.0 & 1.0 & 32 & Off & False & \ref{sec:shatter}\\
       Sphere &10 &  12  &  120 & 3 & 3 & 1.39 & 3.7 & 1.0 & 1.0 & 32 & Off & True  & \ref{sec:shatter}\\
       Sphere &10 &  13  &  130 & 3 & 3 & 1.39 & 3.7 & 1.0 & 1.0 & 32 & Off & True  & \ref{sec:shatter}\\
       Sphere &10 &  21  &  210 & 30& 30& 13.92& 3.7 & 1.0 & 1.0 & 32 & Off & False & \ref{sec:shatter}\\
       Sphere &10 &  22  &  220 & 30& 30& 13.92& 3.7 & 1.0 & 1.0 & 32 & Off & False & \ref{sec:shatter}\\
       Sphere &10 &  23  &  230 & 30& 30& 13.92& 3.7 & 1.0 & 1.0 & 32 & Off & True  & \ref{sec:shatter}\\
       Sphere &10 &  24  &  240 & 90& 90& 41.77& 3.7 & 1.0 & 1.0 & 32 & Off & False & \ref{sec:shatter}\\
       Sphere &10 &  26  &  260 & 90& 90& 41.77& 3.7 & 1.0 & 1.0 & 32 & Off & False & \ref{sec:shatter}\\
       Sphere &10 &  28  &  280 & 90& 90& 41.77& 3.7 & 1.0 & 1.0 & 32 & Off & False & \ref{sec:shatter}\\
       Sphere &10 &  30  &  300 & 90& 90& 41.77& 3.7 & 1.0 & 1.0 & 32 & Off & True  & \ref{sec:shatter}\\
       \hline \hline
       Stream &10 &   3 &   30 & 3 & 2.44 & 0.95 & 293 & 0.03 & 0.1 & 32 & On & False & \ref{sec:metallicity}, \ref{sec:sizes}, \ref{sec:shatter}\\
       Stream &10 &   6 &   60 & 3 & 2.44 & 0.95 & 293 & 0.03 & 0.1 & 32 & On & True/Borderline  & \ref{sec:metallicity}, \ref{sec:sizes}, \ref{sec:shatter}\\
       Stream &10 &  10 &  100 & 3 & 2.44 & 0.95 & 293 & 0.03 & 0.1 & 32 & On & True  & \ref{sec:metallicity}, \ref{sec:sizes}, \ref{sec:shatter}\\
       Stream &10 &  40 &  400 & 3 & 2.44 & 0.95 & 293 & 0.03 & 0.1 & 32 & On & True  & \ref{sec:metallicity}, \ref{sec:sizes}, \ref{sec:shatter}\\
       Stream &10 & 100 & 1000 & 3 & 2.44 & 0.95 & 293 & 0.03 & 0.1 & 32 & On & True  & \ref{sec:metallicity}, \ref{sec:sizes}, \ref{sec:shatter}\\
       Stream &10 &  10 &  100 & 1 & 0.49 & 0.32 & 293 & 0.03 & 0.1 & 32 & On & True  & \ref{sec:sizes}, \ref{sec:shatter}\\
       Stream &10 &  10 &  100 & 1 & 0.49 & 0.32 & 293 & 0.03 & 0.1 & 64 & On & True  & \ref{sec:sizes}\\
       Stream &10 &  10 &  100 & 1 & 0.49 & 0.32 & 293 & 0.03 & 0.1 & 128 & On & True  & \ref{sec:sizes}\\
       Stream &10 &   6 &  60  & 30& 30   & 9.49 & 293 & 0.03 & 0.1 & 32 & On & False & \ref{sec:sizes}, \ref{sec:shatter}\\
       Stream &10 &  10 &  100 & 30& 30   & 9.49 & 293 & 0.03 & 0.1 & 32 & On & False & \ref{sec:sizes}, \ref{sec:shatter}\\
       Stream &10 &  20 &  200 & 30& 30   & 9.49 & 293 & 0.03 & 0.1 & 32 & On & True  & \ref{sec:sizes}, \ref{sec:shatter}\\
       Stream &10 &  100 &  1000 & 30& 30   & 9.49 & 293 & 0.03 & 0.1 & 32 & On & True  & \ref{sec:sizes}, \ref{sec:shatter}\\
       \hline
   \end{tabular*}
   \label{tab:sim_table}
\end{table*}


\subsection{Initial conditions}
\label{sec:ICs}

We initialize the simulations with a static, warm component (sheet/stream/sphere) of density $\rho_\mathrm{s,i}$ in pressure equilibrium with a static, hot background of density $\rho_\mathrm{bg}$. The initial overdensity is thus $\chi_\mathrm{i}\equiv \rho_\mathrm{s,i}/\rho_\mathrm{bg}$. The temperature of the warm gas, $T_\mathrm{s,i}$, is lower than that of the hot background, $T_\mathrm{bg}$, but higher than the temperature floor $T_\mathrm{floor}$. As the warm component cools towards thermal equilibrium at $T_\mathrm{floor}$, a pressure contrast is generated between the cold gas and the hot background. We define  
$\eta\equiv \rho_\mathrm{s,f}/\rho_\mathrm{s,i}=(T_\mathrm{s,i}/\mu_\mathrm{s,i})/(T_\mathrm{floor}/\mu_\mathrm{s,f})$, where $\rho_\mathrm{s,f}$ denotes the final density of cold gas once pressure equilibrium has been reestablished at $T_\mathrm{floor}$, and $\mu_{\rm s,i}$ and $\mu_{\rm s,f}$ are the mean molecular weight in the initial warm and final cold gas, respectively. Consequently the final overdensity between the cold and hot gas is $\chi_\mathrm{f}\equiv \rho_\mathrm{s,f}/\rho_\mathrm{bg}=\eta\chi_\mathrm{i}$, which is expected to be the key parameter determining whether cold gas shatters or coagulates \citep{Gronke.Oh.20b}. 
We adopt $\eta=10$ for most simulations presented in the main text unless otherwise noted, and vary $\chi_\mathrm{f}$ by changing $\chi_\mathrm{i}$. 

\smallskip
\tab{sim_table} summarises the parameters of the simulations analysed throughout this work, and the sections where they are discussed.

\subsection{Perturbations}
\label{sec:shape}

\smallskip
We introduce density perturbations in the initial warm gas component. In units of the initial mean density of warm gas, the density follows a Gaussian distribution with $(\mu,\sigma)=(1,0.01)$, truncated at $3\,\sigma$, similar to \citet{Gronke.Oh.20b}. We further introduce shape perturbations at the interface between the warm and hot components, as described below. Such interface perturbations have been shown to suppress the carbuncle instability \citep{1994IJNMF..18..555Q}, which is a numerical instability affecting strong shock-fronts on the grid scale in multidimensional simulations, by misaligning the interface and the grid, and by generating additional vorticity and turbulence when the shock overtakes the interface. 

\smallskip
For the stream geometry, we adopt the same interface perturbations as implemented in \citet{Mandelker.etal.19a, Mandelker.etal.20a}, 
\begin{equation}
r=r_\mathrm{s,i}\left[1+\sqrt{\frac{2}{N_\mathrm{pert,str}}}\delta r \sum_{j=1}^{N_\mathrm{pert,str}}\mathrm{cos}(k_jz+m_j\varphi+\phi_j)\right].
\end{equation}
Here $\delta r=0.1\,r_\mathrm{s,i}$ is the rms amplitude of perturbations, and $k_j=2\pi n_j$ where $n_j$ is an integer corresponding to a wavelength $\lambda_j=1/n_j$. We include all wavenumbers in the range $n_j=16-64$, corresponding to wavelengths in the range from $r_\mathrm{s,i}/2$ to $2\,r_\mathrm{s,i}$. $m_j$ is the azimuthal mode number of the perturbation, with $m=0$ corresponding to axisymmetric modes, $m=1$ to helical modes, and $m\ge 2$ to high-order fluting modes \citep{Mandelker.etal.16,Mandelker.etal.19a}. For each longitudinal wavenumber, $n_j$, we include two azimuthal modes, $m_j=0,1$. This results in a total of $N_\mathrm{pert,str}=2\times 49=98$ modes. Each mode has a random phase, $\phi_j \in [0,2\pi)$.

\smallskip
For sheets, we use analogous interface perturbations 
\begin{equation}
r=r_\mathrm{s,i}\left[1+\sqrt{\frac{2}{N_\mathrm{pert,sht}}}\delta r \sum_{j=1}^{N_\mathrm{pert,sht}}\mathrm{cos}(k_{x,j}x+k_{y,j}y+\phi_{j})\right], 
\end{equation}
where $k_{x,j}=2\pi n_{x,j}$, $k_{y,j}=2\pi n_{y,j}$. $n_{x,j}$, $n_{y,j}$ are the wavenumbers along the $x$ and $y$ directions, chosen from 16 to 64 with an interval of 6, i.e. $[16,\,22,\,...,\,58,\,64]$. This corresponds a total of $N_\mathrm{pert,sht}=9\times9=81$ modes, each with a random phase $\phi_j \in [0,2\pi)$.

\smallskip
For spheres, we use spherical harmonics to perturb the interface: 
\begin{equation}
r=r_\mathrm{s,i}\left[1+\sqrt{\frac{4\pi}{N_\mathrm{pert,sph}}}\delta r \sum_{l=0}^{l_\mathrm{max}} \sum_{m=0}^{l} Y_l^{m}(\theta,\phi)\right],
\end{equation}
where $Y_l^{m}$ is the spherical harmonic given by 
\begin{equation}
Y_l^{m}(\theta,\phi) = \sqrt{\frac{2l+1}{4\pi}\frac{(l-m)!}{(l+m)!}}P_l^{m}(\mathrm{cos\theta})\mathrm{cos}(m\phi),
\end{equation}
with $P_l^{m}$ the associated Legendre polynomial. We choose $l_{\rm max}=13$, yielding $N_\mathrm{pert,sph}=1/2\times 14\times 15=105$ modes. 

\section{Geometrical effects on shattering and coagulation}
\label{sec:geometry}

In this section we present our results comparing shattering and coagulation in the three geometries at solar metallicity. We begin in \se{shattering} by addressing the initial implosion and explosion phases of the shattering process. Then in \se{coagulation} we discuss the subsequent coagulation of the resulting cloudlets.

\subsection{Cold gas shattering}
\label{sec:shattering}

\subsubsection{The implosion and explosion of cold gas}
\label{sec:explosion}

\begin{figure*}
    \centering	\includegraphics[width=\textwidth]{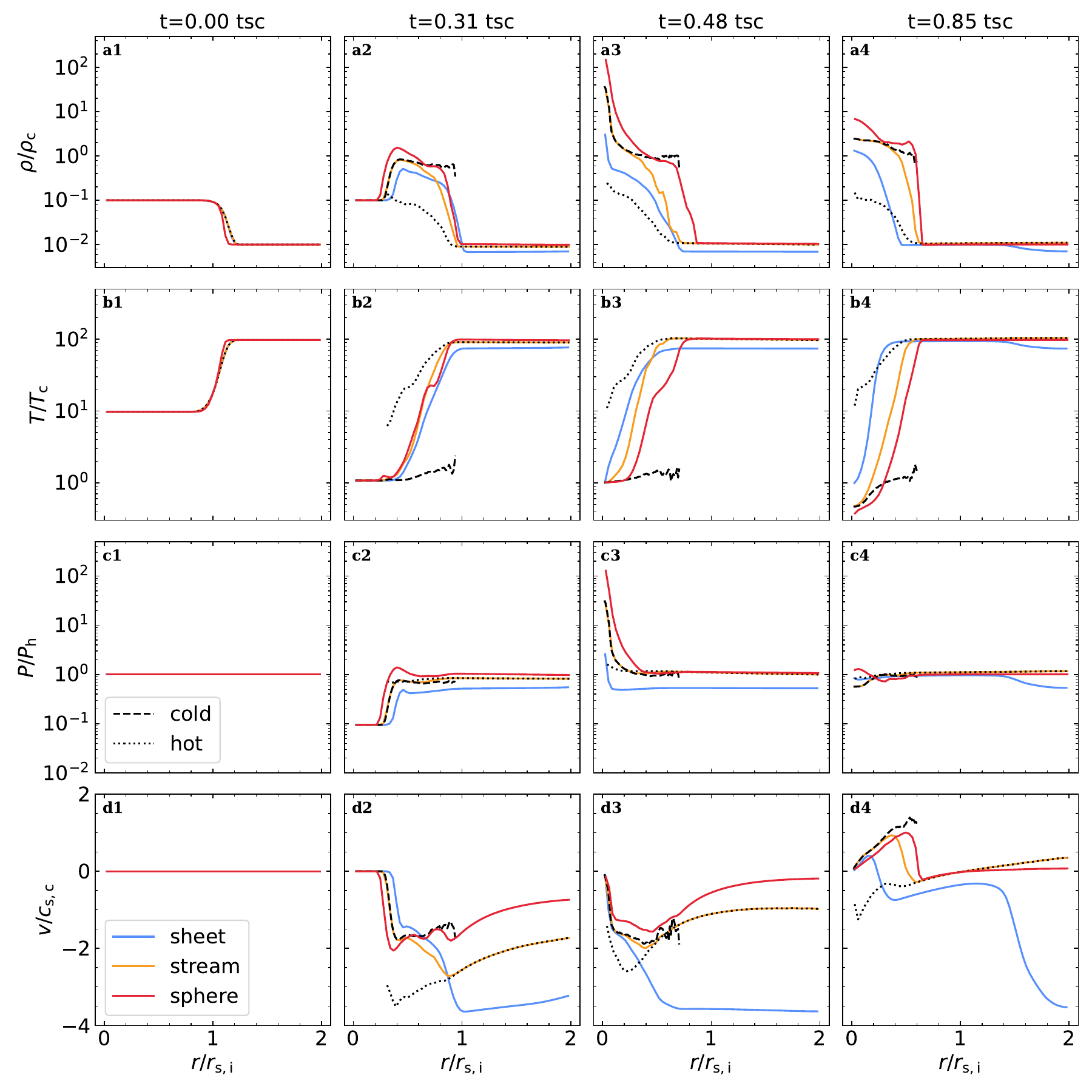}
   \caption{Radial profiles during the implosion and explosion of clouds. From top to bottom, we show radial profiles of density, temperature, pressure, and radial velocity, in simulations with $r_\mathrm{s,i}=3~\mathrm{kpc}$, $\eta=10$, $\chi_\mathrm{f}=100$ (top row of \tab{sim_table}). Different colour lines mark different geometries, sheets in blue, streams in orange, and spheres in red. Black dashed and dotted lines show respectively the cold ($T<10^5\,$K) and hot ($T>10^5\,$K) components of the stream case. The density and pressure profiles are volume-weighted while temperature and velocity are mass-averaged. The four columns represent four different stages of evolution. From left to right these are the initial conditions, the implosion, the shock collision and peak central pressure, and the end of the explosion phase when pressure equilibrium is reestablished and the cold-gas explosion velocity, $v_{\rm ex}$, is at its peak. The time of each phase in the sphere simulation is shown in each column in units of the cold gas sound crossing time of the intial cloud, $t_\mathrm{sc}\equiv r_\mathrm{s,i}/c_\mathrm{s,c}$. For streams and sheets, the time in the last two columns is shifted slightly to correspond to the peak central pressure and peak $v_{\rm ex}$, respectively. During the implosion phase, an isothermal shock propagates into the cloud more rapidly than the contact discontinuity between cold and hot gas, while a rarefaction wave propagates into the hot medium, adiabatically lowering the pressure there. The implosion shock is faster in streams than in sheets and faster still in spheres, as are the post-shock density and velocity. The peak in central density and pressure is only $\sim (2-3)$ times greater than the equilibrium values for sheets, but $1$ and $2$ orders of magnitude larger than that for streams spheres, respectively. Correspondingly, the peak explosion velocity for streams and spheres is $v_{\rm ex}\sim c_{\rm s,c}$ while it is only $\sim 0.4c_{\rm s,c}$ for sheets.}
   \label{fig:profiles}
\end{figure*}

We begin with a detailed physical description of the implosion of the initial cloud, triggered by a lack of pressure support due to cooling, and the subsequent explosion triggered by a reverse shock reflected off the cloud centre. While this process is interesting in its own right, the main outcome relevant for our discussion of coagulation which follows is the `explosion velocity', $v_{\rm ex}$, the characteristic velocity of the cold gas once the reverse shock reaches scales of order the final cloud radius. In Appendix~\ref{sec:imandex} we present a detailed mathematical discussion of these processes and a derivation of $v_{\rm ex}$ for sheets. In this section, we offer more general considerations valid for all three geometries, and demonstrate these using results from one simulation for each geometry (\fig{profiles}). 

\smallskip
Consider a warm gas cloud with cooling time, 
\be 
\label{eq:tcool}
t_{\rm cool} = \frac{kT}{(\gamma-1)n\Lambda},
\ee 
with $n$ the gas number density, $k$ the Boltzmann constant, $\gamma=5/3$ the adiabatic index of the gas, and $\Lambda$ the cooling function. The sound crossing time is 
\be 
\label{eq:tsc}
t_{\rm sc} = \frac{r_{\rm s}}{c_{\rm s}} = r_{\rm s}\left(\frac{\rho}{\gamma P}\right)^{1/2},
\ee 
with $P$ the gas pressure. Mass conservation during the collapse tells us that $r_{\rm s}\propto \rho^{-1/(n+1)}$ with $n=2$, $1$, and $0$ for spheres, streams, and sheets, respectively. Thus, $t_{\rm cool}/t_{\rm sc}\propto \rho^{-(3n+1)/(2n+2)}$, which decreases as the density rises. 
A cloud for which $t_{\rm cool}>t_{\rm sc}$ initially will cool isobarically, contracting and growing denser as it cools, until it reaches a radius $r^{*}$ where $t_{\rm cool}$ becomes shorter than $t_{\rm sc}$. At this stage, the cloud loses sonic contact and proceeds to cool isochorically \citep{Burkert.Lin.00,Gronke.Oh.20b,Waters.Proga.23}. From this point, the pressure within the cloud drops as it cools, driving it further away from equilibrium with the surrounding background. The resulting pressure gradient causes the cold gas to implode, with an isothermal shock (due to strong cooling) propagating inwards and a rarefaction wave outwards. 
This implosion decelerates and eventually reverses when the shock nears the centre and the central pressure becomes large compared to the background pressure, causing the cold gas to explode outwards. Eventually, the cold gas regains pressure equilibrium with its surroundings such that $\rho_\mathrm{s,f}T_\mathrm{floor}/\mu_\mathrm{s,f}=\rho_\mathrm{s,i}T_\mathrm{s,i}/\mu_\mathrm{s,i}$, or $\rho_\mathrm{s,f}=\eta\rho_\mathrm{s,i}$ (\se{ICs}). Assuming that most of the cold gas mass in this final state is in a single cloud, mass conservation yields $\rho_\mathrm{s,f}r_\mathrm{s,f}^{n+1}=\rho_\mathrm{s,i}r_\mathrm{s,i}^{n+1}$, with $r_\mathrm{s,f}$ the final radius of the cold cloud. We thus obtain $r_\mathrm{s,f}=r_\mathrm{s,i}\,\eta^{-1/(n+1)}$, showing that at a given $\eta$, the final contraction radius decreases from spheres to streams to sheets. 

\begin{figure}
    \centering	\includegraphics[width=\columnwidth]{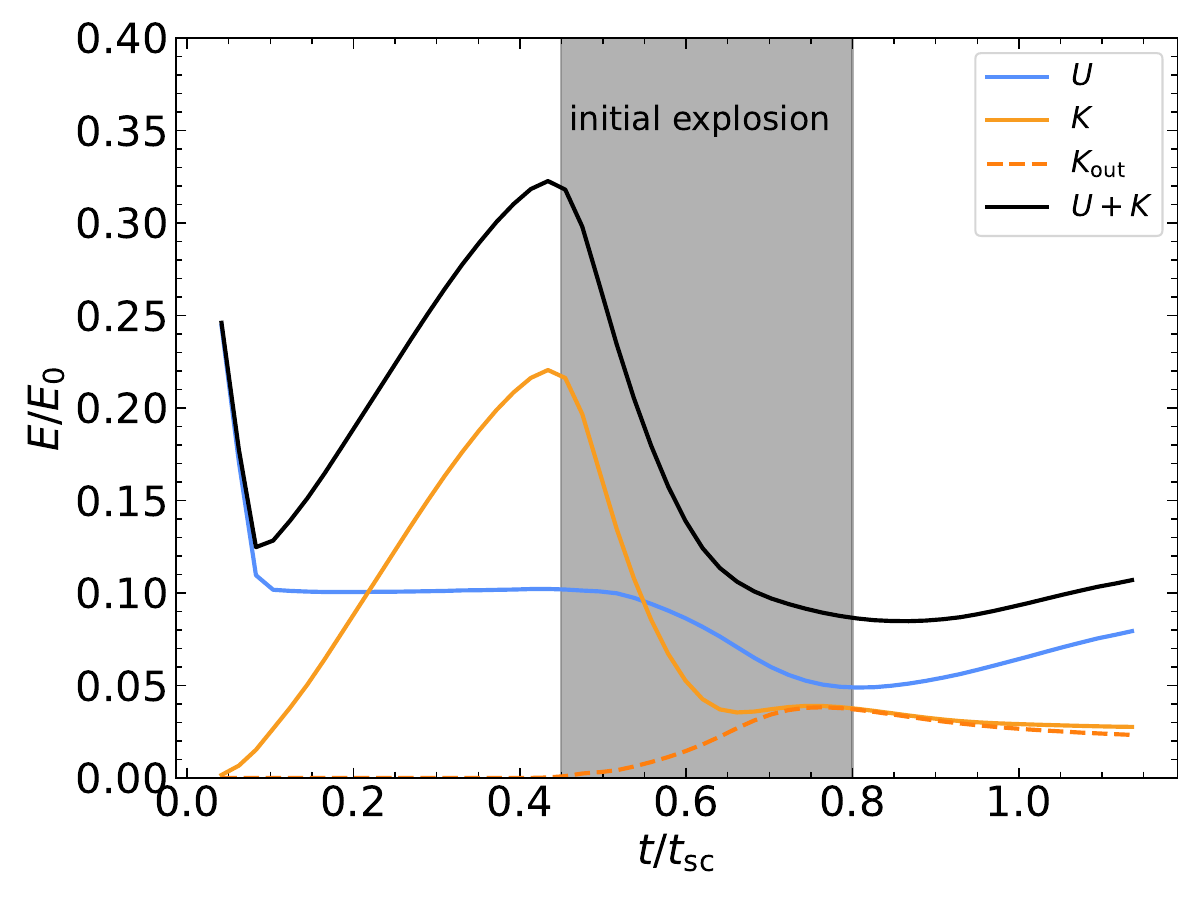}
   \caption{Energy evolution of the cloud during the implosion and explosion phases in the stream simulation with $\rsi=3\,$kpc, $\eta=10$, $\chif=100$. Blue, orange, and black lines show the thermal, kinetic, and total energy of the cold gas (defined as $T<10^5\,$K), respectively. The dashed orange line represents the kinetic energy of outflowing gas specifically. The time is normalized by the cold gas sound crossing time of the initial stream radius, $\tsc\equiv \rsi/c_\mathrm{s,c}$, and the energy is normalized by the initial thermal energy of the stream, $E_0$, with initial temperature $T_0\sim 2\times 10^5\,$K. After $t_\mathrm{cool}$, the internal energy drops to approximately $0.1\,E_0$ corresponding to $\eta=10$. It remains constant until the expansion phase, where roughly half of it is converted to outflowing kinetic energy. Following the explosion (gray region), the internal energy slowly increases again after the explosion velocity peaks. The total energy is not conserved during the expansion due to efficient radiative cooling in the post-reflected shock gas. However, most of the energy loss comes from the kinetic energy of gas which is still imploding, while the sum of the internal energy and the outflowing kinetic energy remains roughly constant.}
   \label{fig:energy}
\end{figure}
\smallskip
If the peak central pressure is sufficiently large compared to the background pressure (see \fig{profiles} panel c3, discussed below), a strong reflected shock propagates outwards from the cold to the hot gas, and a rarefaction wave propagates inwards causing the cold gas to expand. The post-shock cold gas is accelerated outward by the shock, and the peak cold-gas expansion velocity occurs when the central pressure subsides and pressure equilibrium between the cold and hot gas has been regained. Below, we provide three explanations for why this peak velocity, $v_\mathrm{ex}$, is of order the cold gas sound speed, $c_{\rm s,c}$.

\smallskip
We first consider energy conservation as the thermal energy after the implosion gets converted into kinetic energy of the expanding gas. The total internal energy after the implosion is 
\be 
\label{eq:Implosion_energy}
E=\frac{PV}{\gamma-1}=\frac{m_\mathrm{s,i}kT_\mathrm{c}}{\mu m_\mathrm{p}(\gamma-1)},
\ee 
where $m_\mathrm{s,i}$ is the initial mass in the cooling cloud, and $T_{\rm c}$ is the final equilibrium cloud temperature.\footnote{we hereafter use $T_{\rm c}$ and $T_{\rm floor}$ interchangeably, as in our simulations the two are nearly identical.} When 
the shock propagation time is shorter than the cooling time of post-shock gas, we can assume that most of this internal energy becomes the kinetic energy of cold gas, $E\sim 1/2m_\mathrm{s,i}v_\mathrm{ex}^{2}$. We thus obtain 
\be 
\label{eq:vex}
v_\mathrm{ex}\sim \left(\frac{2}{\gamma(\gamma-1)}\right)^{1/2}c_\mathrm{s,c} \sim 1.34 c_{\rm s,c}.
\ee 
Note that once thermal and pressure equilibrium have been reestablished at the very end of the shattering process, the internal energy of the cold gas is the same as it is after the implosion since the same mass of gas is at the same temperature, $T_{\rm c}$. It would thus seem that one cannot convert the internal energy in \equ{Implosion_energy} to kinetic energy. However, the peak pressure following the implosion initially causes the cloud to adiabatically expand, with the central temperature dropping to as low as $\lsim 0.3T_{\rm c}$ before rising back up to $T_{\rm c}$ due to mixing and compression of the hot gas. This drop in temperature seen in our simulations cannot be due to cooling, since we do not allow cooling below $T_{\rm c}$. While energy is not conserved throughout the implosion-explosion process due to strong cooling, the radiative losses primarily come at the expense of the kinetic energy of imploding gas while the internal energy of the post-implosion gas is converted to kinetic energy of exploding gas. It is during this adiabatic expansion phase that the cold gas is accelerated to $v_{\rm ex}$. The energy evolution of the cloud during the first $\tsc$ is illustrated in \fig{energy}.

\smallskip
An alternative way to see that the explosion velocity must be of order the cold gas sound speed is to consider the shock-jump conditions. 
For isothermal shocks, we know that $v_1 v_2 \sim c_{\rm s,c}^2$, where $v_1$ and $v_2$ are the pre- and post-shock velocities in the shock frame. $v_1$ must be several times larger than $c_{\rm s,c}$ due to the shock speed exceeding $c_{\rm s,c}$ and the negative velocity of the still imploding pre-shock cold gas. Thus, $v_2$ must remain small, indicating that the post-shock gas should have a velocity close to $c_{\rm s,c}$ in the lab frame.

\smallskip
Yet a third way to understand why $v_{\rm ex}\sim c_{\rm s,c}$ is as follows. One can think of the contracting cloud as a spring which is compressed and then released. Thus, from energy conservation, the explosion velocity cannot exceed the implosion velocity (in general, it will be smaller, because of radiative losses). The implosion velocity, or velocity of the cloud-crushing shock $v_{\rm s}$, is given by $\rho_{\rm c} v_{\rm s}^2 \sim \delta P \sim P \sim \rho_{\rm h} c_{\rm s,h}^2$, or $v_{\rm s} \sim c_{\rm s,h}/\sqrt{\chi} \sim c_{\rm s,c}$ \citep{Klein.etal.94}. Hence, the cold gas velocity is a characteristic expansion velocity. In detail, the implosion velocity is somewhat larger than we have estimated (since the overdensity during contraction is less than $\rho_{\rm c}/\rho_{\rm h}$), and the expansion velocity is somewhat smaller (due to radiative losses). However, the estimate $v_{\rm expand} \sim c_{\rm s,c}$ is robust.

\smallskip
All these estimates suggests that $v_\mathrm{ex}\sim c_\mathrm{s,c}$ for all geometries, regardless of the initial conditions (see Appendix~\ref{sec:imandex} for a detailed derivation of $v_{\rm ex}$ for sheet geometries).

\smallskip
In \fig{profiles} we demonstrate the key features of the implosion and explosion processes in each of our three geometries. From top to bottom, we show radial profiles of density, temperature, pressure and radial velocity, taken from simulations with $r_\mathrm{s,i}=3\,\mathrm{kpc}$, $\eta=10$, and $\chi_\mathrm{f}=100$ (first row of \tab{sim_table}). The density and pressure profiles are volume-weighted averages within each radial bin, while the temperature and velocity profiles are mass-weighted. We show the profiles at four times, from left to right these are at the initial condition, during the implosion, at the central shock collision, and near the end of the explosion phase when pressure equilibrium has been reestablished and the explosion velocity has reached its peak value. Different colour lines mark the different geometries, while black dashed and dotted lines show results for cold and hot gas respectively (separated at $T=10^5\,{\rm K}$) in stream geometry. Separating cold and hot gas for spheres yields similar results as for streams and is not shown, while the results for sheets are presented in Appendinx~\ref{sec:imandex}.

\smallskip
Initially, the cold gas is in pressure equilibrium with the hot background, with a density contrast of $\chi_{\rm i}=10$, and the fluid velocity is zero everywhere. Note that the density and temperature profiles exhibit a smooth transition between the cloud and the background. This is due to the shape perturbations we include (\se{shape}) and not to any explicit smoothing or ramp function in the initial conditions. 

\smallskip
The initial cloud properties yield $t_\mathrm{cool}<t_\mathrm{sc}$, so the cold gas loses sonic contact as soon as it starts cooling, meaning $r^{*}=r_\mathrm{s,i}$. Consequently, the central pressure rapidly drops, within a cooling time, forming a pressure gradient between the cold cloud and the hot background. This results in an isothermal shock propagating inwards and a rarefaction wave outwards. The shock is visible in panel c2 of \fig{profiles} as a jump in pressure. Note that the contact discontinuity, where the gas temperature begins rising in panel b2, is outside the implosion shock, while cold gas both inside and outside the shock has $T\sim T_{\rm c}$. In sheets, the implosion shock propagates inwards with a roughly constant Mach number of $\sim 2$, reaching the centre in roughly $0.5\tsc$ (see Appendix~\ref{sec:imandex}). However, in streams and spheres the Mach number increases as the shock radius decreases due to geometrical effects \citep{guderley1942starke,Modelevsky.Sari.21}. The implosion shocks thus reach the centre faster in these geometries, as can be seen by the shock positions in panels a2 and c2. 

\smallskip
This also causes the density and velocity in the post-shock region, between the shock radius and the contact discontinuity, to increase from sheets to streams to spheres due to geometrical effects. At the contact discontinuity, hot gas mixes with cold gas through a turbulent mixing layer, causing an entrainment flow to develop in the hot gas outside the cloud. This is why the hot gas inflow velocity is roughly twice as large as that of the cold gas. The mass flux of the entrainment flow, ${\dot{M}} \propto \rho_{\rm h} v_{\rm h} r^{n}$, is constant, with $n=0$, $1$, and $2$ for sheets, streams, and spheres, respectively. Since the density in the hot gas is roughly constant with radius (panel a2), this implies that the hot gas velocity scales as $v_{\rm h}\propto r^{-n}$, which is broadly consistent with the hot gas velocities seen in panel d2, which increase in magnitude from spheres to streams to sheets. Finally, we note that the density, temperature, and pressure in the hot gas at $r>r_{\rm s,i}$ all decrease during the implosion phase, due to the outward propagating rarefaction wave. 

\smallskip
At $t\lsim 0.5\,t_\mathrm{sc}$ the implosion shock reaches the cloud centre. As this occurs, the central density and pressure reach very large values, though the central temperature remains at $T_{\rm c}$. We note that the boost in central pressure and density is much larger in streams compared to sheets, and much larger still in spheres (panels a3 and c3). Formally, one can show that for self-similar collapse the central pressure in spheres and streams diverges \citep{guderley1942starke}, while it reaches a finite value in sheets   \citep[][see also Appendix~\ref{sec:imandex}]{toro2013riemann}. In practice, however, the finite resolution of our simulations, or any additional physical length scale such as the viscous length scale, prevent the central pressure from actually diverging. 

\smallskip
The right-most column shows the situation once pressure equilibrium between the hot and cold phases has been reestablished, and the outward velocity of the post-shock cold-gas, $v_{\rm ex}$, has reach its peak value. In both spheres and streams we find $v_{\rm ex}\sim c_{\rm s,c}$ as expected from \equ{vex} (panel d4). However, in sheets we find $v_{\rm ex}\sim 0.4\,c_{\rm s,c}$, due to the relatively small peak central pressure in sheets ($\sim 3$ times larger than the equilibrium pressure) compared to streams and spheres ($\sim 30-200$ times larger than the equilibrium pressure). This implies that only a small fraction of the peak internal energy in sheets goes into kinetic energy before the system regains pressure balance, leading to an explosion velocity smaller than $c_{\rm s,c}$. At the same time, the central temperature in spheres and streams is $\sim 0.3\,T_{\rm c}$ due to the adiabatic expansion phase described above, while it is $\sim T_{\rm c}$ for sheets implying no adiabatic expansion of the cold phase (panel b4). As discussed in Appendix \ref{subsec:mixing_2}, this seems to be due to some fragmentation occuring in the sheet already during the implosion, thus decreasing the central overpressure and strength of the collision shock. For cylinders and spheres, even if such fragmentation occurs during the implosion, geometrical focusing will still enhance the collision shock and the corresponding central overpressure. We note that this result does not appear to be an artifact of limited resolution, as we obtain the same result for sheet simulations with cell sizes 2 and 4 times smaller (see Appendix~\ref{subsec:peakpres_sheet}).


\begin{figure*}
    \centering	\includegraphics[width=\textwidth]{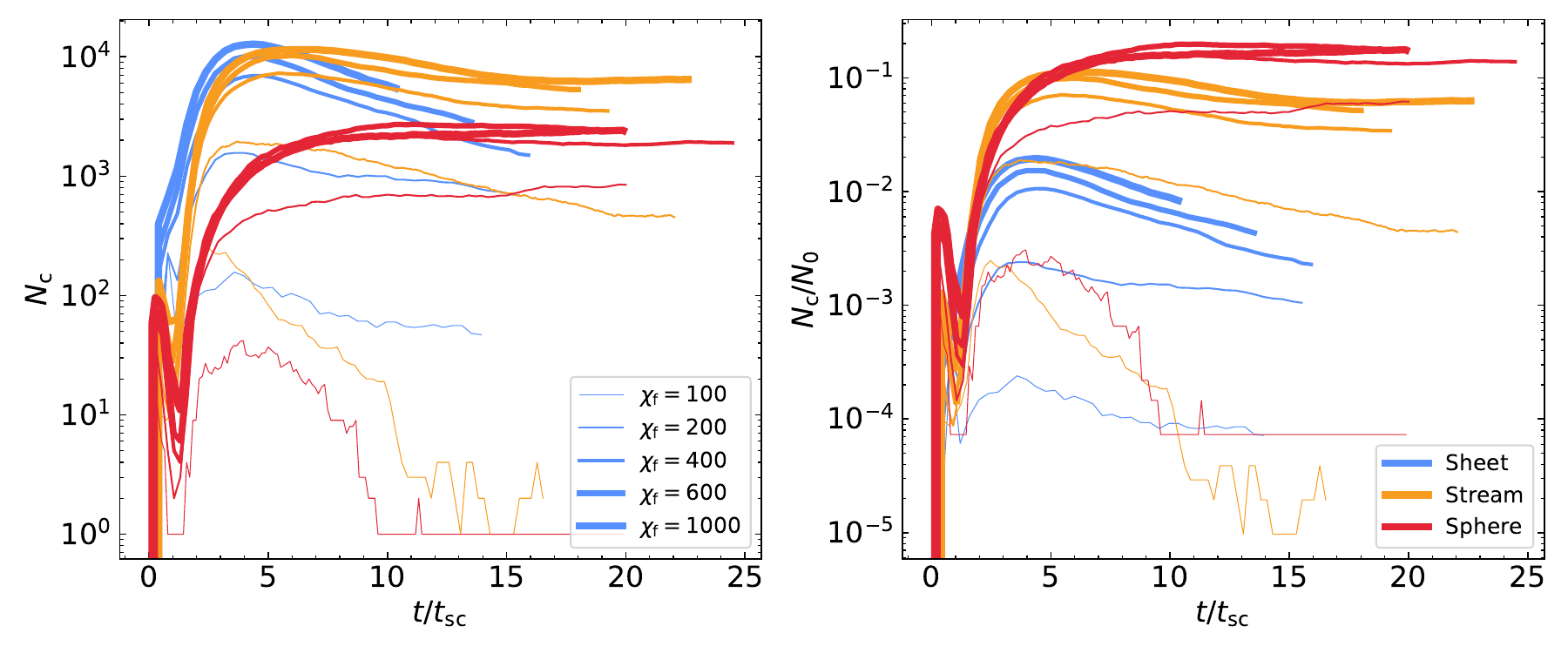}
   \caption{The number of clumps identified in simulations as a function of time, normalized by the initial sound crossing time, $t_{\rm sc}$. Different colours show different geometries, spheres (red), streams (orange), and sheets (blue). The line thickness represents the final overdensity, with thicker lines having larger $\chi_{\rm f}$. On the left, we show the absolute number of clumps, $N_{\rm c}$, while on the right we normalize this by the maximal possible number of clumps given the initial cold gas mass, $N_0$, to better highlight the differences in the degree of fragmentation between different geometries, which we see increases from sheets to streams to spheres. All cases show coagulation at late times for $\chi_{\rm f}=100$, with $N_{\rm c}$ decreasing to 1 for spheres, a few for streams, and a few tens for sheets, due to coagulation being suppressed along the stream axis and in the sheet plane. At larger overdensities, $N_{\rm c}$ monotonically increases or saturates for spheres, while it reaches a peak and decreases for streams and sheets, due to stronger radial coagulation in these geometries. 
   } 
   \label{fig:Nclump}
\end{figure*}
\begin{figure*}
    \centering	\includegraphics[width=\textwidth]{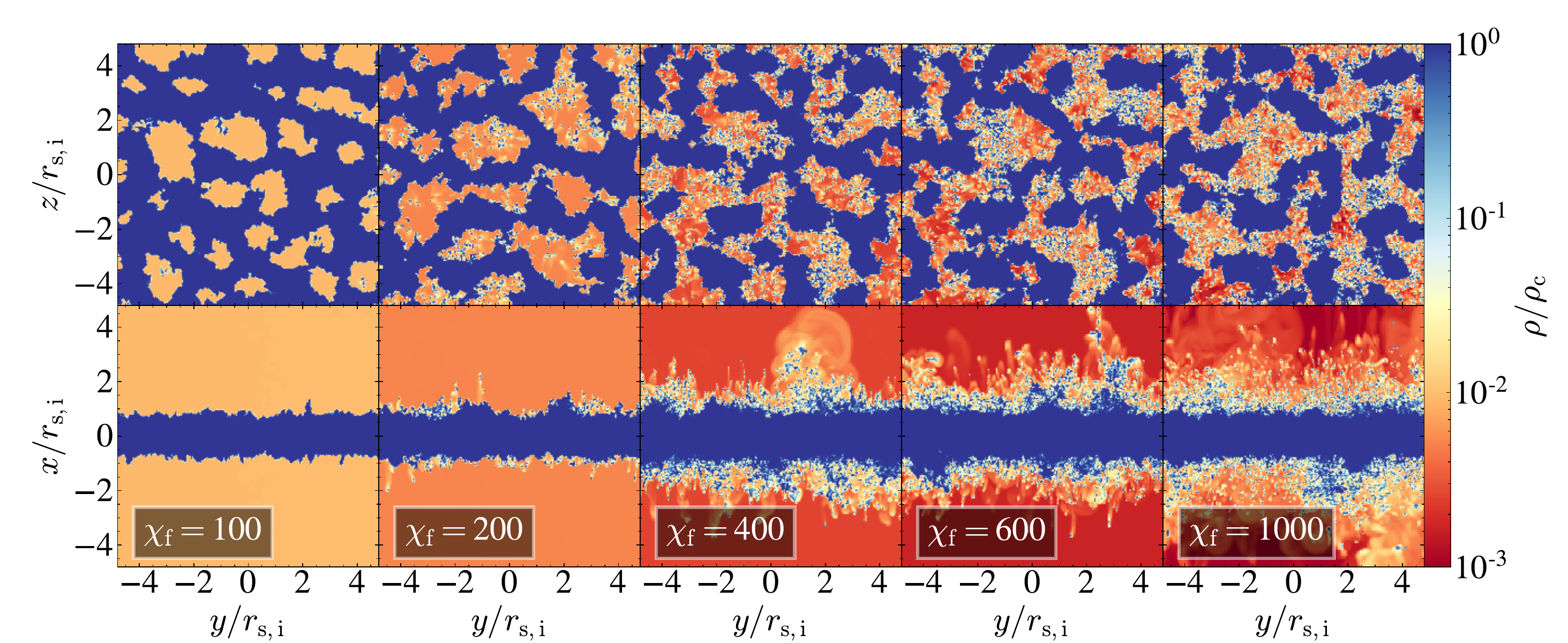}
   \caption{
   Density maps in sheets. Face-on (\textit{top}) and edge-on (\textit{bottom}) projections at $t=10\tsc$ for sheet simulations with different overdensities, as marked. The colour represents the maximal density along the line-of-sight, over the full length of the analysis region, $|x|\le16\rsi$, $|y|,|z|\le 6\rsi$. While significant radial coagulation is always present, even for $\chif=1000$, the number of small clumps at larger radial distances increases with $\chif$. We further see that coagulation within the plane of the sheet is strongly suppressed relative to the radial direction, even for $\chif=100$ (see also \fig{Sheet_map_avg}).
   }
   \label{fig:Sheet_map}
\end{figure*}

\begin{figure*}
    \centering	\includegraphics[width=\textwidth]{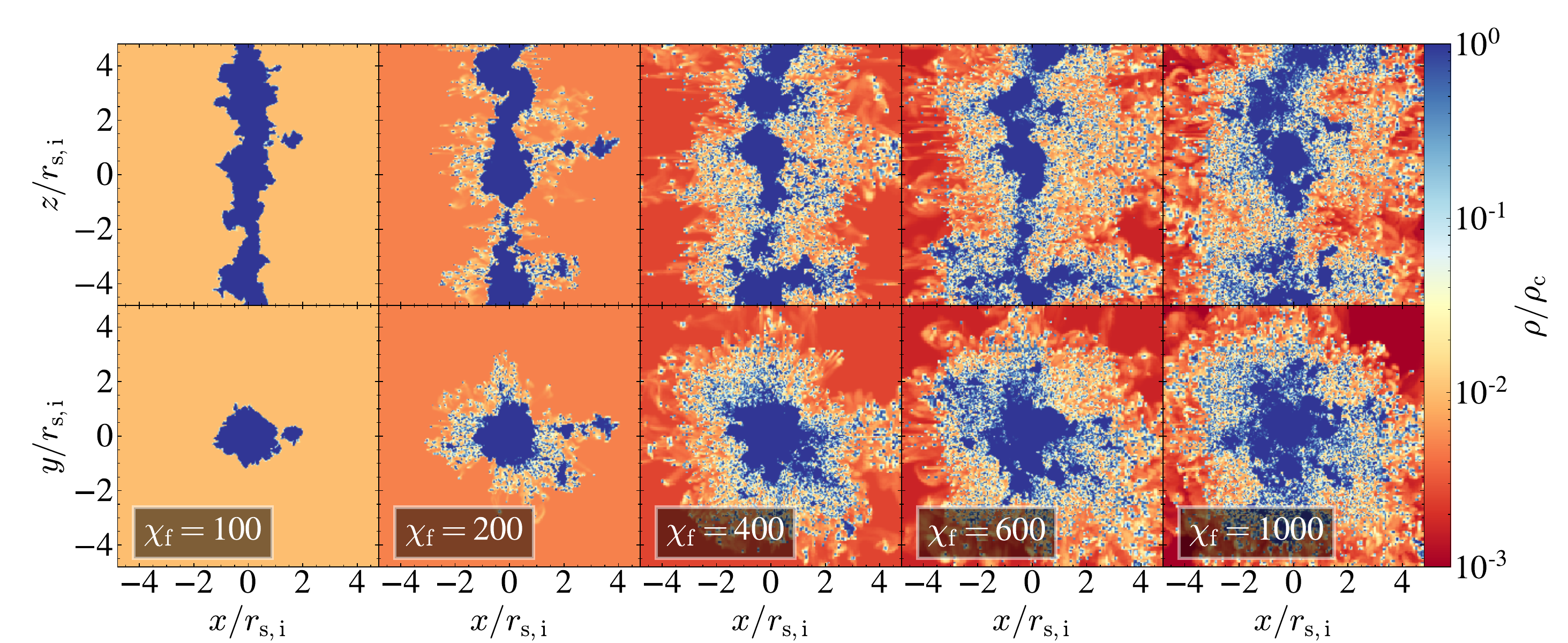}
   \caption{Density maps in streams. Edge-on (\textit{top}) and face-on (\textit{bottom}) projections at $t=10\tsc$ for stream simulations with different overdensities, as marked. The colour-bar is the same as in \fig{Sheet_map}. While some radial coagulation is always present, even for $\chif=1000$, both the number of small clumps and their radial distances noticeably increase with increasing $\chif$, especially for $\chif\ge 400$. Along its axis, the stream is broken into several large clumps (see also \fig{Stream_map_avg}). 
   }
   \label{fig:Stream_map}
\end{figure*}

\begin{figure*}
    \centering	\includegraphics[width=\textwidth]{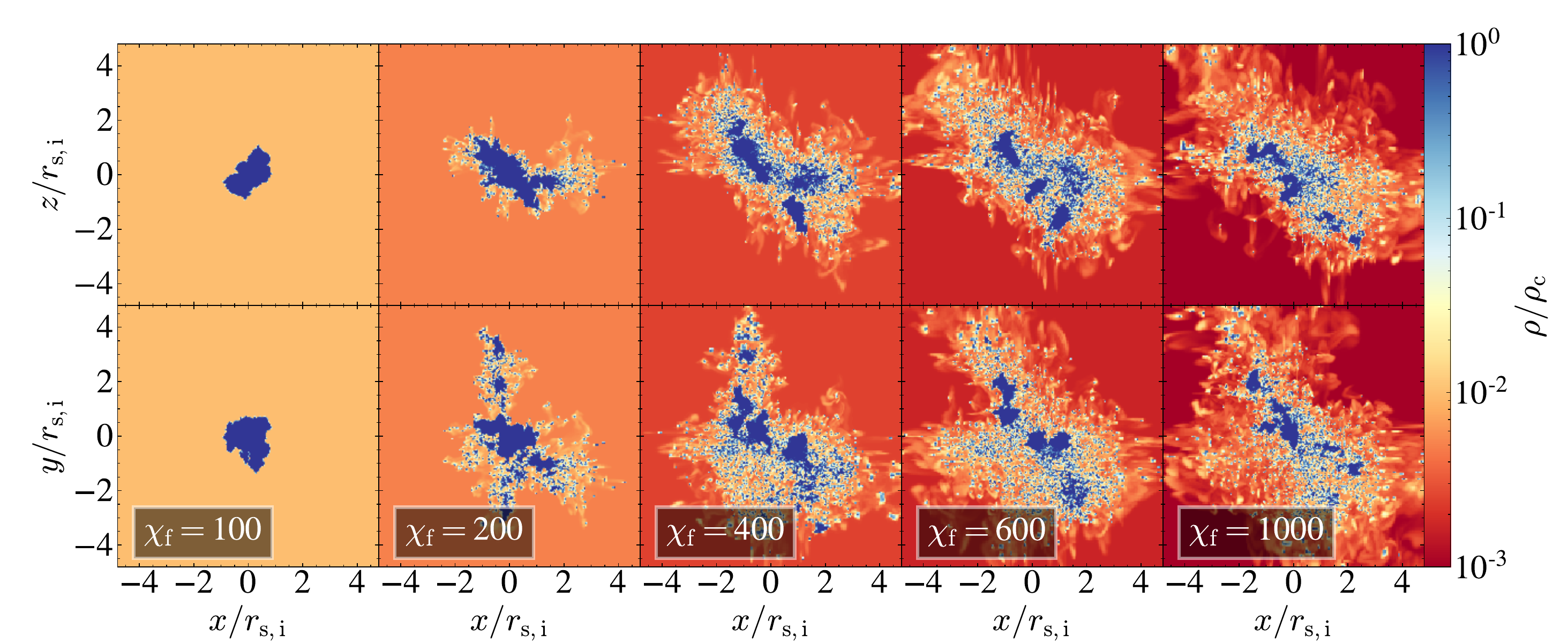}
   \caption{
   Density maps in spheres. Two orthogonal projections at $t=10\tsc$ for sphere simulations with different overdensities, as marked. The colour-bar is the same as in \figs{Sheet_map} and \figs{Stream_map}. While some radial coagulation is always present, even for $\chif=1000$, both the number of small clumps and their radial distances noticeably increase with increasing $\chif$, especially for $\chif\ge 400$ (see also \fig{Sphere_map_avg}). 
   }
   \label{fig:Sphere_map}
\end{figure*}

\subsubsection{Cold Gas Fragmentation}
\label{sec:number}

As the reflected shock sweeps over density inhomogeneities at the interface of the two phases created during the contraction, the local density and pressure gradients become misaligned leading to RMI which drives the fragmentation of cold gas. In the weak shock limit, RMI can be modeled as a form of Rayleigh-Taylor instability (RTI) in which the gravitational force is impulsive, i.e. $g\sim \Delta v\delta(t-t_{\rm shock})$, where $\delta(t)$ is the Kronicker delta function, $t_{\rm shock}$ is the time when the shock overtakes the interface, and $\Delta v$ denotes the interface velocity jump \citep{Zhou.2017a}. We assume $\Delta v \sim v_\mathrm{ex} \sim c_\mathrm{s,c}$, and further assume that due to density inhomogeneities and shape perturbations throughout the cold gas and across the interface, the effective gravitational acceleration is better modeled as $g\sim \Delta v/t_{\rm cross}$, where $t_{\rm cross} \sim \ell/\Delta v$ with $\ell\sim r_{\rm s,f}$, the characteristic cloud size after cooling and contraction, and likely the dominant perturbation wavelength. The growth timescale of the RM instability is thus $t_\mathrm{RM}\sim (\ell / g)^{1/2}\sim r_{\rm s,f}/c_\mathrm{s,c}$, comparable to the sound-crossing time of the collapsed cloud. However, the constant of proportionality can deviate from unity and depends on the initial cloud geometry, as discussed in \se{fastvsslow}. The fragmentation timescale, over which the number of clumps/cloudlets rapidly increases, is proportional to $t_{\rm RM}$. 

\smallskip
The left panel of \fig{Nclump} shows the number of clumps as a function of time, normalized by the initial cold cloud sound crossing time, $t_{\rm sc}\equiv r_{\rm s,i}/c_{\rm s,c}$. Different colour lines represent different geometries, while the line thickness grows with increasing final overdensity, $\chif$. In streams and spheres, the number of clumps peaks at $N_\mathrm{c}\sim100$ immediately after the simulation starts, due to a combination of RMI and thermal instabilities during the implosion. The number of clumps then decreases due to coagulation enhanced by further contraction, before rapidly rising again to values of a few $10^3-10^4$ during the explosion. In sheets, the coagulation during the implosion is much weaker due to the lack of geometrical focusing, so the number of clumps monotonically increases until the end of the explosion phase. In all cases, $N_\mathrm{c}$ stops growing rapidly by $t\sim 4\tsc$, when fragmentation stops and/or coagulation begins. 

\smallskip
While the peak number of clumps increases from spheres to streams to sheets, this 
is proportional to the total amount of cold material. To factor this out, we present in the right-hand panel of \fig{Nclump} the evolution of $N_{\rm c}/N_{\rm 0}$, where $N_{\rm 0}=m_{\rm s,i}/m_{\rm cell}$ with $m_{\rm s,i}$ the initial cold gas mass in the analysis region (more than $10\rsi$ from any boundary, \se{BCs}) and $m_{\rm cell}=\rho_{\rm s,f}\Delta^{3}$ the mass of a cell at the equilibrium density of cold gas at $T_{\rm floor}$. $N_0$ thus represents the maximal number of cold gas clumps possible if the cold gas mass does not increase due to entrainment. With this normalization, we see that the efficiency of shatterring increases from sheets to streams to spheres. The rate of fragmentation and clump formation is similar in all three geometries, while the timescale for $N{\rm c}$ to reach its peak and saturate increases from sheets to streams to spheres. This will be further discussed \se{coagulation} in the context of coagulation.

\smallskip
The number of clumps increases with $\chi_{\rm f}$ for all geometries, though it tends to converge at $\chi_{\rm f}\gsim 600$ for streams and spheres\footnote{This convergence may be numerical, due to our resolution decreasing away from the initial cloud, suppressing further fragmentation and causing clumps to artificially disrupt once they move too far from the centre. This is discussed further in \se{dmax}. Regardless, it does not affect our main conclusions regarding whether a cloud remains shattered or recoagulates.}. In spherical geometry, the case of $\chif=100$ recoagulates into a single large cloud after roughly $10\tsc$, while cases with $\chif\ge 200$ remain shattered with $N_{\rm c}$ either continuing to grow or saturating at $N_{\rm c}\sim 3000$. This is qualitatively similar to the results found in \citet{Gronke.Oh.20b}, only with a lower threshold in $\chif$ for shattering. This will be discussed further in \se{fastvsslow}. Streams exhibit qualitatively similar behaviour, with $N_{\rm c}$ decreasing to order unity for $\chif=100$ and remaining at large values for $\chif\ge 200$. However, unlike the spherical case, streams with $\chif\ge 200$ do exhibit some coagulation, with $N_{\rm c}$ decreasing after an initial peak. This is particularly noticeable for $\chif=200$, which we consider a borderline case (\tab{sim_table}). Furthermore, unlike the spherical case which coagulated into a single cloud for $\chif=100$, streams with $\chif=100$ maintain several distinct clumps along the stream axis. The coagulation along the stream axis is suppressed compared to the radial direction due to opposing forces pulling clumps in either direction\footnote{This is not true near the edges of a finite stream, which contract along the stream axis towards the centre as described in \se{BCs}}. At late times, the number of clumps fluctuates between $N_{\rm c}\sim (1-4)$ due to centres 
of large clumps along the stream axis moving in and out of our analysis region, $|z| \le 6\rsi$ (see \se{BCs}). 

\smallskip
Sheets exhibit even stronger coagulation than streams for $\chif\ge 200$, with $N_{\rm c}$ decreasing by more than $\sim 50\%$ between its peak and $10\tsc$ even for $\chif=1000$. However, similar to streams, the coagulation is primarily in the radial direction and is suppressed in the plane of the sheet. As a result, several tens of clumps remain at the end of the $\chif=100$ simulation.

\begin{figure}
    \centering	\includegraphics[width=\columnwidth]{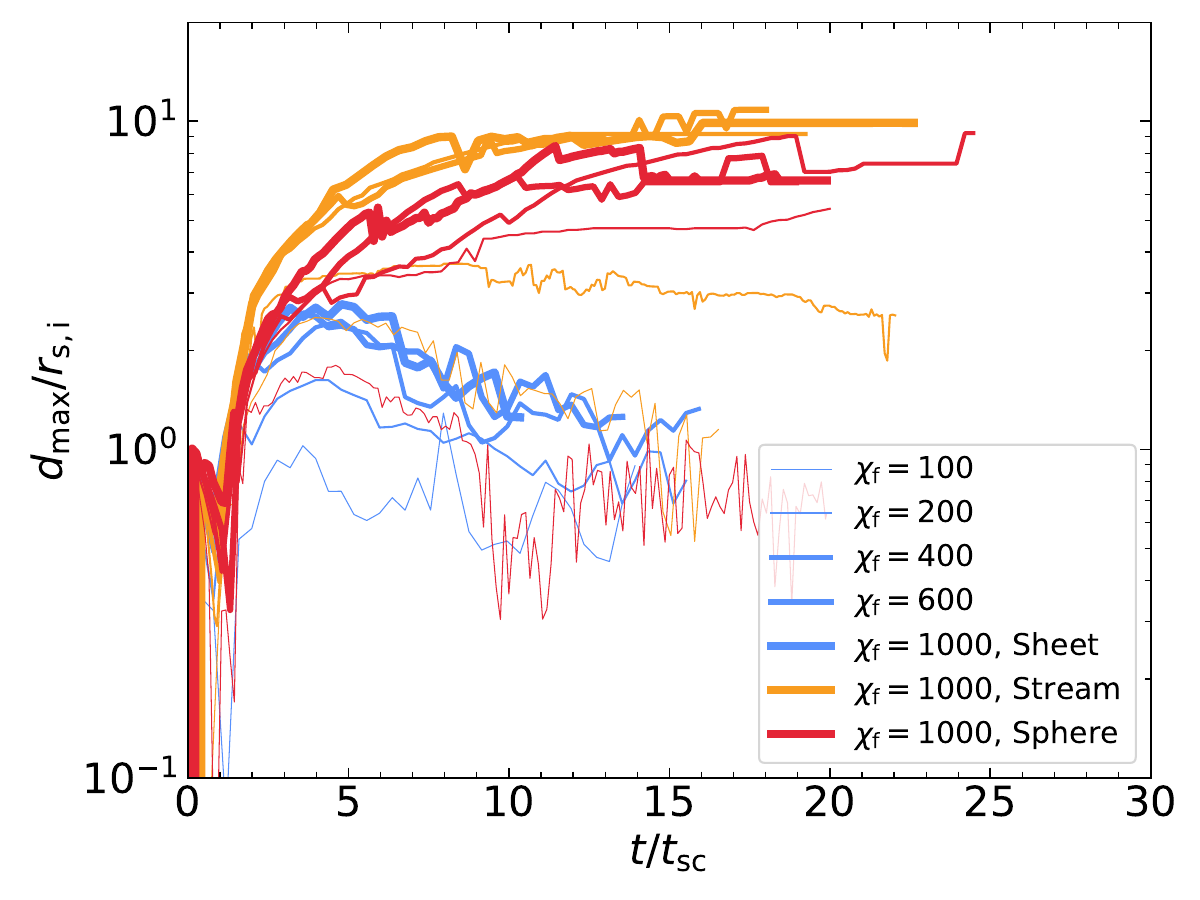}
   \caption{The distance of the farthest clump as a function of time, limited to clumps with mass at least 8 times the mass of a high-resolution cell at the equilibrium density of the cold phase, $m_{\rm c}\ge 8\rho_{\rm c}\Delta^3$. $d_{\rm max}$ represents the distance to the central plane/line/point for sheets/streams/spheres respectively. 
   As in \fig{Nclump}, different colours represent different geometries, while line thickness increases for increasing $\chif$. For spheres and streams with $\chif=100$, and for sheets regardless of $\chif$, $\dmax$ reaches a maximum at $\sim (1-3)\rsi$ after the explosion phase and then decreases, indicative of strong coagulation. In streams and spheres with higher overdensities, $\dmax$ continues rising or saturates at a finite value of order $\sim 10\rsi$. 
   }
   \label{fig:dmax}
\end{figure}

\smallskip
We show density maps at $t=10t_{\rm sc}$ in two orthogonal projections and for different values of $\chif$, for sheets, streams, and spheres in \figs{Sheet_map}, \figss{Stream_map}, and \figss{Sphere_map} respectively. For sheets and streams, the projections correspond to face-on and edge-on, while for the spheres we simply show two orthogonal orientations. These maps show the maximal density along the line of sight, which highlights the small clumps resulting from shattering. Complementary to these, we show in \figs{Sheet_map_avg}-\figss{Sphere_map_avg} the average density along the line of sight for the same projections, which better highlights coagulation and the geometry of large clouds. Qualitatively, one sees radial coagulation grow stronger from spheres to streams to sheets, with the edge-on projection revealing strong coagulation in sheets even when $\chif=1000$ (bottom-right panel of \figs{Sheet_map} and \figss{Sheet_map_avg}). While some radial coagulation is always apparent in each geometry, we find more small clumps at larger radial distances as $\chif$ increases. On the other hand, coagulation is suppressed along the stream axis and within the plane of the sheet even for $\chif=100$, where $N_{\rm c}=1$ for spheres, a few for streams, and a few tens for sheets. While this is hard to see in \figs{Sheet_map} and \figss{Stream_map}, it becomes clear when examining \figs{Sheet_map_avg} and \figss{Stream_map_avg}

\subsubsection{The distance of clump propagation}
\label{sec:dmax} 

One of the key features of shattering is that cold gas clumps get spread over an area which grows larger with time as the cloudlets spread out and fragmentation continues, whereas if coagulation is important the region occupied by cold gas reaches a maximum and then begins to shrink. In \fig{dmax}, we show the time evolution of $\dmax$, the maximal radial distance of any clump whose mass is at least $8\,{\rm m_{cell}}\equiv8\,\rho_{\rm c}\Delta^3$, where $\Delta$ is the minimal cell size (valid in the region $|x_1|<3\rsi$, \se{BCs}) and $\rho_{\rm c}$ is the equilibrium density in the cold phase. We implement this threshold to reduce our sensitivity to resolution effects by removing small clumps 
at the grid scale. 
Using a threshold of 16 cells gives very similar results, while taking all clumps gives qualitatively similar results. $\dmax$ is measured as the radial distance from the centre of simulation domain for spheres, the radial distance from the $z$ axis for streams, and the vertical distance from the $yz$ plane for sheets, and is normalized by the initial cloud radius, $r_{\rm s,i}$. 

\smallskip
Initially, $\dmax=0$ because there is only one ``clump'' whose centre is at the centre of the simulation volume. However, immediately following that we get $\dmax\sim r_{\rm s,i}$ for all cases, as fragmentation begins near the initially perturbed cloud interface. $\dmax$ then proceeds to shrink during the implosion before rapidly rising during the explosion phase. The growth rates during the explosion are similar for all cases, 
peaking at $({\rm d}/{\rm d}t)\dmax\sim (0.5-2.0)c_{\rm s,c}\sim v_{\rm ex}$ (\se{explosion}).  

\smallskip
For sheets with $\chi_{\rm f}= 100$, $\dmax$ never exceeds $r_{\rm s,i}$ due to very strong coagulation. In all other cases $\dmax$ peaks at values several times the initial cloud radius. For sheets with $\chi_{\rm f}\ge 200$, $\dmax$ peaks at $\sim (2-3)\rsi$ at $t\sim 4\tsc$ and then noticeably decreases, indicative of strong coagulation and consistent with the decline in $N_{\rm c}$ seen in \fig{Nclump}. The same is true for streams and spheres with $\chi_{\rm f}=100$. In all coagulating cases, $\dmax$ appears to oscillate at late times, consistent with the pulsations observed at relatively low values of $\chi_{\rm f}$ in previous work \citep{Gronke.Oh.20b,Gronke.Oh.23}. On the other hand, streams and spheres with $\chif\ge 200$ exhibit $\dmax$ values that either continue to rise or saturate until the end of the simulation. This suggests a critical $\chi_{\rm f}\sim 200$ for shattering in streams and spheres, consistent with \fig{Nclump}. In the stream simulation with $\chif=200$, $\dmax$ slightly decreases at $t>10\tsc$ consistent with this being more of a borderline case as noted above. We note that the strong saturation observed at $\dmax\sim 10\rsi$ is likely numerical, because the cell size at $d>9\rsi$ grows to $8\Delta$, causing even large clumps to artificially disrupt. However, this does not change the qualitative distinction between cases where $\dmax$ decreases due to strong coagulation and cases where it does not. 

\subsection{Cold gas coagulation}
\label{sec:coagulation}
In the previous section, we saw that sheets were prone to strong coagulation at all overdensities, while spheres and streams were prone to coagulation for $\chi_{\rm f}\lsim 200$. This was found to be the case based on the number of clumps (\fig{Nclump}), their morphology (\figs{Sheet_map}-\figss{Sphere_map}) and their radial extent (\fig{dmax}). In this section, we study the coagulation process in detail. 

\subsubsection{Theoretical Overview}
\label{sec:theory}

We first review the theoretical model described in \citet{Gronke.Oh.23}, based on their analysis of quasi-spherical systems of clouds. 
Consider a cold clump, hereafter clump $1$, embedded in a hot gaseous medium. Any perturbations at the interface of the cold and hot phases will induce a turbulent mixing layer, where efficient radiative cooling will drive an entrainment flow of hot background gas onto the clump with velocity 
\begin{equation}
\label{eq:vhot}
v_\mathrm{hot}=v_\mathrm{mix,1}\left(\frac{r_\mathrm{cl,1}}{r}\right)^{2}
\end{equation}
for radii $r\ge r_\mathrm{cl,1}$. Here, 
\begin{equation}
\label{eq:vmix}
v_\mathrm{mix,1}\approx0.4c_\mathrm{s,c}(t_\mathrm{sc,1}/t_\mathrm{cool,c})^{1/4}
\end{equation}
is the entrainment velocity at the surface of the clump (\citealp{Gronke.Oh.18,Gronke.Oh.20,Fielding.etal.20,Sparre.etal.20,Mandelker.etal.20a,Tan.etal.21,Gronke.Oh.23}; see also Appendix~\ref{subsec:mixing} and \fig{EarlyVmix}). 

\smallskip
Now consider a static clump, hereafter clump $2$, at a distance $d$ from the central clump $1$. The force acting on clump $2$ is 
\be 
\label{eq:F21_Gronke}
F_{2,1}=\dot{p}_2=\dot{m}_2\Delta v_2+m_2\Delta\dot{v}_2, 
\ee 
where $\Delta v_2$ is the velocity of clump $2$ relative to the hot backgroud. The first term on the right-hand side represents an effective force due to condensation, which causes the mass of the cold cloud to increase and thus its velocity to decrease due to momentum conservation, 
\begin{equation}
\label{eq:Fcond}
F_\mathrm{cond}=\dot{m}_2\Delta v_2=\rho_\mathrm{h}v_\mathrm{mix,2}4\pi r_\mathrm{cl,2}^{2}\Delta v_2. 
\end{equation}
The second term on the right-hand-side of \equ{F21_Gronke} is the drag force at fixed cloud mass which is given by 
\begin{equation}
F_\mathrm{drag}=C_\mathrm{d}\rho_\mathrm{h}\Delta v_2^{2}\pi r_\mathrm{cl,2}^{2},
\end{equation}
where $C_\mathrm{d}\approx0.47$ is the drag coefficient for a sphere. The ratio of these two forces is thus 
\begin{equation}
\label{eq:Fratio}
\frac{F_\mathrm{cond}}{F_\mathrm{drag}}=\frac{4}{C_\mathrm{d}}\frac{v_\mathrm{mix,2}}{\Delta v_2}.
\end{equation}

\smallskip
Initially, clump 2 is at rest while the hot gas moves at velocity $v_\mathrm{hot}(r=d)$. Over time, clump 2 becomes entrained in the flow and the relative velocity decreases. We thus have in general $\Delta v_2\le v_\mathrm{hot}(r=d)$, yielding 
\begin{equation}
\label{eq:Fratio_ent}
\frac{F_\mathrm{cond}}{F_\mathrm{drag}}\ge\frac{4}{C_\mathrm{d}}\frac{v_\mathrm{mix,2}}{v_\mathrm{mix,1}}\left(\frac{d}{r_\mathrm{cl,1}}\right)^{2}\approx \frac{4}{C_\mathrm{d}}\left(\frac{r_\mathrm{cl,2}}{r_\mathrm{cl,1}}\right)^{1/4}\left(\frac{d}{r_\mathrm{cl,1}}\right)^{2},
\end{equation}
where in the final equation we have used \equ{vmix} and the fact that the two clouds have similar temperatures and densities, though not necessarily similar sizes. 
We conclude that $F_\mathrm{cond}\gg F_\mathrm{drag}$ unless $r_\mathrm{cl,1}/r_\mathrm{cl,2}> 10^{4}$ and $d$ is not much larger than $r_{\rm cl,1}$. 
This is true regardless of the details of the cooling function, and in particular regardless of the gas metallicity, so long as the two clumps are initially static with respect to the background. 

\smallskip
So long as the velocity of clump $2$ with respect to clump $1$ is negligible compared to the entrainment velocity of hot gas onto clump 1, we have $\Delta v_2\sim v_\mathrm{hot}$. The condensation force then becomes 
\begin{equation}
\label{eq:Fcond,grav}
F_\mathrm{cond}\sim\frac{\rho_\mathrm{h}v_\mathrm{mix,1}^{2}}{4\pi}\left(\frac{r_{\rm cl,2}}{r_{\rm cl,1}}\right)^{1/4}\frac{A_\mathrm{cl,1}A_\mathrm{cl,2}}{d^{2}},
\end{equation}
where $A_{\rm cl}=4\pi r_{\rm cl}^2$ is the cloud surface area. Neglecting the weak dependence on the ratio $r_{\rm cl,2}/r_{\rm cl,1}$, this force is similar to gravity if we make the analogy that $G\rightarrow \rho_\mathrm{h}v_\mathrm{mix,1}^{2}/4\pi$ and $m\rightarrow A$, with both forces proportional to $d^{-2}$. 

\smallskip
If the central region is occupied by a collection of similar sized clumps each with $r_{\rm cl}\sim r_{\rm cl,1}$, the entrainment velocity of the background flow towards the centre becomes
\begin{equation}
v_\mathrm{hot}\sim v_\mathrm{mix,1}\frac{A_\mathrm{cl,tot}(<r)}{4\pi r^{2}}\equiv f_\mathrm{A}(r)v_\mathrm{mix,1},
\end{equation}
where $A_{\rm cl,tot}(<r)$ is the total combined surface area of all cold clouds interior to $r$, and $f_\mathrm{A}(r)\equiv A_\mathrm{cl,tot}(<r)/4\pi r^{2}$ is the area modulation factor. This factor can be greater or less than unity, as discussed in \se{fA}.
Consequently, the total force acting on a test clump 2 at a distance $d$ from the centre of clumps is 
\begin{equation}
F_2\sim \rho_\mathrm{h}v_\mathrm{mix,2}A_\mathrm{cl,2} f_\mathrm{A}(d)v_\mathrm{mix,1}\sim\frac{\rho_\mathrm{h}v_\mathrm{mix}^{2}}{4\pi}\frac{A_\mathrm{cl,tot}(<d)A_\mathrm{cl,2}}{d^{2}},
\end{equation}
where we have approximated $v_\mathrm{mix}\sim v_\mathrm{mix,1}\sim v_\mathrm{mix,2}$. The acceleration of clump 2 is thus 
\begin{equation}
a_2=F_2/m_{\rm cl,2} \sim \frac{3 v_{\rm mix}^2 f_{\rm A}(d)} {\chi r_{\rm cl,2}}.
\end{equation}
We can use this to derive a characteristic timescale for coagulation, analogous to the gravitational free-fall time. Assuming that the acceleration is roughly constant, namely that $f_{\rm A}(d)\sim \const$, we have\footnote{If we assume instead that $A_{\rm cl,tot}(<d)\sim \const$ during the collapse, so $a_2\propto d^{-2}$, $t_\mathrm{coag}$ is multiplied by a factor of $\pi/4\sim 0.8$ compared to \equ{tcoag}.} 
\begin{equation}
\label{eq:tcoag}
t_\mathrm{coag}\sim \left(\frac{2d}{a_2}\right)^{1/2} \sim \left(\frac{2\chi}{3f_\mathrm{A}(d)}\right)^{1/2}\frac{(r_\mathrm{cl,2}d)^{1/2}}{v_\mathrm{mix}}.
\end{equation}

\begin{figure*}
    \centering	\includegraphics[width=\textwidth]{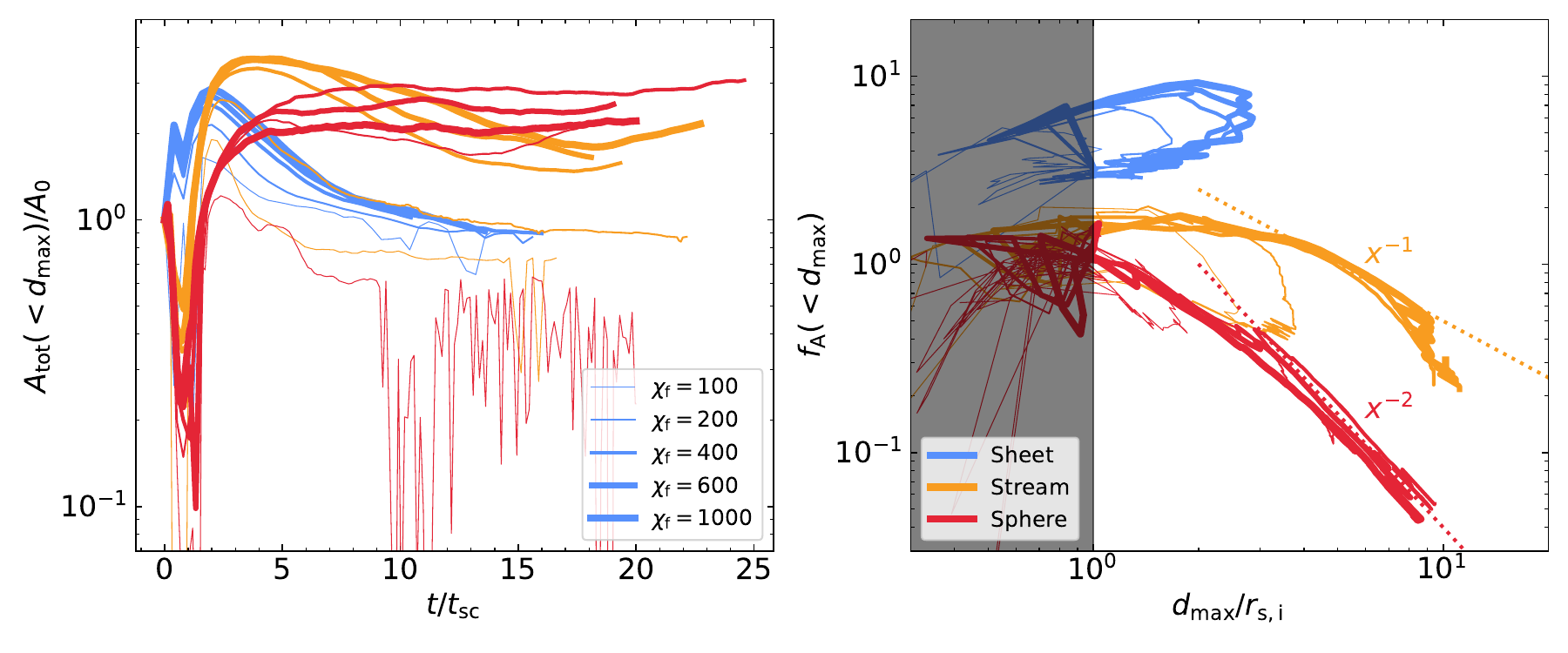}
   \caption{\textit{Left}: Time evolution of the total surface area of cold gas within $\dmax$, normalized to its initial value. \textit{Right}: The area modulation factor, $f_{\rm A}$, as a function of $\dmax$. The shaded region, at $\dmax<r_{\rm s,i}$, is not relevant to our discussion but is shown for completeness. As in previous figures, different colours represent different geometries, while the line thickness is proportional to $\chif$. For coagulating cases, namely all sheets, streams with $\chif\lsim 200$ and spheres with $\chif=100$, the cold gas surface area declines towards values $<A_0$ after an initial peak at the end of the explosion phase. For these cases there is no explicit trend of $f_{\rm A}$ with $\dmax$. For streams and spheres with larger overdensities which remain shattered, the total surface area saturates at values of $(2-3)A_0$. During this phase, $f_{\rm A}\propto d^{-1}$ for streams and $d^{-2}$ for spheres.}
   \label{fig:fA}
\end{figure*}

\smallskip
We can similarly derive the coagulation force and timescale for stream and sheet geometries. It turns out that the only difference\footnote{Note that regardless of the large scale geometry of $A_{\rm tot}$, the test clump 2 is always assumed to be spherical.} is the factor $f_\mathrm{A}$, which can be defined as 
\begin{equation}
\label{eq:fa}
f_\mathrm{A}(d)=
\left\{
\begin{array}{ll}
A_\mathrm{tot}(<d)/4\pi d^{2} &\mathrm{sphere}, \\
\Lambda_\mathrm{tot}(<d)/2\pi d &\mathrm{stream}, \\
\Sigma_\mathrm{tot}(<d) &\mathrm{sheet}
\end{array}
\right.
\end{equation}
$\Lambda_\mathrm{tot}$ is the total area of all cold clouds per unit length, proportional to the cloud diameter, and $\Sigma_\mathrm{tot}$ is the total area of cold clouds per unit area, proportional to the number of clouds. Note the different scaling with $d$ in different geometries, similar to the gravitational force from a spherical, cylindrical, and planar distribution of mass. 

\smallskip
The above considerations, based on \citet{Gronke.Oh.23}, are valid for a ``test clump'' (clump 2) initially at rest with respect to the central collection of cold clouds (clump 1). 
Now let us consider the case where clump 2 is escaping the central region, while at the same time there is an entrainment flow of hot gas towards the centre. This is the situation immediately after the explosion phase described in the previous section, where shattered clumps were escaping the central region with a velocity $v_{\rm ex}\sim c_{\rm s,c}$. 
The total timescale for the escaping clumps to recoagulate is the sum of the deceleration 
timescale, when the interaction of the clumps and the entrainment flow causes the clumps to decelerate and turn around, and the coagulation timescale from \equ{tcoag} which measures the coagulation time of static clumps in the hot entrainment flow. The turnaround/deceleration process can be due either to the condensation force or the drag force, depending the ratio in \equ{Fratio}. By  combining \equ{vmix} and \equ{Fratio}, we have
\begin{equation}
\label{eq:Fratio2}
\frac{F_\mathrm{cond}}{F_\mathrm{drag}}\approx 3.4\frac{c_\mathrm{s,c}}{\Delta v_2}\left(\frac{t_\mathrm{sc,2}}{t_\mathrm{cool,c}}\right)^{1/4}.
\end{equation} 
Considering $t_\mathrm{sc,2}/t_\mathrm{cool}\approx r_\mathrm{cl,2}/\ell_\mathrm{shatter}\gtrsim1$, and $\Delta v_2\sim c_\mathrm{s,c}$ (see \fig{profiles}), we conclude that regardless of geometry and metallicity $F_\mathrm{cond}$ dominates over $F_\mathrm{drag}$ in the deceleration process, as well as in the coagulation process discussed above. 

\smallskip
The deceleration timescale is thus determined by the condensation timescale, which is the timescale for cloud growth by condensation
\begin{equation}
\label{eq:tcond}
t_{\rm decel}\sim t_\mathrm{cond}\sim \frac{m_2}{\dot{m_2}}\sim \frac{\chi r_\mathrm{cl,2}}{3v_\mathrm{mix,2}} \sim \frac{\chi r_\mathrm{cl,2}}{1.2c_{\rm s,c}}\left(\frac{t_{\rm sc,2}}{t_{\rm cool,c}}\right)^{-1/4}.
\end{equation}
The ratio of the deceleration and coagulation timescales is thus
\begin{equation}
\label{eq:tratio}
\frac{t_\mathrm{decel}}{t_\mathrm{coag}}\sim \left(\frac{\chi f_A(d) r_\mathrm{cl,2}} {6d}\right)^{1/2}.
\end{equation}
Since $t_\mathrm{decel}$ has no explicit dependence on $d$ while $t_\mathrm{coag}$ increases with $d$, we obtain that $t_\mathrm{decel}>t_\mathrm{coag}$ at small $d$ where $f_{\rm A}$ is large (\equnp{fa}). However, if a clump finds itself at large $d$ with small $f_{\rm A}$, then $t_\mathrm{coag}$ will dominate the remaining recoagulation timescale even if the clump is still in the process of decelerating. 

\smallskip
Before significantly decelerating, clumps can reach a maximal distance of $d_\mathrm{max}\sim \xi c_\mathrm{s,c}t_\mathrm{decel}$, where $v_{\rm ex}\sim \xi c_{\rm s,c}$ is the explosion velocity and $\xi\sim 0.4$ or $1.0$ for sheets or streams and spheres, respectively (\se{explosion}). The maximum value of $t_\mathrm{coag}$ along the clump's orbit is reached at $d=d_\mathrm{max}$,
\begin{equation}
t_\mathrm{coag,max}\sim \frac{\chi_\mathrm{f}r_\mathrm{cl,2}}{3v_\mathrm{mix}}\left(\frac{2\xi c_\mathrm{s,c}}{f_\mathrm{A}(d_\mathrm{max})v_\mathrm{mix}}\right)^{1/2}. 
\end{equation}
Reaching $d_{\rm max}$ and turning around is a necessary condition for the clump to coagulate. The question of whether the final coagulation is efficient is determined by the ratio in \equ{tratio}, 
\begin{equation}
\label{eq:tratiomax}
\frac{t_\mathrm{decel}}{t_\mathrm{coag,max}}\sim \left(\frac{f_\mathrm{A}(d_\mathrm{max}) v_\mathrm{mix}} {2\xi c_\mathrm{s,c}}\right)^{1/2}\sim \left(\frac{f_\mathrm{A}(d_\mathrm{max})}{5\xi}\right)^{1/2}\left(\frac{t_\mathrm{sc,2}}{t_\mathrm{cool,c}}\right)^{1/8}
\end{equation}
Neglecting the second term with the $1/8$-power, we see that larger values of $f_\mathrm{A}(d_\mathrm{max})$ result in more efficient coagulation. We now turn to quantify $f_\mathrm{A}$ as a function of $d$ in different geometries, to gain a better understanding of the geometrical effects on coagulation.

\subsubsection{Cold gas area}
\label{sec:fA}

We can crudely estimate the efficiency of coagulation by measuring the coagulation forces on the outermost clump. To this end, we estimate $f_{\rm A,dmax}\equiv f_{\rm A}(\dmax)$, with $\dmax$ as in \fig{dmax}. At early times, when $\dmax$ is still small and the fragmentation process is still ongoing, we expect $A_{\rm tot}$ to increase with $d$ and therefore $f_{\rm A}$ to either increase or decrease depending on the details. However, at later times once $\dmax$ grows beyond the central concentration of cold gas, both the number of clumps (\fig{Nclump}) and their radial extent (\fig{dmax}) are roughly constant, consistent with the fragmentation process ending and/or coagulation affecting the inner region. At this stage, we expect $A_{\rm tot}$ to be roughly constant and $f_{\rm A}\propto d_{\rm max}^{-n}$, with $n=0$, $1$, or $2$ for sheets, streams and spheres (\equnp{fa}).

\smallskip
In \fig{fA} we show the total surface area of cold gas, $A_{\rm tot}(<\dmax)$, on the left, and the area modulation factor, $f_{\rm A}$, on the right. Here, $A_{\rm tot}$ refers to the total surface area of cold gas, regardless of the size of the clumps, since this is the relevant quantity governing coagulation. To measure this in the simulations, we first interpolate the gas density onto a uniform grid at the highest resolution of our refined grid using \texttt{yT} \citep{Turk.etal.11}, and then extract a 2D surface mesh from a 3D volume using the python package \textit{scikit-image} \citep{scikit-image}. We extract the density isosurface corresponding to $\rho_{\rm iso}=\rho_{\rm s,i}/1.05$, where the factor $1.05$ is to ensure that this captures all of the initial cloud including any density fluctuations present in the initial conditions (\se{ICs}). We then normalize $A_{\rm tot}$ by its initial value, $A_0$, which lacks an analytical form due to the shape perturbations on the initial cloud interface. As noted in \fig{dmax}, $\dmax=0$ at the initial condition, and is only self-consistently defined once fragmentation begins near the cloud interface. Therefore, $A_0=A_{\rm tot}(t=0)$ simply represents the total surface area of the initial cloud, and $A_{\rm tot}$ is only limited by $\dmax$ from the second snapshot.

\smallskip
During the implosion phase, $A_{\rm tot}$ decreases rapidly for spheres and streams due to contraction, which reduces the surface area of the cloud. This effect is stronger for spheres than for streams because of the different scaling of cloud area with radius. However, for sheets $A_{\rm tot}$ actually increases during the implosion phase, because the overall area of the central sheet is independent of its `radius' (thickness), while fragmentation increases the total cold gas surface area. 

\smallskip
Following the explosion phase, $A_{\rm tot}$ reaches values of $\sim (2-4)A_0$ at $t\sim (2-4)\tsc$. In cases with strong coagulation (based on \fig{Nclump} and \fig{dmax}), namely spheres with $\chif=100$, streams with $\chif\lsim 200$, and all sheet simulations, $A_{\rm tot}$ proceeds to decrease towards values of $\lsim A_0$ at late times. On the other hand, in cases which show strong shattering, namely spheres with $\chif\ge 200$ and streams with $\chif\ge 400$, $A_{\rm tot}$ saturates following the explosion phase at values of $A_{\rm tot} \sim (2-3)A_0$. We note that the borderline nature of the stream simulation with $\chif=200$ is particularly evident here. While a boost in the surface area by a factor of $\sim 3$ may not seem like much, we recall that without shattering the final equilibrium configuration has a radius $\rsf$, which is $\sim 10$, $3$, and $2$ times smaller than $\rsi$ for sheets, streams, and spheres, respectively (\tab{sim_table}), yielding a final equilibrium surface area which is $\sim 1$, $3$, and $5$ times smaller than $A_0$. The actual increase in cold gas surface area with respect to the equilibrium configuration due to shattering is thus a factor of $\sim 6$ for streams and $10$ for spheres. 

\smallskip
The precise late-time value of $A_{\rm tot}$ is somewhat sensitive to the late-time value of $\dmax$ (\fig{dmax}) and therefore to our threshold of $m_{\rm cl}\ge 8m_{\rm cell}$. Likewise, the decrease in the late-time value of $A_{\rm tot}$ with increasing $\chif$ for spheres seems to also be an artifact of our refinement strategy which causes even large clumps to artificially disrupt at distances of $d>9\rsi$ 
thus decreasing the cold gas surface area. Nevertheless, there is a real qualitative distinction between $A_{\rm tot}$ saturating at $\sim (2-3)A_0$ for shattering cases, versus decreasing to $\lsim A_0$ for coagulating cases. 

\smallskip
On the right, we show $f_{\rm A,dmax}$ by combining $A_{\rm tot}(<\dmax)$ shown on the left with \equ{fa}. We show this as a function of $\dmax/r_{\rm s,i}$ itself, so each point on this plot represents the situation of the farthest clump at the corresponding time, and can be inserted into \equ{tcoag} or \equ{tratio}. Note that initially, when $\dmax=\rsi$ prior to the implosion, $f_{\rm A,dmax}=1$ for spheres and streams without shape perturbations, though $f_{\rm A,dmax}=2$ for sheets with no perturbations, because of the definition of $\Sigma$ in \equ{fa} and the fact that a sheet has two surfaces. The initial perturbations increase this value, such that spheres and streams begin at $f_{\rm A,dmax}\sim 1.6$ while sheets begin at $f_{\rm A,dmax}\sim 3.2$, which is consistent with each surface adding $\sim 0.6$ to $f_{\rm A,dmax}$. During the implosion phase $\dmax$ decreases, while $A_{\rm tot}$ increases for sheets and decreases for streams and spheres. This leads to fairly chaotic behavior in the the plane of $\dmax/r_{\rm s,i}$ and $f_{\rm A,dmax}$ (grey shaded region). However, the value of $f_{\rm A}$ during the implosion phase is uninteresting, since the distinction between coagulation and shattering is only meaningful after the explosion phase, once $\dmax>\rsi$. 

\smallskip
For sheets, as well as for streams with $\chif\le 200$ and spheres with $\chif=100$, coagulation is important and $\dmax$ decreases back towards $\rsi$ after peaking at somewhat larger values (\fig{dmax}). This is also seen in the right-hand panel of \fig{fA}, where the curves turn around towards lower $\dmax$ after first reaching values $\dmax/\rsi>1$. For these cases, while the value of $f_{\rm A}$ is smaller during the coagulation then during the initial expansion, there is no clear trend of $f_{\rm A}$ with $d$ during either phase. On the other hand, for cases which shatter, namely spheres with $\chif\ge 200$ and streams with $\chif\ge 400$, $\dmax$ never decreases (\fig{dmax}), while $A_{\rm tot}$ is roughly constant at late times. This results in $f_{\rm A}\propto d^{-1}$ for streams and $f_{\rm A}\propto d^{-2}$ for spheres (\equnp{fa}), as highlighted in the figure. 
Overall, $f_\mathrm{A,dmax}$ at late times is largest in sheets and smallest in spheres, 
as is the coagulation efficiency 
(\equnp{tratiomax}), as demonstrated in previous sections. We will discuss this further in \se{fastvsslow}.
 
%

\begin{figure}
    \centering	\includegraphics[width=\columnwidth]{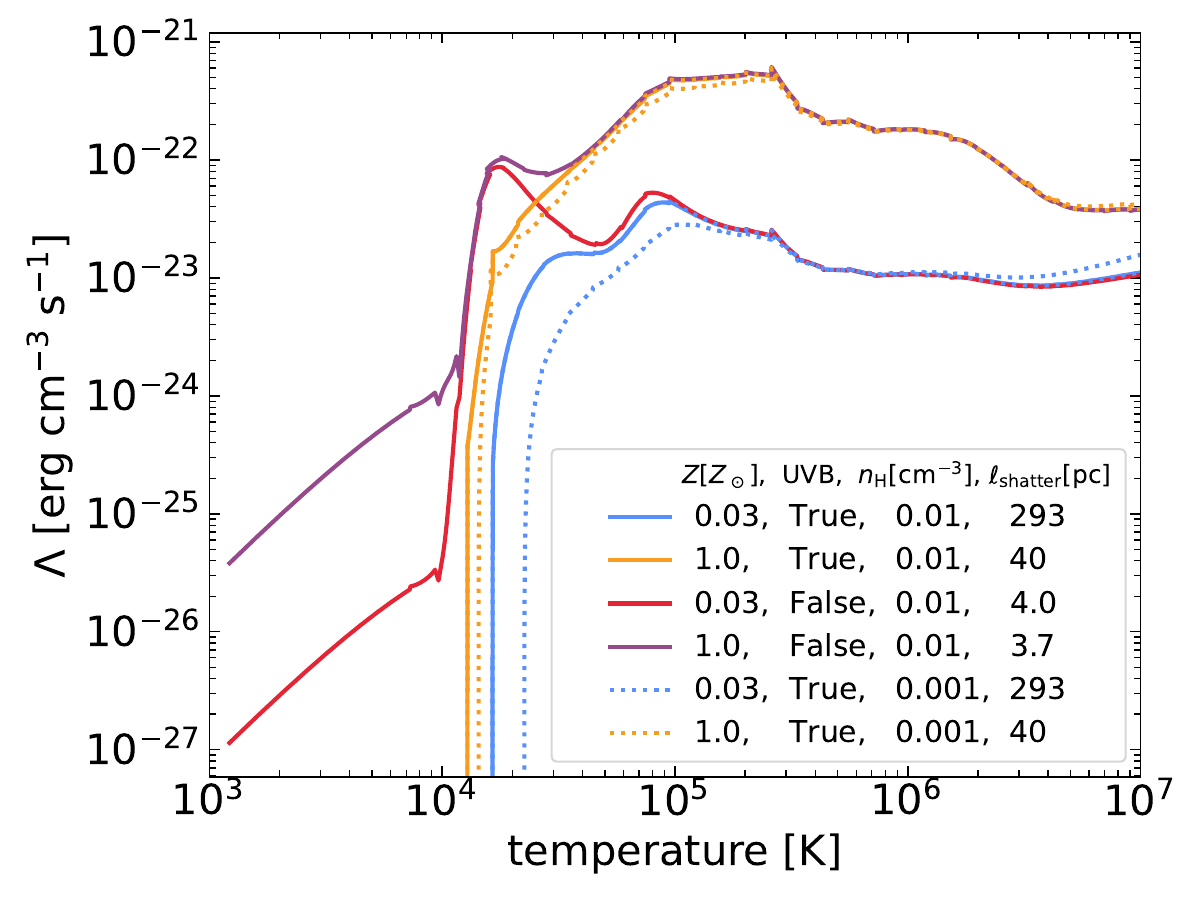}
   \caption{Net cooling rates as a function of temperature. We show results for gas with metallicity $Z=Z_\odot$ and $0.03\,Z_\odot$, with and without the $z=2$ \citet{Haardt.Madau.96} UVB. The values of $\ell_{\rm shatter}$ for each case are written in the legend. For cases with a UVB, solid lines represent gas with density $n_\mathrm{H}=0.01~{\rm cm^{-3}}$, similar to the cold phase in our simulations, while dashed lines represent gas with $n_\mathrm{H}=10^{-3}~{\rm cm^{-3}}$, similar to the initial warm gas for our fiducial $\eta=10$ and $\chif=100$. The purple line ($Z=Z_{\odot}$, no UVB) is relevant for the simulations analyzed in \se{geometry}, while the blue lines ($Z=0.03Z_{\odot}$, with UVB) are relevant for the simulations analyzed in \se{metallicity}.
   }
   \label{fig:CoolingRate}
\end{figure}

\begin{figure*}
    \centering	\includegraphics[width=\textwidth]{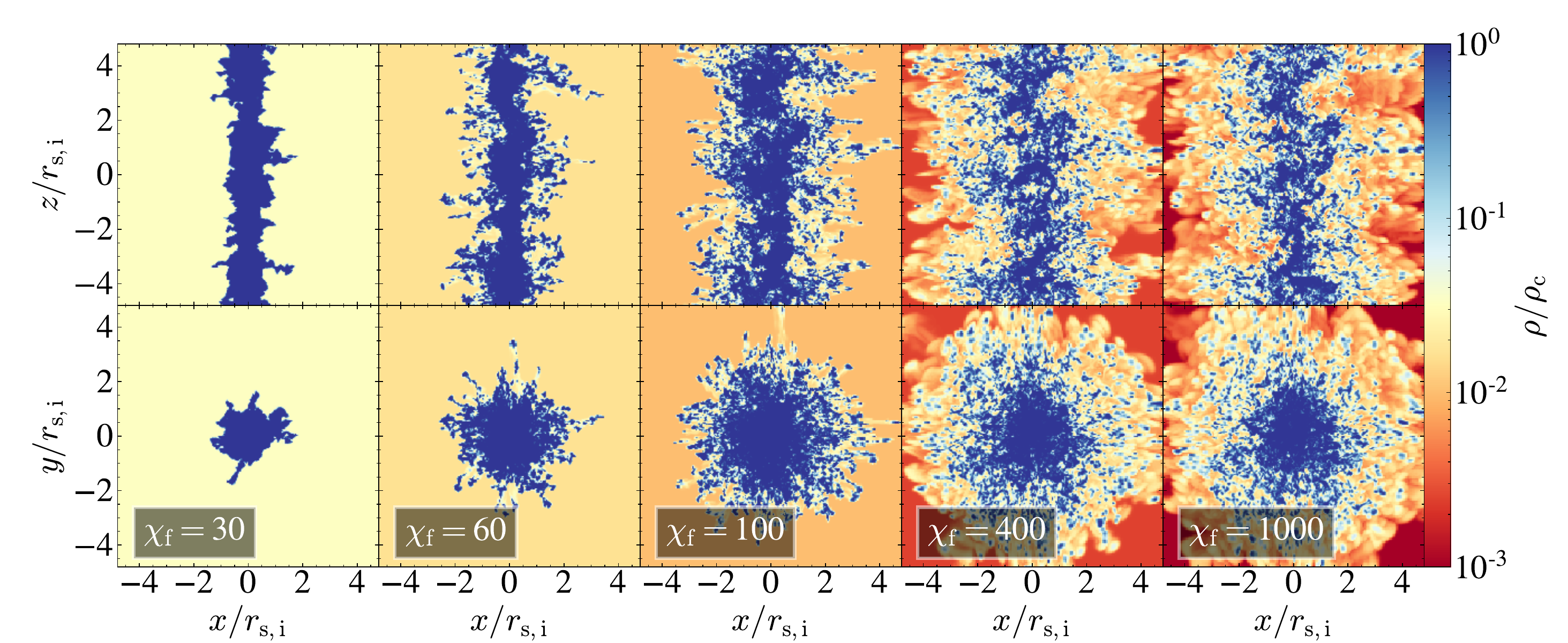}
   \caption{
   Similar to \fig{Stream_map}, but for low-metallicity streams with a UVB (\tab{sim_table}). Low-$Z$ streams with $\chi_\mathrm{f}=60$ exhibit shattering, compared to $\chi_{\rm f}\gsim 200$ for solar metallicity streams without a UVB. See also \fig{Stream_met_map_avg}.
   }
   \label{fig:Stream_met_map}
\end{figure*}

\begin{figure*}
    \centering	\includegraphics[width=\textwidth]{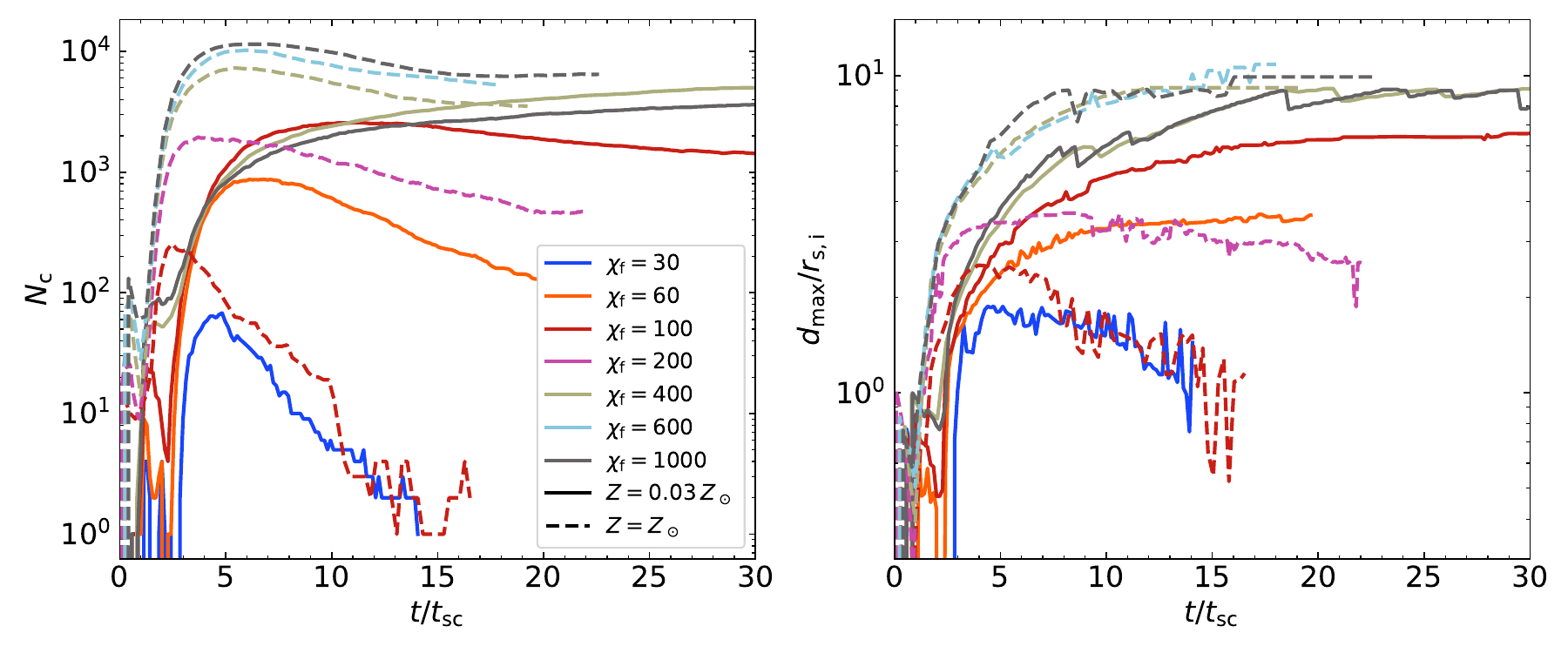}
   \caption{Time evolution of the number of clumps, $N_{\rm c}$ (\textit{left}, compare to \fig{Nclump}), and $d_\mathrm{max}$ (\textit{right}, compare to \fig{dmax}). Different colours represent different $\chi_\mathrm{f}$, while solid (dashed) lines represent low (high)-metallicity streams with (without) a UVB. Low-$Z$ streams have a smaller critical $\chif$ for shattering, $\chif\sim 60$ compared to $\sim 200$ for high metallicity streams. Due to an initial stage of isobaric cooling, the explosion is delayed by $\sim 2\tsc$ in low-$Z$ streams compared to high-$Z$ streams. Following the explosion, shattered low-$Z$ streams 
   produce fewer clumps and have slightly lower $\dmax$ values than high-$Z$ streams, though at late times the values are comparable.}
   \label{fig:Clump_dmax_met}
\end{figure*}

\begin{figure*}
    \centering	\includegraphics[width=\textwidth]{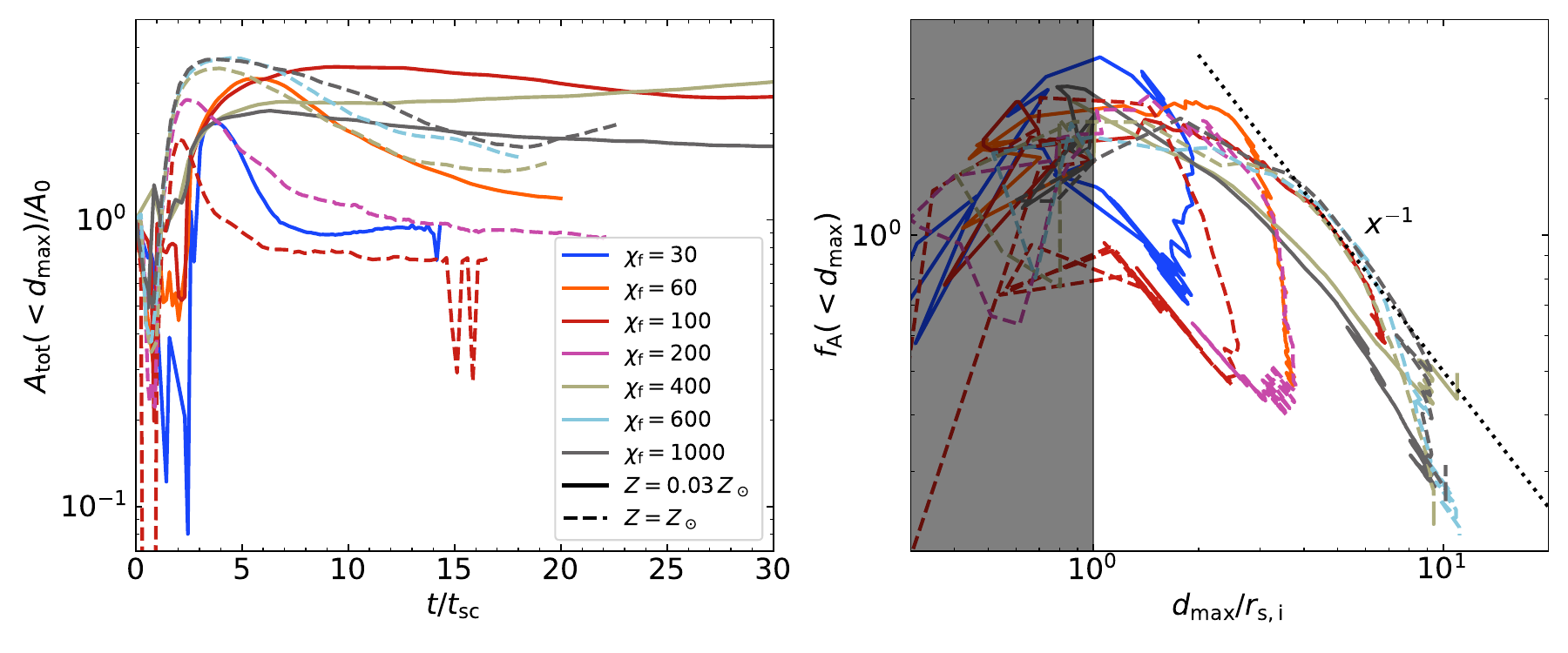}
   \caption{Time evolution of the total surface area of cold gas within $\dmax$, normalized to its initial value (\textit{left}), and the area modulation factor, $f_{\rm A}$, as a function of $\dmax$ (\textit{right}). This is similar to \fig{fA}, but here we compare high- and low-metallicity streams. Line styles and colours are as in \fig{Clump_dmax_met}. Despite certain differences in detail, the overall evolution of the cold-gas area is similar in low and high-$Z$ streams, saturating at values of $A_{\rm tot}\sim (2-3)A_0$ for shattering streams and $A_{\rm tot}\lsim A_0$ for coagulating streams.
   The behaviour of $f_{\rm A}$ is similar between the different metallicities, with shattered streams having $f_{\rm A}\propto d^{-1}$ at late times and large distances. 
   }
   \label{fig:Clump_area_met}
\end{figure*}

\begin{figure}
    \centering	\includegraphics[width=\columnwidth]{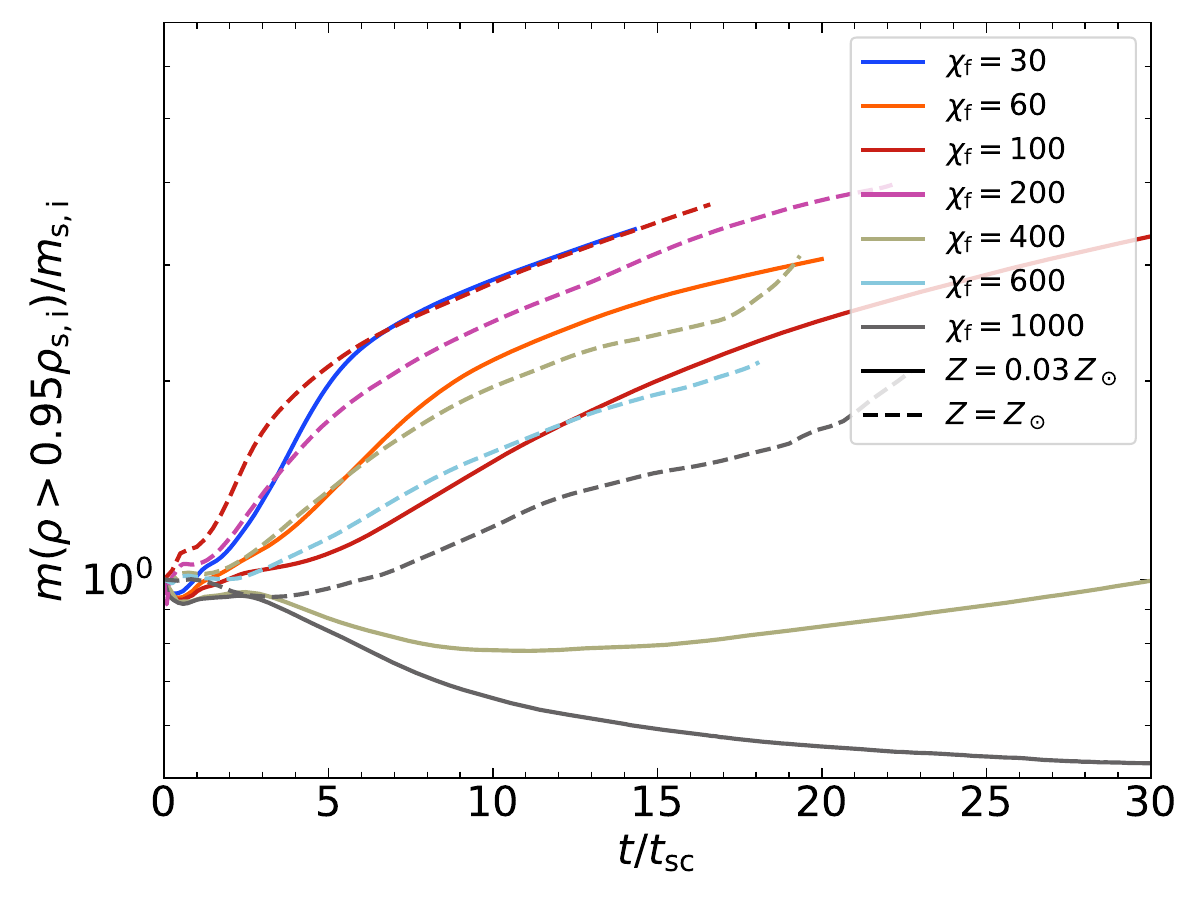}
   \caption{Time evolution of cold gas mass normalized by the initial stream mass. The cold gas is defined by 
   $\rho>\rho_{\rm s,i}/1.05$. Line styles and colours are as in \fig{Clump_dmax_met}. High-$Z$ streams are always in the growth regime, where the cold clumps grow in mass due to entrainment, thus increasing the total cold gas mass. Low-$Z$ streams, on the other hand, are in the disruption regime for $\chif\gsim 400$, causing the total cold gas mass to decrease, at least initially.}
   \label{fig:ColdMass_met}
\end{figure}


\section{Metallicity and UVB effects on shattering and coagulation}
\label{sec:metallicity}

Our analysis in the previous section focused on solar metallicity gas in collisional ionization equilibrium (CIE), 
similar to 
previous studies of shattering versus coagulation in quasi-spherical clouds \citep{Gronke.Oh.20b,Gronke.Oh.23}. While solar metallicity may be reasonable for gas in the inner CGM at $z\sim 0$, the metallicity in the high-$z$ cosmic web which is our primary focus is much lower. This lowers the cooling rates 
for intermediate temperature gas 
in the turbulent mixing layers 
between the cold clouds and hot background, 
thus lowering the efficiency of phase mixing and entrainment (\fig{CoolingRate}). Moreover, intergalactic gas is 
affected by the ionizing UVB, and photoionization is more important than collisional ionization over a wide temperature range \citep[e.g.][]{Strawn.etal.21,Strawn.etal.23}. This 
lowers the cooling rates near the cold phase, $T\gsim 10^4\K$ (\fig{CoolingRate}). These changes to the cooling curve affect both the shattering lengthscale, $\ell_{\rm shatter}$, and the strength of coagulation forces. Therefore, in this section we revisit some of our previous analysis focusing on low-metallicity gas exposed to a UVB. We focus here specifically on stream geometry and in \se{application} below we apply these results to cosmic web filaments at high-$z$. 

\smallskip
As stated in \se{methods}, we assume metallicity values of $Z_{\rm bg}=0.1Z_{\odot}$ for the background and $Z_{\rm s}=0.03Z_{\odot}$ for the initial stream, and apply the ionizing UVB of \citet{Haardt.Madau.96} at $z=2$. We do not include self-shielding of dense gas. At our fiducial densities and 
metallicities, 
the equilibrium temperature of the cold stream is $1.68\times 10^4 \K$, which was our chosen $T_{\rm floor}$ in the previous section. In this section, we do not use an artificial temperature floor since the UVB sets an effective cooling floor. Hereafter, when we refer to simulations of low metallicity (low-$Z$) streams it is understood that these are also exposed to a UVB, while high metallicity (high-$Z$) streams are not. 

\smallskip
While our initial conditions are the same as in \se{geometry} in terms of stream size, density, and temperature, the initial cooling rates are much lower (\fig{CoolingRate}). Therefore, unlike our simulations in \se{geometry} where the streams immediately cool isochorically, in our current simulations the streams initially contract isobarically and only loose sonic contact when they reach a radius of $r^{*}\sim2.5\,\mathrm{kpc}$ (see \tab{sim_table}). After this point, the implosion and explosion processes occur similarly at low- and high-$Z$. 
We note that despite the weaker radiative cooling, the explosion velocities are still roughly $v_{\rm ex}\sim c_\mathrm{s,c}$, and the dominant deceleration force as clumps escape outward is still $F_\mathrm{cond}$ (\equnp{Fratio2}). However, the corresponding $t_{\rm decel}$ at a fixed $\chif$ is longer for low-$Z$ streams, due to $v_\mathrm{mix}$ being smaller (\equnp{tcond}).

\smallskip
\Fig{Stream_met_map} shows maps of the maximal density along the line-of-sight in two orthogonal projections for low-$Z$ streams with different values of $\chif$, similar to \fig{Stream_map}. Complementary to this, we show in \fig{Stream_met_map_avg} the average density along the line-of-sight in the same projections. These simulations exhibit shattering at $\chif\gsim 60$, in contrast to $\chif\gsim 200$ at solar metallicity with no UVB, with $\chif=60$ at low-$Z$ being similarly borderline to $\chif=200$ at high-$Z$. This is further demonstrated in \fig{Clump_dmax_met}, which shows the number of clumps, $N_{\rm c}$, on the left and the maximal clump distance, $\dmax$, on the right for different $\chif$ values. Solid lines show results for low-$Z$ streams, while dashed lines show the same high-$Z$ simulations as \figs{Nclump} and \figss{dmax}. Note that the low-$Z$ runs extend to lower $\chif$ values. 
In general, the explosion phase and subsequent rapid rise of both $N_{\rm c}$ and $\dmax$ takes $\sim2\,t_\mathrm{sc}$ longer for the low-$Z$ streams because of the longer cooling times and the initial phase of isobaric cooling.
We find a lower $\chi_\mathrm{f,crit}$ for shattering in these simulations, with the $\chif=60$ run demonstrating weak to no coagulation, and the $\chif=100$ run displaying strong shattering. We will discuss this further in \se{fastvsslow}.

\smallskip
The shattering lengthscale in the low-$Z$ runs is $\sim 80$ times larger than in the high-$Z$ runs, as is the cooling time near the cold phase. From \equ{tcond}, we thus expect 
low-$Z$ clumps to propagate to distances $\sim 3$ times greater than high-$Z$ clumps.
This is indeed the case for $\chif=100$, as can be seen by comparing the solid and dashed red lines in the right-hand panel of \fig{Clump_dmax_met}. However, this is not evident in streams with higher $\chif$ where $\dmax>10\rsi$, due to artificial clump disruption in low resolution regions, as discussed in the context of \fig{dmax}. 
We also note the similarity between the evolution of $\chif=30$ ($60$) at low-$Z$ and $\chif=100$ ($200$) at high-$Z$. These simulations have a similar ratio of $\chi_\mathrm{f}/v_\mathrm{mix}$ (\equnp{vmix}) and a similar distribution of clump sizes (\se{sizes}) yielding a similar deceleration timescale (\equnp{tcond}).

\smallskip
\Fig{Clump_area_met} shows the total surface area of cold gas, $A_{\rm tot}(<\dmax)$ on the left, and the area modulation factor $f_{\rm A,dmax}$ (\equnp{fa}) on the right, as in \fig{fA}. We compare simulations with different $\chif$ at both low (solid) and high (dashed) metallicity. Despite some differences during the initial isobaric contraction phase, the overall behaviour of the cold gas surface area is similar at low- and high-$Z$. Following the explosion, the area increases rapidly, saturating at 
$A_{\rm tot}\sim (2-3)A_0$ for shattered streams, and at $A_{\rm tot}\lsim A_0$ for coagulated streams. As discussed following \fig{fA}, this corresponds to a cold-gas surface area $\sim 6$ times larger than the expected surface area in the final equilibrium state without shattering, namely a single stream of radius $\rsf$. 

\smallskip
Note that $A_{\rm tot}$ does not grow monotonically with $\chi_\mathrm{f}$ at low-$Z$, but rather reaches a maximum around $\chif\sim (100-400)$. This is due to the disruption of cold gas clumps with large $\chif$ by hydrodynamic instabilities caused by the interaction with the surrounding hot wind (i.e. cloud crushing). Radiatively cooling clouds can survive these instabilities and grow in mass by entrainment if their overdensity obeys \citep{Gronke.Oh.18}
\begin{equation}
\label{eq:chi_growth}
\chi_\mathrm{f}\lesssim1000\frac{P_{2.3}\Lambda_\mathrm{mix,-22.4}}{T_\mathrm{cl,4.3}^{5/2}M_{0.1}}\frac{r_\mathrm{cl}}{560\,\mathrm{pc}}{\tilde {\alpha}},
\end{equation}
where $P_{2.3}=P/(200{\rm cm^{-3}\K})$ is the thermal pressure of the hot gas, $T_{\rm cl,4.3}=T_{\rm cl}/(2\times 10^4\K)$ is the temperature of the cold cloud, $\Lambda_{\rm mix,-22.4}=\Lambda_{\rm mix}/(10^{-22.4}{\rm erg\,sec^{-1}\,cm^{3}})$ is the cooling rate of gas in the mixing layer with $T_{\rm mix}\sim (T_{\rm c}T_{\rm h})^{1/2}$ and $n_{\rm mix}\sim (n_{\rm c}n_{\rm h})^{1/2}$, $M_{0.1}=v_{\rm cl}/(0.1c_{\rm s,h})$ is the cloud Mach numberwith respect to the hot gas, $r_{\rm cl}$ is the cloud radius normalized to $560\pc$ which is roughly six high resolution cells, and ${\tilde {\alpha}}\sim 1$ in the `wind-tunnel' setup. To demonstrate this, we show the evolution of the total cold gas mass in \fig{ColdMass_met}. The cold mass increases with time for low-$Z$ streams with $\chif \lsim 1000$, while it decreases for larger $\chif$ as clouds move into the disruption regime\footnote{$\chif=400$ seems to be a borderline case between survival and disruption.}. For high-$Z$ streams, the cold mass increases for all values of $\chif$ due to the order-of-magnitude larger cooling rates.

\smallskip
The area modulation factor, $f_\mathrm{A}$ (right panel of \fig{Clump_area_met}), behaves similarly in the low- and high-$Z$ runs, and is as expected based on \fig{fA}. At early times when $\dmax\lsim \rsi$ (grey region), $f_{\rm A}$ displays chaotic behaviour, stemming from a competition between a decrease in cold gas surface area due to the contraction and an increase due to fragmentation. Shattering and coagulation are easily distinguishable by the behaviour of $f_{\rm A}$ at later times, once $\dmax > \rsi$. Coagulation manifests as a turnaround or saturation in the distance at $\dmax\sim (1-2)\rsi$ with no clear trend of $f_{\rm A}$ with $d$. Shattered cases, on the other hand, display $f_{\rm A} \propto d^{-1}$ in the range $\dmax\sim (3-10)\rsi$. Similar to $A_{\rm tot}$, $f_{\rm A}$ in low-$Z$ streams decreases at overdensities $\chif > 100$ due to cold gas disruption. However, $f_{\rm A}$ for $\chif=100$ at low-$Z$ is as large as it is for shattered high-$Z$ streams. 

\smallskip
One of the key observational indicators of shattering, and an important parameter in estimating the cold gas content of the CGM / cosmic web, is the area covering fraction of cold gas, $f_{\rm C}$. In shattered gas, this can be very large even if the volume filling fraction of the cold gas is small \citep{McCourt.etal.18,FG.Oh.2023}. While $f_{\rm C}$ overall behaves similarly to $A_{\rm tot}$, two clumps aligned along the same line of sight would both contribute to $A_{\rm tot}$ but not to $f_{\rm C}$, which can cause large differences if there are many small clumps in the background/foreground of large clumps. Nevertheless, we crudely estimated $f_{\rm C}$ perpendicular to the stream axis in a square region of side $10\rsi$, and found similar enhancements as seen in \fig{Clump_area_met} for $A_{\rm tot}$. Namely, typical values at late times roughly $\sim 2$ times larger than the covering fraction of the initial stream, and $\sim 6$ times larger than the covering fraction of a stream with radius $\rsf$ in thermal and pressure equilibrium. This was found to be true for both low- and high-$Z$ shattered streams. We defer a more detailed study of the cold gas covering fraction and clumping factor, particularly in the context of cold streams feeding massive halos at high-$z$ (see \se{application}), to future work with more realistic simulation setups.


\begin{figure*}
    \centering	\includegraphics[width=\textwidth]{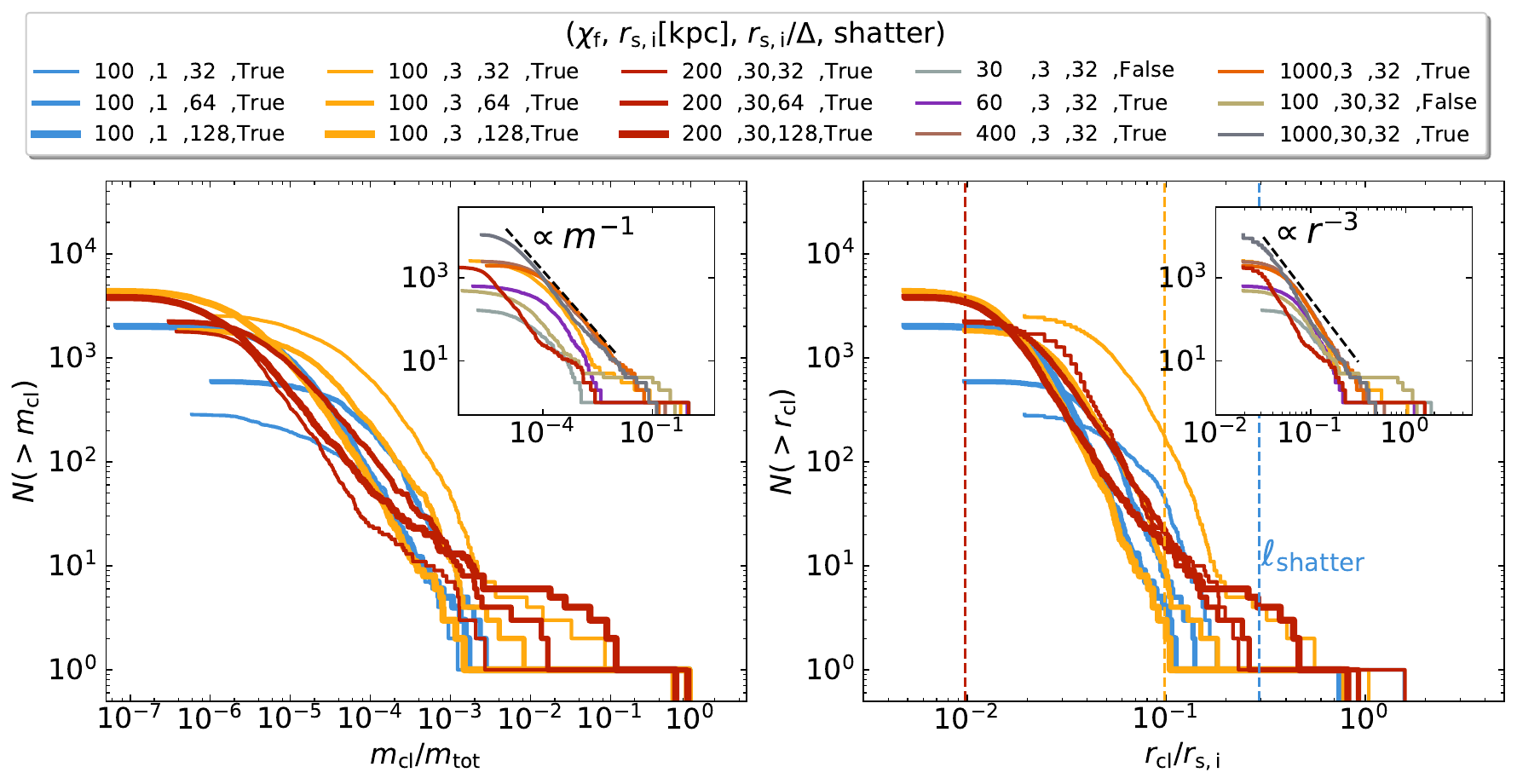}
   \caption{
   Cumulative distribution functions of clump masses (\textit{left}, normalized to the total cold gas mass at the relevant snapshot) and sizes (\textit{right}, normalized to the initial cloud radius) in simulations of low-metallicity streams with different values of $\chif$, $r_{\rm s,i}$, and $r_{\rm s,i}/\Delta$ as shown in the legend. The main panels show convergence tests for three different initial radii, $r_\mathrm{s,i}=1, 3, 30\,\mathrm{kpc}$, while the insets show all simulations with $r_\mathrm{s,i}/\Delta=32$. We show the distributions at $t\sim 10\,t_\mathrm{sc}$ for shattering cases, and at $t\sim 5\,t_\mathrm{sc}$ for coagulating cases when $N_{\rm c}$ is at its peak. The sizes of clumps are derived assuming spherical geometry, $r_\mathrm{cl}=[3m_\mathrm{cl}/(4\pi \rho_\mathrm{cl})]^{1/3}$, where $m_\mathrm{cl}$ and $\rho_\mathrm{cl}$ are obtained from the clump finder. For shattered streams, the clump mass function follows a Zipf's law-like distribution, $N(>m)\propto m^{-1}$ \citep{Gronke.etal.22,Fielding.etal.23}. The size function is slightly steeper than $r_\mathrm{cl}^{-3}$, due to a broader scatter of clump density for small clumps. The vertical lines in the right-hand panel show the values of $\ell_{\rm shatter}/r_{\rm s,i}$ for $r_{\rm s,i}=1\kpc$, $3\kpc$ and $30\kpc$ from right to left respectively. Even when $\ell_{\rm shatter}$ is resolved, the size distribution remains a power law down to the resolution scale without a feature at $\ell_{\rm shatter}$.}
   \label{fig:Clump_mass_dist}
\end{figure*}

\section{Clump Sizes and Masses}
\label{sec:sizes}

\smallskip
Our estimate of the coagulation time (\equnp{tcoag}) and the deceleration time (\equnp{tcond}), as well as their ratio (\equnp{tratiomax}, which determines the efficiency of coagulation in our model) all depend on the radii of clumps resulting from the initial shattering process. It therefore behooves us to quantify the distribution of clump sizes. In the initial formulation of the shattering model, the final scale of cold gas clouds was predicted to be $r_{\rm c}\sim \ell_{\rm shatter}\equiv {\rm min}(c_{\rm s}t_{\rm cool})$ \citep{McCourt.etal.18}. The emergence of this as a characteristic lengthscale for cold gas 
was also discussed in previous analytic models of non-linear thermal instability \citep{Burkert.Lin.00,Waters.Proga.19a}. Motivated by these insights, some cosmological simulations have implemented cooling-length based refinement models in order to resolve cold gas in the CGM \citep{Peeples.etal.19}, while others have implemented sub-grid models for the existence of unresolved cold-gas clouds of size $\ell_{\rm shatter}$ \citep{Butsky.etal.24}. Numerical simulations of non-linear thermal instability in 1D find that while the minimum cloud size was of order $\ell_{\rm shatter}$, larger clouds were common due to merging of smaller clouds \citep{Das.etal.21}. However, the size distribution of post-shattering cloudlets has not been constrained in 3D simulations of thermal instability which resolve $\ell_{\rm shatter}$. 

\smallskip
Several works have studied the size distribution of cold clumps in multiphase gas in other contexts. 
\citet{Gronke.etal.22} studied the mass distribution of cloudlets forming not through an implosion-explosion process as in this work, but rather by placing large clouds that are only slightly out of thermal equilibrium (by a factor of $\sim 2$) in a (driven) turbulent box. They found that when the clouds were in the growth regime (see \equnp{chi_growth}), the clump mass function in their simulations converged to $dN/dm\propto m^{-2}$ over a wide range of simulation parameters. This implies roughly constant mass per logarithmic mass bin, and was found to extend down to the resolution limit of the simulation. Similar results were found in larger scale simulations of AGN jets in cool-core clusters \citep{Li.Bryan.14} and in MHD simulations of a multiphase ISM \citep{Fielding.etal.23}. However, these works did not resolve $\ell_\mathrm{shatter}$ so they could not comment on whether this would serve as a minimal or characteristic size of cold gas cloudlets (see also \citealp{Jennings.etal.23}). Cloud-crushing simulations of large, thermally stable cold clouds interacting with a hot wind that marginally resolve $\ell_\mathrm{shatter}$ find that the size distribution of cold clumps forming in turbulent mixing layers downstream does not appear converged, with clouds smaller than $\ell_\mathrm{shatter}$ common \citep{Sparre.etal.19, Gronke.Oh.20}. On the other hand, \citet{Liang.Ian.20} performed 2D cloud crushing simulations where the initial cloud was thermally unstable with a fractal structure consisting of gas with temperatures ranging from $(10^{4.3}-10^{6.5})\,{\rm K}$ and a median of $\sim 10^5\,{\rm K}$. They found that the resulting cold gas clouds had a characteristic column density of $N_{\rm H}\sim 10^{17.5}\cms$, consistent with predictions for a characteristic cloud size of $\ell_{\rm shatter}$ \citep{McCourt.etal.18}. However, $\ell_{\rm shatter}$ was only marginally resolved in their simulations and the characteristic scale they discovered may have in fact been tied to the grid scale. Moreover, the cloud crushing setup is in general quite different from our own, and its relevance for the pure thermal instability picture discussed in \citet{McCourt.etal.18} is unclear. 
To summarise, while it appears clear that in order for thermally unstable clouds to shatter they must be much larger than $\ell_\mathrm{shatter}$ \citep{McCourt.etal.18, Sparre.etal.19, Gronke.Oh.20b, Farber.Gronke.23}, and that both the total mass and clumpiness of cold gas in multiphase media increase as the resolution approaches $\ell_\mathrm{shatter}$ \citep{Peeples.etal.19, Mandelker.etal.21}, it remains unclear whether this is indeed a lower limit or a characteristic value of the cloud size distribution. 

\smallskip
Our simulations offer a unique opportunity to study this issue, since unlike most previous works we focus on low metallicity gas at relatively low pressures of $P/k_{\rm B}\sim 100\,{\rm K}\,{\rm cm}^{-3}$ exposed to a strong $z=2$ UVB (\se{metallicity}). These result in $\ell_{\rm shatter}\sim 300\pc$ (\tab{sim_table}), a factor of $\sim (100-1000)$ larger than typical in most other works. For instance, in our simulations with solar metallicity presented in \se{geometry}, $\ell_{\rm shatter}\sim 4\pc$. In \fig{Clump_mass_dist}, we study the distribution of clump sizes in simulations of low-$Z$ streams with $\eta=10$ exposed to a $z=2$ UVB, as in \se{metallicity}. All these simulations have $\ell_{\rm shatter}\sim 300\pc$.
We present results for streams with initial radii $r_{\rm s,i}=1\kpc$, $3\kpc$ (these are the simulations presented in \se{metallicity}), and $30\kpc$. At our fiducial resolution of $\Delta=r_{\rm s,i}/32$, these correspond to cell sizes of $\sim 31\pc$, $94\pc$, $938\pc$ in the high-resolution region, corresponding to $\Delta/\ell_{\rm shatter}\sim 0.1$, $0.32$, and $3.2$, respectively. The simulations with $r_{\rm s,i}=30\kpc$ thus do not quite resolve $\ell_{\rm shatter}$, while our fiducial simulations with $r_{\rm s,i}=3\kpc$ marginally resolve it. In the simulations with $\rsi=1\kpc$, $\ell_{\rm shatter}$ is well resolved, though these simulations have 
$r^*\sim 2\ell_{\rm shatter}$ (\tab{sim_table}) due to the initial phase of isobaric cooling discussed in \se{metallicity}, so shattering is expected to be weak \citep{Gronke.Oh.20b}.

\smallskip
For each value of $r_{\rm s,i}$, we perform additional higher resolution simulations with $\Delta=r_{\rm s,i}/64$ and $r_{\rm s,i}/128$, to test convergence. In these simulations, $\ell_{\rm shatter}$ is well resolved for $r_{\rm s,i}=3\kpc$ and marginally resolved even for $r_{\rm s,i}=30\kpc$. We note that in these high-resolution simulations we simulate a smaller portion of the stream with uniform resolution throughout the entire box. The stream radius here is only $1/4$ of the box size as opposed to $1/32$ in our fiducial simulations. The total number of clumps is thus not directly comparable between our fiducial resolution and the two higher resolution runs. However, we have verified that this change does not affect the physics of shattering versus coagulation nor the distribution of clump properties.

\smallskip
In the left panel of \fig{Clump_mass_dist}, we show the cumulative clump mass distribution for these simulations, obtained directly from the clump finder. The main panel shows results from simulations with varying resolution, while the inset displays results from all simulations with $\rsi/\Delta=32$ listed in the legend. We show the distributions at $t=10\tsc$ for shattered streams, and at $t\sim 5\tsc$ for coagulating streams which is when $N_{\rm c}$ is at its peak. In all cases, the number of clumps is clearly dominated by low mass clumps, though a single massive clump contains anywhere from $\sim (15-95)\%$ of the total mass, with more strongly shattered cases having a smaller maximal clump mass. Recall, however, that the total cold gas mass can be a factor of $\lsim 2$ larger or smaller than the original cloud, depending on its size and overdensity (\fig{ColdMass_met}). In shattered streams, a power law distribution of $N(>m_{\rm cl})\propto m_{\rm cl}^{-1}$ develops, consistent with previous studies \citep{Gronke.etal.22,Fielding.etal.23}. Such a scale-free distribution is emblematic of a much more general phenomenon known as Zipf's law \citep{Gronke.etal.22}. It requires a large dynamic range of clump sizes to develop, and has been argued to stem from the fact that for a highly fragmented cold gas distribution, the cold gas mass grows as ${\dot{m}}_{\rm cold,tot}\propto m_{\rm cold,tot}$ while the growth of the lowest mass clumps is dominated by mergers \citep{Gronke.etal.22}. 

\smallskip
At the high-mass end, we see deviations from the power-law behaviour as the distribution becomes dominated by a few large clumps along the stream axis onto which many of the intermediate mass clumps have coagulated. The turnover at low masses is simply due to the resolution limit which sets a minimal mass for clumps, and is partly affected by artificial disruption of small clumps in low resolution regions far from the stream. Even in cases which don't shatter, $(r_{\rm s,i},\chif)=(3\kpc,30),\,(30\kpc,100)$, this power-law holds over a narrow range of intermediate masses where fragmentation is important. However, the global distribution in these cases is not well described by Zipf's law since they are not highly fragmented and their distribution is predominanly shaped by coagulation. 

\smallskip
Larger streams, with larger $r^*/\ell_{\rm shatter}$, exhibit more violent fragmentation with many more clumps (compare the runs with $\chif=1000$ and $\rsi=3\kpc$ and $30\kpc$ in the insets). Comparing the curves with different resolutions in the main panels, we find that higher resolution runs produce more low-mass clumps indicative of stronger fragmentation. This is consistent with previous studies of the formation of multiphase media in other contexts \citep{Sparre.etal.19,Mandelker.etal.21}. However, the shape of the distribution is quite similar in all cases, developing a power law of $N(>m)\propto m^{-1}$ over a wide range of clump masses, which turns over at the resolution scale, and not at some fixed scale despite $\ell_{\rm shatter}$ being resolved. 

\smallskip
However, whether the clump mass distribution is truly an example of Zipf's law is not completely clear. We note that the full distribution down to the minimal clump mass is much better described by a log-normal distribution, rather than a power-law with a turnover (see Appendix~\ref{sec:lognormal}). The mode of the distribution is always at a clump mass slightly larger than the mass of a single cell at the cold density, even when $\ell_{\rm shatter}$ is well resolved, supporting our conclusion that $\ell_{\rm shatter}$ does not set a characteristic size for clumps.

\begin{figure}
    \centering	\includegraphics[width=\columnwidth]{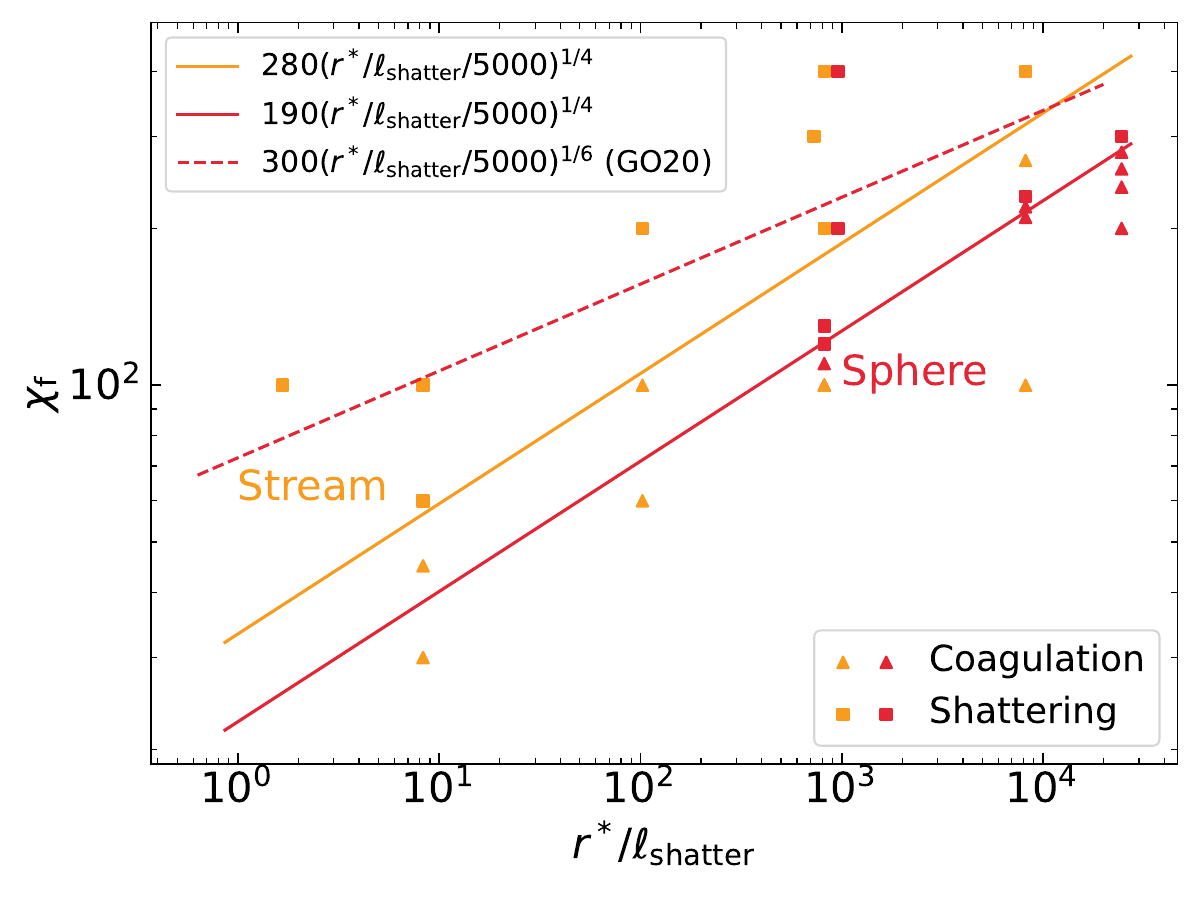}
   \caption{Shattering versus coagulation in the plane of cloud size and overdensity, $(r^*/\ell_{\rm shatter}, \chif)$, for simulations in spherical (red) and cylindrical (orange) geometries, with different initial sizes and metallicities and different values of $\eta$ and $\chif$ (see \tab{sim_table}). Squares represent those simulations with strong shattering, indicated by both a large number of clumps and a non-decreasing $d_\mathrm{max}$, while triangles mark simulations that coagulate. We introduce a small horizontal offset for those overlapping with each other. The orange/red solid line denotes the transition of coagulation for streams/spheres. The dashed line represents the shattering criterion in \citet{Gronke.Oh.20b}.}
   \label{fig:ShatteringCriterion}
\end{figure}

\smallskip
The right-hand panel of \fig{Clump_mass_dist} shows the cumulative distribution of clump radii. These are estimated as\footnote{Note that this implies that at our fiducial resolution of $\Delta=r_{\rm s,i}/32$ a clump consisting of a single cell has a radius of $r_{\rm cl}\sim 0.02r_{\rm s,i}$.} $r_{\rm cl}=[3m_{\rm cl}/(4\pi\rho_{\rm cl})]^{1/3}$, where $\rho_{\rm cl}$ is the mean density in the clump, provided by the clump finder. The vertical dashed lines mark $\ell_{\rm shatter}/r_{\rm s,i}$ for $r_{\rm s,i}=1,\,3,\,30\kpc$, from right to left. Focusing on simulations with $r_{\rm s,i}=1\kpc$ and $3\kpc$ where $\ell_{\rm shatter}$ is resolved, we see that there is no feature in the distribution of clump sizes at $\ell_{\rm shatter}$ for shattered streams. Rather, the distribution is a power-law with $N(>r_{\rm cl})\propto r_{\rm cl}^{-\beta}$ with $\beta\sim(3-4)$, which breaks at the resolution limit. Considering that $N\propto m_\mathrm{cl}^{-1}$, this indicates a mild size-dependence of clump density of $\rho_\mathrm{cl}\propto r_\mathrm{cl}^{\beta-3}$, which is driven by smaller clumps containing more mixed gas at intermediate densities. As we increase the resolution, the break in the power-law and the minimal clump size tend towards smaller sizes, rather than being tied to $\ell_{\rm shatter}$. This suggests that clump sizes are not set by a hierarchical process where clumps continuously fragment to the local cooling-length, $l_{\rm cool}\sim c_{\rm s}t_{\rm cool}$, until they reach $\ell_{\rm shatter}={\rm min}(l_{\rm cool})$ as was originally proposed by \citet{McCourt.etal.18} (see also \citealp{Das.etal.21}). Rather, while $\rsi> \ell_{\rm shatter}$ may be a necessary condition for thermally unstable clouds to shatter, the actual formation of clumps is not itself driven by thermal instability. Rather, these are formed by a combination of RMI during the initial implosion and explosion processes \citep{Gronke.Oh.20b}, and shredding \citep{Jennings.Li.21} and/or rotational fragmentation (`splintering') at late times \citep{Farber.Gronke.23}. The clump sizes are thus determined by these processes and clumps can always exist at the smallest possible scales in a highly perturbed environment. Without additional processes such as thermal conduction \citep{Koyama.Inutsuka.04,Sharma.etal.2010} or strong external turbulence \citep{Tan.etal.21,Gronke.etal.22}, we will always have clumps down to the grid-scale when shattering is present.

\smallskip
Consistent with the fact that $\ell_{\rm shatter}$ is not a characteristic size for clumps, we also find that $n_{\rm c}\ell_{\rm shatter}$ is not a characteristic column density for clumps, where $n_{\rm c}$ is the number density in the cold phase. Rather, the distribution of column densities always peaks at the grid scale, $N\sim n_{\rm c}\Delta$ (see \fig{Clump_column_density}). While this may seem to contradict the results of \citet{Liang.Ian.20}, we again note that $\ell_{\rm shatter}$ was only marginally resolved in their simulations, so it is difficult to disentangle this from the grid scale, and moreover their simulation setup is sufficiently different from ours to render a detailed comparison difficult and beyond the scope of this paper.

\smallskip
Finally, we note that the power-law describing the clump size distribution seems to break at $r_{\rm cl}\sim r_{\rm s,i}/3\sim r_{\rm s,f}$. This seems to be the characteristic size of large clumps along the stream axis onto which many intermediate sized clumps have coagulated (\figs{Stream_map}, \figss{Clump_dmax_met}, \figss{Stream_map_avg}, and \figss{Stream_met_map_avg}). A similar feature is seen in the cumulative distribution of clump sizes in sheets, where a power-law is present from the resolution scale up to $r_{\rm cl}\sim r_{\rm s,f}$ above which it flattens, decreasing again at $r_{\rm cl}\gsim r_{\rm s,i}$. The small clumps are visible in the plane of the sheet, within islands of hot gas in between large cold clumps (\figs{Sheet_map} and \figss{Sheet_map_avg}). The large cold clumps surrounding these hot regions have typical sizes of order $r_{\rm s,i}$ rather than $r_{\rm s,f}$. The reason for this is unclear, though further investigation of it is beyond the scope of this paper and is left for future work.

\section{The shattering criteria for streams and spheres}
\label{sec:shatter}

\begin{figure*}
    \centering	\includegraphics[width=\textwidth]{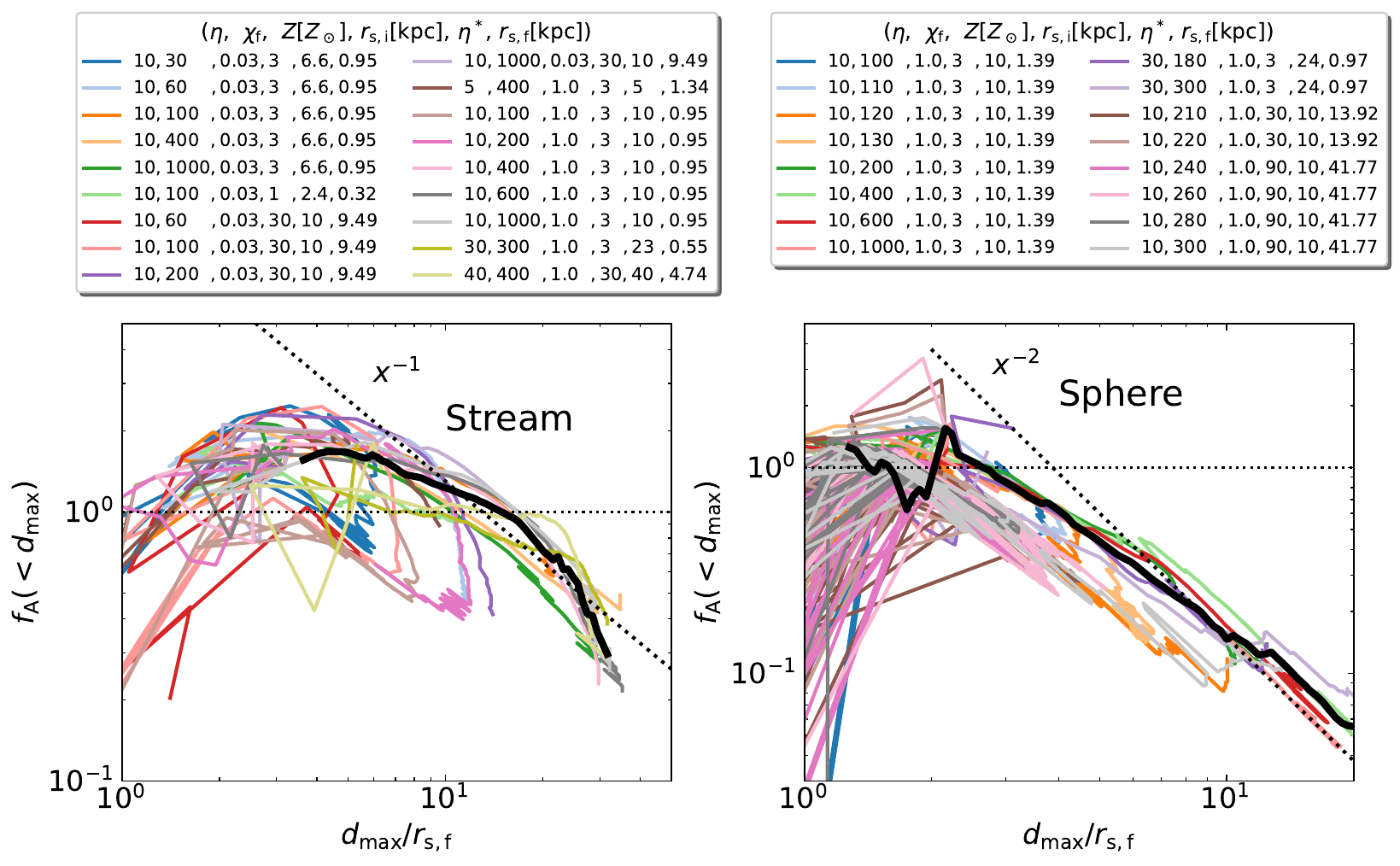}
   \caption{The area modulation factor $f_\mathrm{A}$ as a function of $d_\mathrm{max}$ normalized by $r_\mathrm{s,f}$ for all stream simulations (\textit{left}) and sphere simulations (\textit{right}). Different colours represent different parameters as shown in the legend. The thick black lines in each panel show the median profiles, considering shattering cases only. We define $r_\mathrm{fA}$ to be the radius where $f_\mathrm{A,dmax}$ starts declining, which is $\gsim6\,r_\mathrm{s,f}$ for streams and $\gsim2\,r_\mathrm{s,f}$ for spheres.}
   \label{fig:rsf}
\end{figure*}

\subsection{The shattering criterion and slow coagulation}
\label{sec:fastvsslow}
Our results in Section~\ref{sec:geometry} and Section~\ref{sec:metallicity} revealed a critical overdensity for shattering in streams and spheres, $\chi_{\rm f,crit}$, based on both the number of clumps, $N_{\rm c}$, and their radial extent, $d_\mathrm{max}$. Similar to previous work \citep{Gronke.Oh.20,Farber.Gronke.23}, we find that $\chi_{\rm f,crit}$ depends on the initial cloud size, or more specifically on $r^*/\ell_{\rm shatter}$. In \fig{ShatteringCriterion}, we present a wide range of simulations with different initial sizes and metallicities and different values of $\eta$ and $\chif$, in both spherical and stream geometries (\tab{sim_table}), in the plane of $(r^*/\ell_{\rm shatter}, \chif)$. We distinguish those cases which shattered from those that coagulated as in previous sections, and in Appendix~\ref{subsec:shatteringcriterion} we show that this distinction is converged with resolution. We find that for both streams and spheres, $\chi_{\rm f,crit}\propto (r^*/\ell_{\rm shatter})^{1/4}$, and that the normalization for streams is $\sim 1.5$ times larger than for spheres. Below, we attempt to explain these trends and derive expressions for $\chi_{\rm f,crit}$ in different geometries.

\smallskip
We recall that following the explosion (\se{shattering}), the reflected shock reaches the interface between the cold and hot gas when the cloud radius is roughly $r\sim r_{\rm s,f}$. At this point, linear RMI begins growing at the interface, seeded by perturbations in the collapsed cloud and at its interface, while on larger scales the cloud continues expanding isotropically with its surface area growing as $d_\mathrm{max}^n$, with $n=1$ and $2$ for streams and spheres respectively. Thus, $f_\mathrm{A}$ remains roughly constant during this phase, or may even increase due to perturbations and some fragmentation further increasing the surface area. Eventually, RMI grows to non-linear amplitudes characterised by familiar mushroom-shaped plumes and other non-isotropic features which quickly fragment into small clumps. The surface area then quickly becomes dominated by small clumps which do not further expand as they flow out, and the tight correlation between the total cloud surface area and $d_{\rm max}$ breaks down. Once the gas has sufficiently fragmented, the total cold gas surface area saturates and $f_\mathrm{A}$ decreases as the fragmented clumps escape outwards, asymptotically approaching $d_\mathrm{max}^{-n}$ . Consequently, the coagulation timescale increases significantly $t_\mathrm{coag}\propto (d_\mathrm{max}/f_\mathrm{A})^{1/2} \propto d_\mathrm{max}^{(n+1)/2}$ (\equnp{tcoag}). 

\smallskip
We can now formulate a condition for slow, inefficient, coagulation as follows. We assume that the fragmentation of the initial cloud into small clumps occurs at $r\sim r_{\rm fA}$, the radius where $f_{\rm A}$ begins monotonically declining following the explosion. If the cloud reaches this radius before significantly decelerating, then the clumps will continue expanding outwards with a velocity of $\sim c_{\rm s,c}$ and will not coagulate, following the arguments presented in \se{theory}. If the cloud has significantly decelerated prior to fragmentation, then the resulting clumps will coagulate. In other words, for coagulation to become inefficient we demand 
\begin{equation}
\label{eq:shatteringcriterion}
    t_\mathrm{decel,s}>r_\mathrm{fA}/c_\mathrm{s,c},
\end{equation}
where $t_\mathrm{decel,s}$ is the deceleration timescale of the expanding cloud. If there is a turbulent mixing layer at the surface of the cloud, which is always the case in the presence of shape and/or density perturbations and in the absence of thermal conduction, then the cloud deceleration is dominated by condensation rather than ram pressure because the expansion velocity is $v_{\rm ex}\sim c_{\rm s,c}$ (\equnp{Fratio}). 
We therefore have $t_\mathrm{decel,s}=m_\mathrm{s,i}/\dot{m}$, where $\dot{m}=4\pi r_\mathrm{avg}^2\rho_\mathrm{h}v_\mathrm{mix,s}$ for spheres and $\dot{m}=2\pi r_\mathrm{avg}L\rho_\mathrm{h}v_\mathrm{mix,s}$ for streams, with  $v_\mathrm{mix,s}\sim0.4c_\mathrm{s,c}(r_\mathrm{avg}/\ell_\mathrm{shatter})^{1/4}$ and $r_\mathrm{avg}\sim r_\mathrm{fA}$.

\smallskip
All that remains is to estimate $r_{\rm fA}$. In \fig{rsf} we show $f_{\rm A}(<d_{\rm max})$, similar to \figs{fA} and \figss{Clump_area_met}, for a large number of simulations with a wide range of different properties in both spherical and stream geometries (see \tab{sim_table}). Unlike previous figures, we here plot $f_{\rm A}$ as a function of $r/r_{\rm s,f}$ in order to distinguish cases with different $\eta$. For this wide range of simulation parameters, we find that $r_\mathrm{fA}\sim \alpha r_\mathrm{s,f}$, where $\alpha\sim 6$ for streams and $\alpha\sim 2$ for spheres. This can be more easily seen by examining the median profiles of $f_{\rm A}$, which are shown with thick black lines. 
The larger $\alpha$ for streams may be due in part to the Bell-Plesset effect \citep{bell1951taylor,plesset1954stability}, which is a geometrical factor in the linear growth rate of RMI for converging or expanding shock-waves, which can be $\lsim 2$ times faster in spherical compared to cylindrical geometry. Additionally, if the total surface area saturates during the non-linear phase of RMI in both geometries, then $f_{\rm A}\propto d_{\rm max}^{-n}$ declines faster in spherical geometry than in cylindrical geometry, yielding a larger $r_{\rm fA}$ for the latter. 

\smallskip
Inserting our expressions for $t_\mathrm{decel,s}$ and $r_\mathrm{fA}$ into the equation~\ref{eq:shatteringcriterion}, we obtain a critical overdensity for shattering, 
\begin{equation}
\label{eq:chi_crit_sphere}
\chi_\mathrm{f}>190\left(\frac{\alpha}{2.6}\right)^{13/4}\left(\frac{\eta^*}{10}\right)^{-1/12}\left(\frac{r^*/l_\mathrm{shatter}}{5000}\right)^{1/4}
\end{equation}
for spheres, and 
\begin{equation}
\label{eq:chi_crit_stream}
\chi_\mathrm{f}>280\left(\frac{\alpha}{6}\right)^{9/4}\left(\frac{\eta^*}{10}\right)^{-1/8}\left(\frac{r^*/l_\mathrm{shatter}}{5000}\right)^{1/4}
\end{equation}
for streams, where $\eta^*\equiv\rho_\mathrm{s,f}/\rho(r=r^*)$
and we have chosen a characteristic value of $\alpha$ to set the normalization. 
These two relations are plotted in \fig{ShatteringCriterion}, ignoring the weak dependence on $\eta^*$, and are in very good agreement with the 
shattering thresholds for spheres and streams 
seen in the simulations. In fact, the characteristic values of $\alpha$ in \equs{chi_crit_sphere}-\equm{chi_crit_stream} were determined based on a fit to the threshold seen in \fig{ShatteringCriterion}, but are also nicely consistent with \fig{rsf}.

\smallskip
We stress that the condition derived above for rapid coagulation does not explicitly depend on metallicity, or on the cooling rate more generally. This is because, as highlighted in \se{theory}, the condensation force dominates the deceleration process for all relevant parameters, including low-metallicity clouds/streams exposed to a UVB. 
However, the slower cooling and corresponding larger $\ell_{\rm shatter}$ yield a lower critical overdensity for a given cloud/stream size.

\begin{figure}
    \centering	\includegraphics[width=\columnwidth]{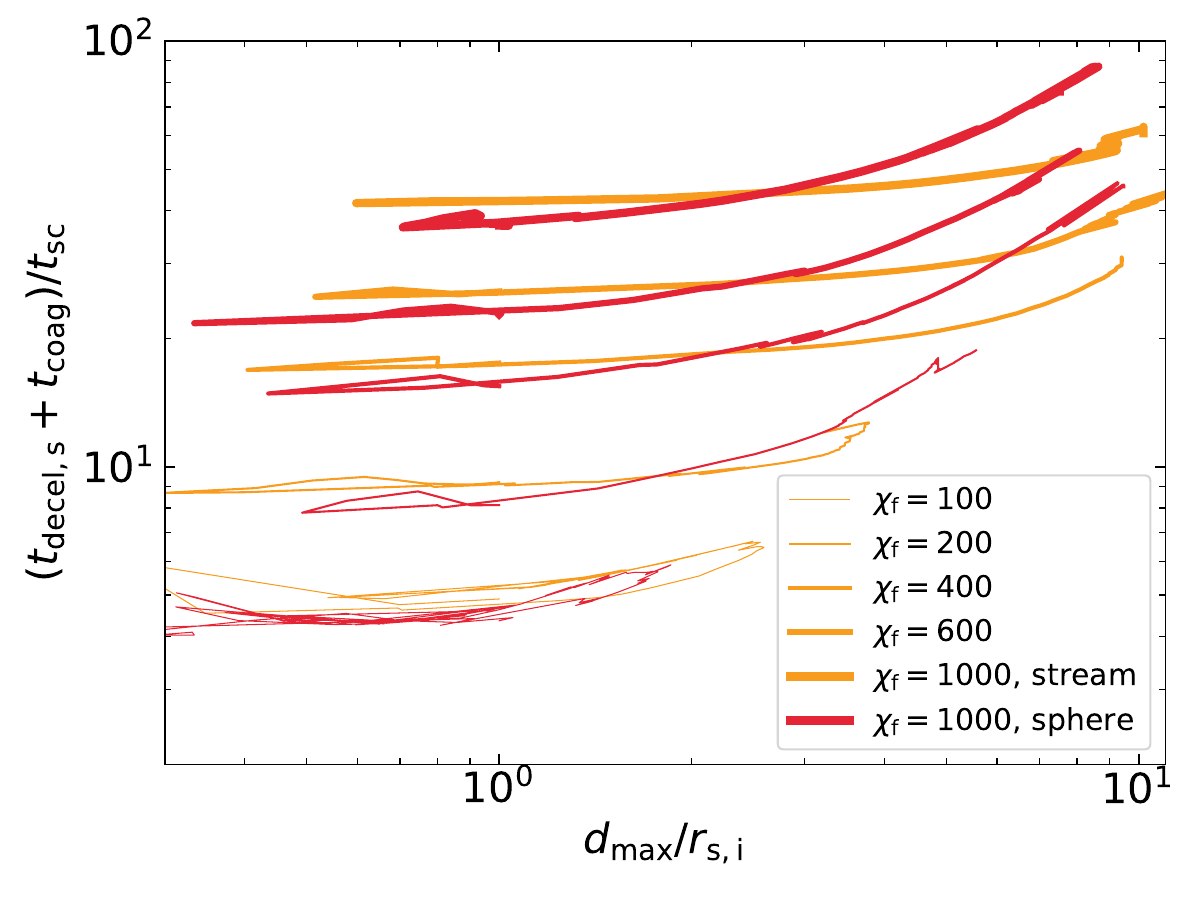}
   \caption{We show the total time for clumps to turnaround and coagulate, $t_{\rm decel,s}+t_{\rm coag}$, as a function of their maximal distance, $d_\mathrm{max}$. $t_{\rm decel,s}$ is the deceleration timescale for a uniformly expanding cloud (\equnp{shatteringcriterion}), while $t_{\rm coag}$ is computed from \equ{tcoag} assuming a clump radius of $r_{\rm cl}\sim 0.1r_{\rm s,i}$. 
   Red (orange) lines represent spheres (streams) with $\eta=10$, $r_{\rm s,i}=3\kpc$, $Z=Z_{\odot}$, and no UVB (as in \se{geometry}). Thicker lines denote higher $\chif$. The total time is dominated by $t_{\rm decel}$ at small $d_\mathrm{max}$ and by $t_\mathrm{coag}\propto d_\mathrm{max}^{n/2+1}$ at large radii.
   }
   \label{fig:Timescales}
\end{figure}

\smallskip
To emphasize the transition between fast and slow coagulation, we plot in \fig{Timescales} the total time for outgoing clumps to turn around and coagulate, namely $t_{\rm decel,s}+t_{\rm coag}$, as a function of $d_{\rm max}$, for high-$Z$ spheres and streams with $\eta=10$ and $r_{\rm s,i}=3\kpc$ (as in \se{geometry}). 
These are all measured directly from the simulation, but $d_{\rm max}$ here can be interpreted as the hypothetical maximal distance reached by a clump prior to turnaround, with the $y$ axis showing the total time to coagulation given this $d_{\rm max}$. At small distances, this timescale is dominated by $t_{\rm decel}$ which is independent of $d$ (\equnp{tcond}). However, once the maximal distance exceeds $r_{\rm max}\sim r_{\rm fA}$, the timescale quickly becomes dominated by $t_{\rm coag}\propto d_{\rm max}^{(n+1)/2}$ and coagulation becomes inefficient. Sheets only have a fast coagulation regime since $f_\mathrm{A}$ never systematically decreases. Note that in this estimate of timescales we neglect the continuous evolution of the clump velocity, which may slightly alter the actual time to turnaround. 

\smallskip
As a final note, we stress that our distinction between `fast' and `slow' coagulation implies that given a long enough time and a large enough simulation volume, all cases should eventually coagulate. However, this is an artificial conclusion based on our highly idealized numerical setup. For example, external turbulence in the surrounding hot gas would drive the small clouds further away from the central cloud and/or outright disrupt them \citep{Gronke.etal.22}. We thus expect that cases which are in our `slow coagulation` regime will in practice not coagulate in realistic scenarios.

\subsection{Comparison to Previous Work}
\label{sec:comparison}

\citet{Gronke.Oh.20b} studied shattering versus coagulation of thermally unstable clouds using simulations similar to ours. Their initial conditions consisted of four identical spherical clouds, with radius $r_{\rm s,i}$, overdensity $\chi_{\rm i}$, thermal imbalance factor $\eta$, and solar metallicity (without a UVB). They found a critical final overdensity for shattering of  $\chi_\mathrm{f}\gtrsim300(r^{*}/\ell_\mathrm{shatter}/5000)^{1/6}$, while in our spherical simulations we find  $\chi_\mathrm{f}\gtrsim190(r^*/\ell_\mathrm{shatter}/5000)^{1/4}$ (\equnp{chi_crit_sphere}, assuming a fixed $\alpha$ and ignoring the weak dependence on $\eta^*$). These two criteria are compared in \fig{ShatteringCriterion}, where it is clear that our data are not well fit by the \citet{Gronke.Oh.20b} criterion. We predict a slightly lower $\chi_\mathrm{f,crit}$ for $r^{*}/\ell_\mathrm{shatter}\lesssim10^{4}$, and a slightly steeper dependence on cloud sizes. Furthermore, their criterion depends only on the cloud radius when it loses sonic contact, $r^*$, while ours has an additional dependence on the pressure contrast at this time, $\eta^*$. In practice, our criterion depends on the final cloud radius, $r_{\rm s,f}=(\eta^*)^{-1/2}r^*$.

\smallskip
We suspect that the lower normalization is primarily due to our adopting a temperature floor of $T_\mathrm{floor}=1.68\times10^{4}\,\mathrm{K}$ compared to $4\times10^{4}\,\mathrm{K}$ in their simulations. Consequently, our $\ell_\mathrm{shatter}$ is a factor of $\sim 14$ smaller at the same density and metallicity, which results in a very similar normalization of $\chi_\mathrm{f,crit}$ given the same initial cloud parameters. Moreover, since \citet{Gronke.Oh.20b} initialized their simulations with four clouds compared to our one, they have four times the cold gas mass given the same initial cloud parameters, which makes coagulation more likely. 
More difficult to explain is the different dependence of $\chi_\mathrm{f,crit}$ on cloud size. We note that in Fig.~5 of \citet{Farber.Gronke.23}, who performed similar experiments to \citet{Gronke.Oh.20b}, the slope of their atomic shattering criterion is closer to $1/4$ than $1/6$. Regardless, these scalings are very weak, and other complications we have ignored (e.g., non-spherical geometry, background turbulence) could have a stronger impact on $\chif$.

\begin{figure}
    \centering	\includegraphics[width=\columnwidth]{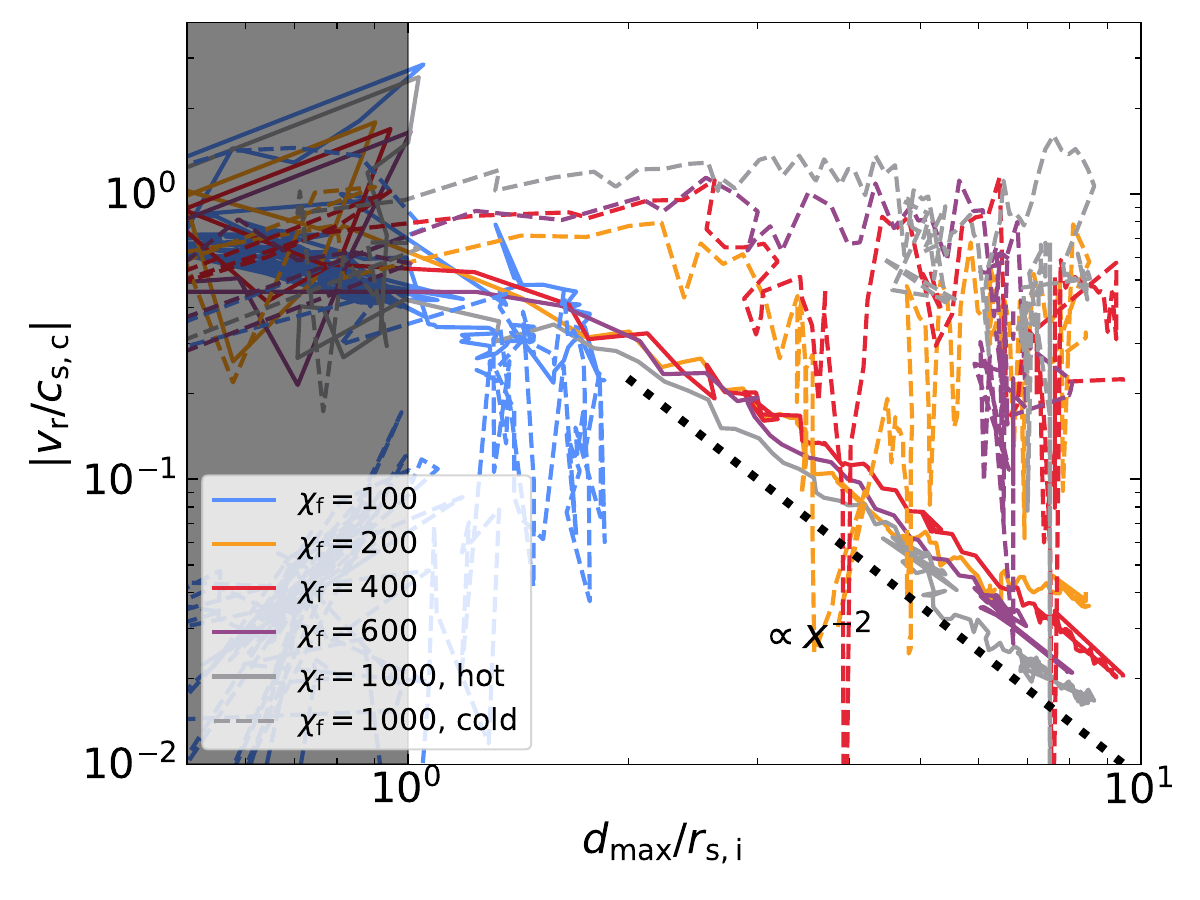}
   \caption{Radial velocities of the outermost cold clump (at $d=d_{\rm max}$) and the surrounding hot gas as a function of distance to the centre for simulations of high-$Z$ spherical clouds with $r^*=r_\mathrm{s,i}=3\kpc$ and $\eta=10$. Different colours denote different $\chi_\mathrm{f}$. Solid lines represent hot background velocities and dashed lines represent cold gas velocities. We show the absolute values of the velocities, though note that the hot gas is flowing in while the cold clumps are flowing out. At $d>r_{\rm s,i}$ the relative velocity between the clumps and the hot background is $\Delta v\sim v_{\rm cold}\sim c_{\rm s,c}$.}
   \label{fig:Clump_velocity}
\end{figure}

\smallskip
\citet{Gronke.Oh.23} described an analogy between coagulation forces and gravity, summarised in \se{theory} above (see \equnp{Fcond,grav}). They used this to calculate the effective binding energy of a collection of clumps, and then evaluated a coagulation criterion by comparing this to the clumps' kinetic energy, in analogy to the virial parameter of a self-gravitating system. While this analogy is compelling and has intriguing implications, it is intended to provide physical intuition rather than a precise quantitative criterion, as acknowledged in their paper. Firstly, their model does not easily lend itself to calculating the dependence of $\chi_{\rm f,crit}$ on $r^*/\ell_{\rm shatter}$ without knowing the number of clumps, which is very difficult to model. Secondly, the analogy between coagulation forces and simple point-mass gravity may not strictly hold in our case. This analogy is expected to be valid when the velocities of clumps with respect to the central cloud are negligible compared to the hot background velocity, so that $\Delta v_2\sim v_\mathrm{hot}\propto d^{-n}$ in \equ{Fcond}. However, in the case of imploding/exploding thermally unstable clouds as studied in this work, the clumps escape at a velocity $\sim c_\mathrm{s,c}$, which can be much larger than the hot gas velocity at large radii. \fig{Clump_velocity} shows the radial velocities of cold clumps and hot gas near $d_{\rm max}$ as a function of $d_\mathrm{max}$, for high-$Z$ spheres with $\eta=10$ and $r_{\rm s,i}=3\kpc$. For all $\chi_\mathrm{f}$, hot gas velocities exhibit similar profiles of $\propto d^{-2}$, while the velocities of cold clumps are roughly constant at $c_\mathrm{s,c}$ during their escape. 
The larger relative velocity means that clumps feel a stronger coagulation force than if they were static, $F=\dot{m}v$, facilitating `turn-around' and eventual coagulation. 
Nevertheless, we agree with the qualitative conclusion of the \citet{Gronke.Oh.23} energy argument that the coagulation efficiency increases from spheres to streams to sheets. A more detailed analysis of the effective virial parameter argument in shattered systems would be interesting to pursue in future work.

\section{Cold streams penetrating virial shocks}
\label{sec:application}

\begin{figure}
    \centering	\includegraphics[width=\columnwidth]{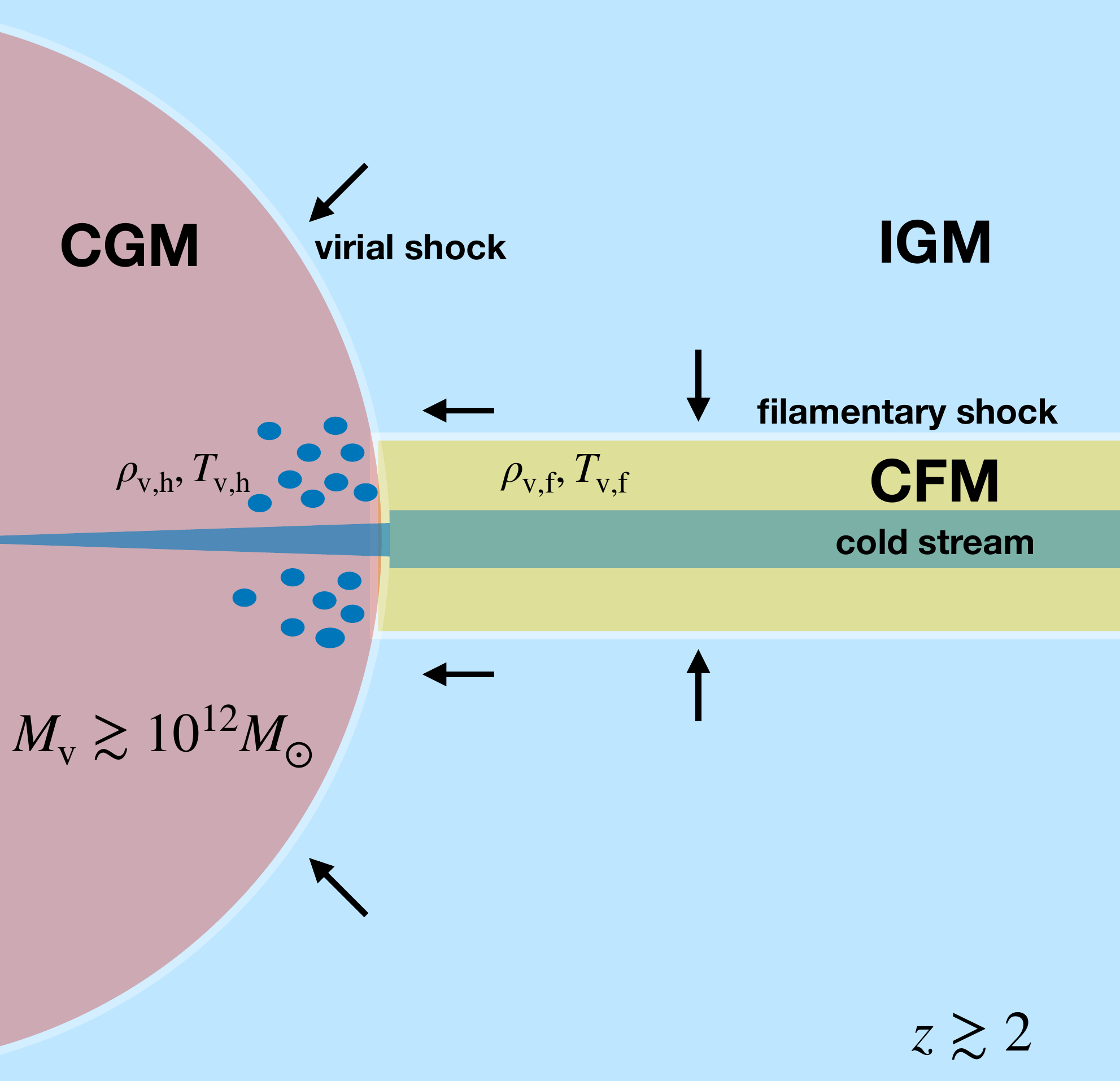}
   \label{fig:Cartoon_stream}
\caption{A schematic cartoon showing a cold stream penetrating the virial shock around a massive halo ($\Mv\gsim 10^{12}\msun$) at high-$z$ ($z\gsim 2$). In the IGM, the cold stream is the dense, isothermal core at the center of a cosmic web filament filled with hot gas at the filament virial temperature and density, $T_{\rm v,f}$ and $\rho_{\rm v,f}$. The cold stream is initially in pressure equilibrium with this circumfilamentary medium (CFM). However, as it penetrates the virial shock, the stream becomes confined by CGM gas at the halo virial temperature and density, $T_{\rm v,h}$ and $\rho_{\rm v,h}$. This results in a pressure imbalance of $\eta=P_{\rm v,h}/P_{\rm v,f}=(\rho_{\rm v,h}T_{\rm v,h})/(\rho_{\rm v,f}T_{\rm v,f})$. If pressure equilibrium in the CGM is established the stream will be narrower and denser, though it may instead shatter. 
}
\label{fig:filament}
\end{figure}
As described in \se{intro}, one of our main motivations for studying shattering in low-$Z$ streams is the application to cold streams feeding massive dark matter halos at high-$z$ from the cosmic web. In this section, we attempt to apply our results from previous sections to such streams, by constructing a toy model for how their interaction with virial shocks around dark matter halos may cause them to shatter. The basic picture is that even though the cold streams themselves are not expected to shock as they enter the virial radius \citep{Dekel.Birnboim.06,Dekel.etal.09a}, their confining pressure can increase by a large factor \citep{Lu.etal.24,Aung.etal.24}, and the resulting pressure contrast that develops between the hot CGM and the cold streams can cause streams to shatter. While this pressure contrast is not caused by radiative cooling of an intermediate temperature stream as assumed in previous sections, we show in Appendix \ref{sec:rad} below (\fig{EarlyVarwithChi_coolsmooth}) that, at least in sheets, the implosion and explosion processes are similar in cases with initially cold and underpressureized gas. In any event, this section serves as a proof-of-concept and a prelude to a more in depth study where we will explore this interaction using simulations (Yao et al., in preparation). 

\smallskip
Cosmological simulations show that at high-$z$, cosmic web filaments in the IGM have a three-zone structure \citep{Lu.etal.24}. Their outer regions contain hot, diffuse gas in virial equilibrium within the potential well set by the dark matter filament. Interior to this is a zone of multiphase gas with high turbulence and vorticity. The innermost region is a dense, isothermal core. \citet{Lu.etal.24} dubbed the outer two zones the circumfilamentary medium (CFM), while the innermost region is the cold stream that penetrates the hot CGM around massive galaxies. The relative size of each zone, and in particular how much of the CFM mass is hot at the virial temperature, depend on the profiles of the cooling and free-fall times within the filament \citep{Birnboim.etal.16, Stern.etal.20,Aung.etal.24}. Important for our purposes is that cosmological simulations suggest that these three zones are in approximate pressure equilibrium, with gas cooling isobarically from the hot CFM towards the cold stream \citep{Ramsoy.etal.21,Lu.etal.24}. Therefore, 
the pressure in cold streams prior to their entering the halo is roughly the virial pressure of cosmic web filaments, while after entering the halo they become confined by the halo virial pressure. 
A schematic cartoon of this setup is presented in \fig{filament}. We can thus estimate the pressure contrast between the cold stream and the hot CGM by comparing the virial pressure in cosmic web filaments versus dark matter halos\footnote{We assume here that the streams are fully pressure confined, ignoring their self-gravity. While self-gravity may be important in setting the structure of cold streams \citep{Mandelker.etal.18,Aung.etal.19}, this seems to be sub-dominant in the IGM \citep{Lu.etal.24} and we neglect it here for simplicity and consistency with other models of filament properties \citep{Mandelker.etal.20b}. In future work we will relax this assumption and account for the effect of self-gravity on stream shattering and evolution more generally.}.

\smallskip
The virial temperature and density of hot CGM gas are given by 
\begin{align}
\label{eq:Tvir_h}
T_{\rm v,h}&\sim 1.5\times 10^6~M_{12}^{2/3}(1+z)_3,\\
\label{eq:Dvir_h}
\rho_{\rm v,h}&\sim \Delta_{\rm v,h}\times f_{\rm b}\rho_{\rm u}(z)
\end{align} 
\citep[e.g.][]{Dekel2013}, where $M_{12}=M_{\rm v}/10^{12}\msun$ is the halo virial mass, $(1+z)_3=(1+z)/3$ is the redshift, $f_{\rm b}\sim 0.17$ is the Universal baryon fraction, $\rho_{\rm u}(z)$ is the mean matter density in the Universe at redshift $z$, and $\Delta_{\rm v,h}\sim 200$ is the halo virial overdensity in the spherical collapse model. In practice, the post-shock values of the temperature and density at the virial radius will differ from these values, which represent averages over the whole halo and neglect non-thermal pressure support of hot CGM gas \citep{Komatsu.Seljak.01,Lochhaas.etal.21}. Nevertheless, $P_{\rm v,h}\propto \rho_{\rm v,h}\,T_{\rm v,h}$ using \equs{Tvir_h}-\equm{Dvir_h} is a good representation of the characteristic thermal pressure of hot CGM gas. 

\smallskip
The corresponding values for filaments are \citep{Lu.etal.24}, 
\begin{align}
\label{eq:Tvir_f}
T_{\rm v,f}&\sim 0.4\times 10^6~M_{12}^{0.77}(1+z)_3^2f_{\rm acc,3}\mathcal{M}_{\rm v}^{-1},\\
\label{eq:Dvir_f}
\rho_{\rm v,f}&\sim \Delta_{\rm v,f}\times f_{\rm b}\rho_{\rm u}(z).
\end{align}
Here, $f_{\rm acc,3}=f_{\rm acc}/(1/3)$ is the fraction of total accretion onto the halo flowing along the given filament normalized to the typical value for a halo fed by three prominant streams \citep{Dekel.etal.09a,Danovich.etal.12}, $\mathcal{M}_{\rm v}=V_{\rm s}/V_{\rm v,h}$ is the inflow velocity along the stream normalized by the halo virial velocity, and $\Delta_{\rm v,f}\sim (15-35)$ is the virial overdensity assuming cylindrical collapse \citep{Mandelker.etal.18,Lu.etal.24}\footnote{While \citet{Mandelker.etal.18} estimated $\Delta_{\rm v,f}\sim 36$ based on the self-similar collapse models of \citet{Fillmore.Goldreich.84}, \citet{Lu.etal.24} found values of $\Delta_{\rm v,f}\sim (15-20)$ in their simulations.}. We adopt $\Delta_{\rm v,f}\sim 20$ as our fiducial value. 

\smallskip
For completeness, we note that \equ{Tvir_f} stems from the fact that 
\be 
\label{eq:Tvir_f_2}
T_{\rm v,f}\propto G\Lambda_{\rm tot,f},
\ee 
where $\Lambda_{\rm tot,f}$ is the total line-mass of the dark matter filament, 
\be 
\label{eq:Lambda_f}
\Lambda_{\rm tot,f}\sim \pi R_{\rm v,f}^2 \rho_{\rm v,f}\sim \dot{M}_{\rm f}/V_{\rm s} \sim f_{\rm acc}\mathcal{M}_{\rm v}^{-1}\,\dot{M}_{\rm v}/V_{\rm v,h},
\ee 
where the specific accretion rate onto halos is roughly \citep{Neistein2008,Fakhouri.etal.10,Dekel2013}
\be 
\label{eq:Mdot_h}
\dot{M}_{\rm v}/\Mv\sim 0.45\,{\rm Gyr}^{-1}\,M_{12}^{0.14}(1+z)_3^{2.5}.
\ee

\smallskip
Combining \equs{Tvir_h}-\equm{Dvir_f}, we obtain the pressure contrast between the hot CGM and the cold stream, 
\be 
\label{eq:Pratio_stream}
\eta\equiv \frac{P_{\rm v,h}}{P_{\rm v,f}}\simeq \frac{\rho_{\rm v,h}T_{\rm v,h}}{\rho_{\rm v,f}T_{\rm v,f}}\sim 35\,M_{12}^{-0.11}(1+z)_3^{-1} \frac{\Delta_{\rm v,h}}{10 \Delta_{\rm v,f}} f_{\rm acc,3}^{-1}\mathcal{M}_{\rm v}.
\ee 
At $z=4$ this results in $\eta\sim 20$, which seems consistent with some cosmological simulations \citep{Lu.etal.24}. In practice, neither the CGM nor the CFM are perfectly isobaric. The cold stream at the center of the filament will have pressures slightly larger than $P_{\rm v,f}$, while the hot CGM near the outskirts of the halo will have pressures slightly smaller than $P_{\rm v,h}$. \citet{Lu.etal.24} find the pressure to increase by a factor of $\sim 2$ from the outer filament to the cold stream at the center. Assuming a similar variation in the CGM pressure, $\eta$ in \equ{Pratio_stream} can be overestimated by a factor of $\gsim 4$. 

\smallskip
\citet{Gronke.Oh.20b} found that clouds can shatter if $\eta>2$, while for smaller values they simply pulsate. 
Even if $\eta$ is overestimated as described above, it is thus still expected to be large enough to cause cold streams to shatter upon entering the CGM of massive halos, and this tendency is expected to increase towards lower redshifts.
This may explain the large observed covering factions and clumping factors of cold gas in the CGM of $\Mv\gsim 10^{12.5}\msun$ halos at $z\gsim 3$ \citep{Cantalupo.etal.14,Borisova.etal.16,Cantalupo.etal.19}. 
It can also have important implications for gas accretion onto the central galaxy, and may lead to galaxy quenching by shutting off the cold-gas supply if the stream remains shattered. Whether the stream remains shattered or recoagulates depends on its size and overdensity compared to the critical overdensity for shattering. 

\smallskip
Assuming pressure equilibrium between the cold stream and the hot CGM, their density ratio is \citep{Mandelker.etal.20b} 
\be 
\label{eq:Dratio_stream}
\chif \simeq 100 \, M_{12}^{2/3}(1+z)_3\frac{\Theta_\mathrm{h}}{\Theta_\mathrm{s}},
\end{equation}
where $\Theta_\mathrm{h}= T_\mathrm{h}/T_\mathrm{v}$ is the hot CGM temperature in units of the halo virial temperature, and $\Theta_\mathrm{s}= T_\mathrm{s}/1.5\times 10^{4}\K$ is the stream temperature normalized to approximate thermal equilibrium with the UVB. For $\Mv>10^{12}\msun$ halos with a hot CGM, 
the radius of the stream in pressure equilibrium with the CGM is \citep{Mandelker.etal.20b} 
\be 
\label{eq:Rsf_stream}
R_{\rm s,f}\simeq 16\kpc (1+z)_3^{-1}\left(\frac{f_{\rm acc,3} f_{\rm c,s} \Theta_\mathrm{s}}{f_{\rm h,0.3}\mathcal{M}_{\rm v} \Theta_\mathrm{h}}\right)^{1/2},
\ee 
with no explicit dependence on halo mass. Here, $f_{\rm c,s}$ is the cold gas mass fraction in the filament and $f_{\rm h,0.3}$ is the 
hot gas mass fraction in the halo 
normalized to $0.3$. Inserting \equ{Rsf_stream} into \equ{chi_crit_stream} we obtain the critical overdensty for shattering of cold streams, 
\be 
\label{chi_crit_cold_stream}
\chi_{\rm f,crit}\simeq 120\,(1+z)_3^{-1/4}\,\left(\frac{\ell_{\rm shatter}}{300\pc}\right)^{-1/4}\,\left(\frac{f_{\rm acc,3} f_{\rm c,s} \Theta_\mathrm{s}}{f_{\rm h,0.3}\mathcal{M}_{\rm v} \Theta_\mathrm{h}}\right)^{1/8}, 
\ee 
where we have used $(\eta^*)^{-1/2}r^*=r_{\rm s,f}$. 
Comparing this to \equ{Dratio_stream} we find for cold streams 
\be 
\frac{\chi_{\rm f}}{\chi_{\rm f,crit}}\simeq 0.85\,M_{12}^{2/3}\,(1+z)_3^{5/4}\,\left(\frac{\ell_{\rm shatter}}{300\pc}\right)^{1/4}\,\left(\frac{f_{\rm h,0.3}\mathcal{M}_{\rm v}}{f_{\rm acc,3} f_{\rm c,s}}\right)^{1/8}\,\left(\frac{\Theta_\mathrm{h}}{\Theta_\mathrm{s}}\right)^{9/8}. 
\ee
We thus find that cold streams marginally meet the shattering criterion for halos with mass $\Mv\gsim 10^{12}\msun$ at redshifts $z\gsim 2$, moreso in more massive halos and higher redshift. Coagulation becomes more likely towards lower redshift, which is further enhanced as the streams become more metal enriched thus decreasing $\ell_{\rm shatter}$. On the other hand, $\ell_{\rm shatter}\propto \rho^{-1}\propto (1+z)^{-3}$ \citep{McCourt.etal.18,Mandelker.etal.20b}, which mitigates this effect somewhat. 

\smallskip
We note that in more massive filaments, $f_{\rm c,s}$ can be significantly lower than unity as larger virial temperatures and longer cooling times in the CFM increase the hot component of cosmic web filaments at the expense of the cold streams \citep{Lu.etal.24,Aung.etal.24}. This becomes important in more massive halos at a given redshift, and higher redshift for a given halo mass, and will increase $\chif/\chi_{\rm f,crit}$ making shattering even more likely in these cases. 

\smallskip
The final question we ask is whether shattering will have time to manifest before streams reach the central galaxy. In our simulations, the shattering process consisting of the implosion, explosion, and fragmentation, takes $\sim (1-2)t_{\rm sc}$ to manifest, where $t_{\rm sc}$ is the sound crossing time of the initial stream radius (see \figs{Nclump} and \figss{Clump_dmax_met}). The stream sound speed is \citep{Mandelker.etal.20b}
\be
\label{eq:stream_cs}
c_\mathrm{s,c}\simeq 18.5\,\mathrm{km s}^{-1}\Theta_{\rm s}^{1/2}\mu_\mathrm{s,0.6}^{-1/2},
\ee
where $\mu_\mathrm{s,0.6}=\mu_\mathrm{s}/0.6$ is the mean molecular weight. The uncertainties in $\eta$ from \equ{Pratio_stream}, as well as our having ignored self-gravity and stream rotation, both of which are important for setting the size of streams in the IGM \citep{Lu.etal.24}, make the initial stream radius prior to contraction difficult to constrain. We thus take as a strict lower limit the final stream radius, $R_{\rm s,f}$ from \equ{Rsf_stream}, to obtain a lower limit for the stream sound crossing time near $\Rv$ and thus the timescale for shattering, 
\be
\label{eq:stream_tsc}
t_\mathrm{sc,v}\simeq 900\,\mathrm{Myr}(1+z)_3^{-1}\left(\frac{f_\mathrm{acc,3} f_{\rm c,s} \mu_\mathrm{s,0.6}}{f_\mathrm{h,0.3}\mathcal{M}_\mathrm{v}\Theta_\mathrm{h}}\right)^{1/2}.
\ee
The streams reach the central galaxy in roughly a virial crossing time, $t_{\rm v}\sim R_{\rm v}/V_{\rm v}$, which is given by \citep[e.g.][]{Mandelker.etal.20b}
\be
\label{eq:tvir}
t_\mathrm{v}\simeq 500\,\mathrm{Myr}(1+z)_3^{-3/2}.
\ee
We thus have 
\be
\label{eq:tvir_ratio}
\frac{t_\mathrm{v}}{t_\mathrm{sc,v}}\simeq0.6(1+z)_3^{-1/2}\left(\frac{f_\mathrm{acc,3} f_{\rm c,s} \mu_\mathrm{s,0.6}}{f_\mathrm{h,0.3}\mathcal{M}_\mathrm{v}\Theta_\mathrm{h}}\right)^{-1/2}.
\ee
This ratio decreases towards lower redshift making stream shattering more likely to manifest in the CGM at later times. However, since the ratio is typically less than 1 it is unclear if streams will have time to shatter prior to reaching the central galaxy. Before we neglect this possibility, we list several issues that may help. First, the stream sound crossing time decreases rapidly as it flows towards the central galaxy due to its shrinking radius, while the stream velocity increases by a factor of $\lsim 2$ from $R_\mathrm{v}$ to $0.1\,R_\mathrm{v}$ \citep{Aung.etal.24}. As a result, the sound crossing time decreases faster than the inflow time, and at some point they should become comparable. Furthermore, the virial shock can extend up to $\sim2\,R_\mathrm{v}$ \citep{Zinger.etal.18,Aung.etal.21}. Finally, as shown in Fig.~1 of \citet{Mandelker.etal.20b}, there is an order of magnitude uncertainty in $R_\mathrm{s,f}$ due to the various additional parameters in the equations. Thinner streams, which also tend to be denser, will be more prone to shattering and disruption. Considering other physical mechanism neglected here, such as the halo potential, a stratified CGM, turbulence and shear, the dynamics of shattering and the fate of streams as they penetrate the CGM remain unclear. This will be explored in detail in future work.

\section{Caveats and Additional Physics}
\label{sec:caveats}

While our analysis has been thorough in terms of the impact of cloud geometry and metallicity on shattering and coagulation, we have neglected a number of important physical processes which can impact these issues, as well as our application to cold streams penetrating the hot CGM. We briefly discuss these here, leaving more in depth analysis to future work. 

\begin{itemize}
    \item \textit{Turbulent environments.} In all our simulations, the background was initially static. However, the CGM and the high-$z$ cosmic web are both highly turbulent environments, driven by inflows, outflows, and galaxy interactions. Turbulence plays a critical role in the dynamics of multiphase gas, contributing to shattering of large clouds and disruption of small clumps. This is particularly significant because the clump velocities in our simulations, $\sim c_\mathrm{s,c}$, are much smaller than the expected turbulent velocities \citep[e.g.][]{Mandelker.etal.21,Gronke.etal.22,Fielding.etal.23,Das.Gronke.24}. Momentum coupling between the cold clumps and the turbulent hot environment quickly entrains the clumps in the turbulent flows, if it does not first destroy them due to mixing. This supports our claim from \se{fastvsslow} that our 'slow coagulation' regime is likely an idealization. Such clouds will likely either remain shattered with the resulting clumps highly dispersed and with a minimal size set by the turbulence \citep{Gronke.etal.22}, or else the resulting small clumps will all be destroyed making it a matter of definition whether we consider this cloud shattered or not. Turbulence may also impact the critical overdensity for shattering, reducing the efficiency of coagulation. 
    
    \item \textit{Magnetic fields.} Most astrophyisical plasma is magnetized. While the outer CGM and cosmic web at high-$z$ are not expected to be highly magnetized, the magnetic fields can amplify during cloud contraction and fragmentation, and small-scale shattered clumps can therefore be magnetically dominated and out of thermal pressure equilibrium with their surroundings \citep{Nelson.etal.20}. This non-thermal pressure support leads to streaming motions along field lines and filamentary clumps, as the total pressure gradient becomes unbalanced along the field direction (Wang, Oh \& Jiang, 2024, in preparation). Magnetic draping affects the kinematics and morphology of clumps \citep{Hidalgo-Pineda.etal.24,Ramesh.etal.24}. Magnetic fields also suppress mixing in the turbulent mixing layer, reducing the entrainment rate \citep{Ji.etal.19,Gronke.Oh.20,Sparre.etal.20,Gronnow.etal.22}. 
    However, if turbulence is externally driven, then for a given rms turbulent velocity hydrodynamic and MHD mixing rates are similar \citep{Das.Gronke.24}. 

    \item \textit{Thermal conduction.} Thermal conduction is a universal dissipation process that smooths temperature gradients along magnetic field lines, leading to the evaporation of clouds smaller than the Field length when competing with radiative cooling \citep{Bruggen.Scannapieco.16,Armillotta.etal.17,Li.etal.20}. As a result, the Field length can act as a lower limit for clump size, similar to turbulence, and may change the extent of small-scale clumps resulting from shattering.
    
    \item \textit{Self-gravity.} Self-gravity affects the dynamics of cold clouds in hot gas only when the cloud mass exceed the Jeans mass \citep{Li.etal.20}, which can occur in large-scale cosmic web structures. It increases the binding energy of explosive clouds, thereby potentially inhibiting shattering and promoting coagulation in these larger clouds. More over, strong implosions like those we see prior to shattering, could directly trigger star formation, offering insights into star formation processes far from galaxies. Self-gravity is also important for cold streams \citep{Mandelker.etal.18, Lu.etal.24}, and contributes to their radial structure as well as to their interactions with the CGM \citep{Aung.etal.19}. 
    
    \item \textit{Shear.} Shear plays a critical role in cold streams penetrating virial shocks. The shear between the cold stream and the hot background gas can drive Kelvin-Helmholtz instabilities and enhance mixing. This will either promote stream growth through entrainment of hot gas in the mixing layer, or lead to stream disruption, depending on the ambient conditions \citep{Mandelker.etal.20a,Aung.etal.24,Ledos.etal.24}. Additionally, shear can induce streaming motions in explosive clumps, as well as induce additional fragmentation within the turbulent mixing layer, leading to the formation of filamentary clump structures. However, the effect of shear on the shattering and coagulation processes, particularly in a stratified medium as found in the hot CGM in galaxy halos, is unknown.
    
\end{itemize}


\begin{figure*}
    \centering	\includegraphics[width=\textwidth]{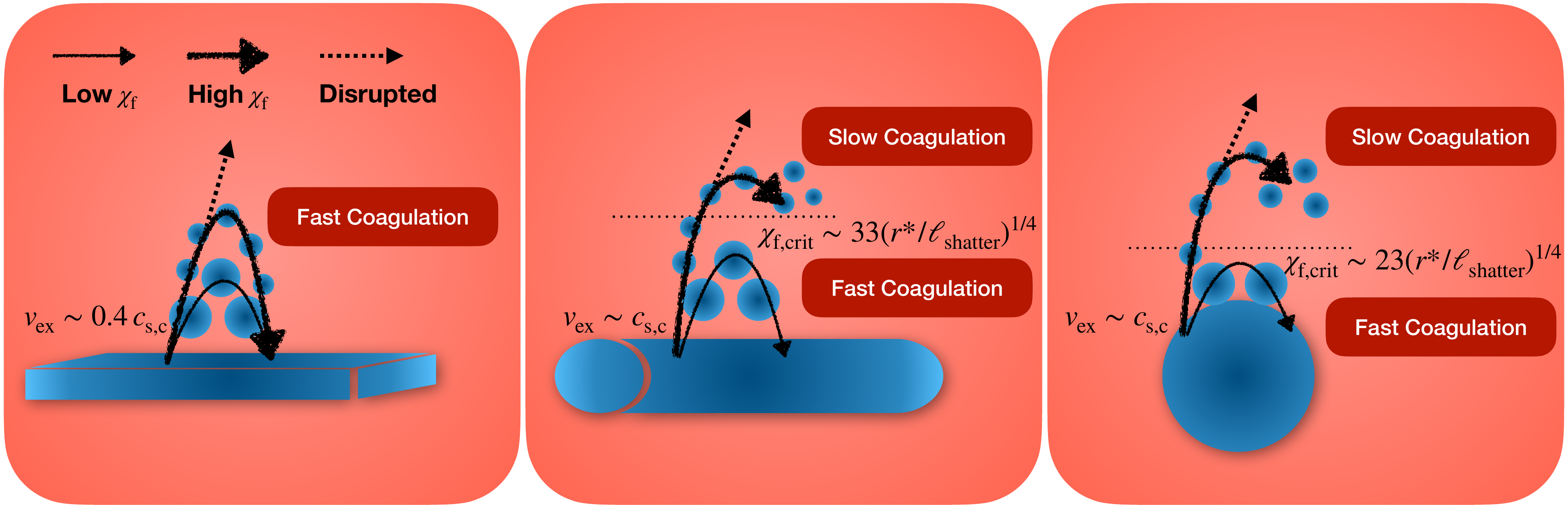}
   \caption{A diagram showing the effects of geometry and metallicity on shattering and coagulation. From left to right we show the shattering of sheets, streams, and spheres. After the initial implosion caused by a strong pressure gradient, cold gas bounces back with an explosion velocity $v_{\rm ex}\lsim c_\mathrm{s,c}$ for sheets (e.g., $\sim 0.4\,c_\mathrm{s,c}$ at an overdensity of $\chi_\mathrm{f}=100$), and $v_\mathrm{ex}\sim c_\mathrm{s,c}$ for streams and spheres. Radial coagulation is always efficient for sheets, while it becomes less efficient at large distance for streams and spheres. The corresponding $\chi_\mathrm{f,crit}$ shown in the diagram suggests a 1/4-power dependence on $r^*/\ell_\mathrm{shatter}$, where $r^*$ is the cloud radius where sonic contact is lost and the implosion phase begins, and $\ell_\mathrm{shatter}\sim \mathrm{min}(c_\mathrm{s}t_\mathrm{cool})$ is sensitive to metallicity. As $\chi_\mathrm{f}$ increases, shattered clumps may enter the disruption regime, which leads to cold gas mass loss during the shattering process.}
   \label{fig:Cartoon}
\end{figure*}
\section{Summary and Conclusions}
\label{sec:summary}

We studied the effects of cloud geometry and metallicity on the shattering and coagulation of cold gas clouds in the circumgalactic medium (CGM) or the high-$z$ cosmic web by utilizing 3D idealized hydro simulations. We initialize a thermally unstable warm cloud embedded in a hot medium in pressure equilibrium. We explored three different cloud geometries (planar sheets, cylindrical streams, and spherical clouds), varying the final overdensity of the cloud with respect to the background after thermal and pressure equilibrium have been reestablished, $\chif\equiv \rho_{\rm s,f}/\rho_{\rm bg}$, with $\rho_{\rm s,f}$ the final density of the cold cloud (sheet, stream, or sphere) and $\rho_{\rm bg}$ the density of hot background. We also varied the size of the initial cloud, $r_{\rm s,i}$, and the ratio of the initial (warm) cloud density to the final (cold) cloud density after thermal and pressure equilibrium have been reestablished, $\eta\equiv \rho_{\rm s,f}/\rho_{\rm s,i}$, which also represents the ratio of the background pressure to the pressure in the cloud after thermal equilibrium. Initially, we explored a setup where both the cloud and background have solar metallicity and there is no ionizing UV background (our `high-$Z$' case, \se{geometry}). We then extended our study of streams specifically to the low-metallicity regime where the streams and background have metallicities $Z_{\rm s}=0.03Z_{\odot}$ and $Z_{\rm bg}=0.1Z_{\odot}$, respectively, and both are exposed to a $z=2$ \citet{Haardt.Madau.96} UVB (our `low-$Z$' case, \se{metallicity}). We examine the distribution of clump sizes in a regime where the shattering lengthscale, $\ell_{\rm shatter}\equiv {\rm min}(c_{\rm s}t_{\rm cool})$, is well resolved to explore its importance as a characteristic size for cold gas (\se{sizes}) and derived analytical criteria for clouds to shatter into small fragments that disperse (\se{shatter}). Finally, we applied our results for low-$Z$ streams to the case of cold streams feeding massive halos ($\Mv\gsim 10^{12}\msun$) at high redshift ($z\gsim 2$) from the cosmic web through virial accretion shocks (\se{application}).

\smallskip
The general evolution of such thermally unstable clouds consists of two stages: 1) the initial implosion and explosion processes, triggered by a rapid loss of thermal pressure in the cloud followed by a converging and subsequent reflected shock that induces fragmentation of cold gas through Richtmyer-Meshkov instability (RMI); 2) the deceleration and recoagulation processes as escaping clumps begin comoving with the entrainment flow of hot gas converging onto the central cloud. We schematically summarize our main results and key insights into these two stages in the cartoon presented in \fig{Cartoon}. In detail, our results can be summarised as follows:

\begin{itemize}
\item The explosion velocity, which determines the growth timescale of RMI, the fragmentation timescale of the initial cloud, and (eventually) the spatial distribution of shattered clumps, is close to the sound speed of cold gas for spheres and streams, regardless of metallicity, while for sheets it is $\sim0.4\,c_\mathrm{s,c}$ (\figs{profiles} and \figss{EarlyVarwithChi_coolsmooth}).

\item Due to this relatively low explosion velocity ($v_{\rm ex}\sim c_\mathrm{s,c}$), the dominant deceleration mechanism for escaping clumps is always momentum exchange with the hot gas through condensation rather than drag or ram pressure. This is true regardless of initial cloud geometry and metallicity (\equnp{Fratio}).

\item We distinguish cases where clouds remain shattered from cases where they recoagulate based on both the number of clumps and their maximal radial distance from the centre, $\dmax$. Based on these metrics, both streams and spheres shatter only above a critical final overdensity ($\chif \gsim 200$ at high-$Z$). Sheets, on the other hand, show coagulation at all $\chif$ (\figs{Nclump} and \figss{dmax}).

\item At the end of their evolution, spherical clouds either remain shattered into many small clumps or else coagulate into a single cloud at the centre. This is not the case for streams and sheets. Rather, even if such clouds coagulate radially, coagulation is suppressed along the stream axis and within the plane of the sheet. As a result, the final number of clumps increases with $\chi_\mathrm{f}$, even if the farthest clump falls back to the central plane/line (\figs{Sheet_map}-\figss{Sphere_map} and \figs{Sheet_map_avg}-\figss{Sphere_map_avg}). 

\item The coagulation timescale is longest for spheres and shortest for sheets. This is determined primarily by the area modulation factor, $f_{\rm A}$, which is the ratio of the total surface area of cold gas to the area of the cloud-centric shell at the distance of the furthest clump (\equnp{fa}). This decreases with the distance of escaping clumps as $d^{-n}$, where $n=0,1,2$ for sheets, streams, and spheres, respectively (\fig{fA}). 

\item Low-$Z$ streams behave similarly to high-$Z$ streams with a similar ratio of $\chif/v_\mathrm{mix}$, which determines the deceleration timescale (\equnp{tcond}). For our chosen metallicities and UVB, this translates into a factor $\sim 3$ increase in $\chif$ from low-$Z$ to high-$Z$. This similar behaviour is seen in the morphology, number of clumps, maximum clump distance, and total cold gas area (\figs{Stream_met_map}-\figss{Clump_area_met}). 

\item As $\chif$ increases, the clumps resulting from shattering can enter the `disruption' regime and proceed to mix with the hot background, thus decreasing the total cold gas mass. For our low-$Z$ runs, this occurs at $\chif \sim 1000$, though for our high-$Z$ runs clumps survive at all simulated values of $\chif$ (\fig{ColdMass_met} and \equnp{chi_growth}).

\item For shattered clouds, the cumulative clump mass distribution, $N(>m)$, closely follows a power-law with an index of $\sim (-1)$, in accordance with Zipf's law (\fig{Clump_mass_dist}, left). This power-law extends from roughly the resolution scale, even if this is much smaller than $\ell_{\rm shatter}$, until the scale of the $r_{\rm s,f}$, the size the initial cloud would have had assuming monolithic collapse to thermal and pressure equilibrium. The distribution of clump radii is a power-law over the same range, showing that $\ell_{\rm shatter}$ is \textit{not} a characteristic size for cold gas clumps resulting from thermal-instability induced shattering (\fig{Clump_mass_dist}, right). The full distribution of clump masses down to the smallest and largest scales is better fit by a log-normal distribution (Appendix \ref{sec:lognormal}). 

\item 
Exploring a wide range of simulations in both spherical and stream geometries, we find a critical overdensity for shattering that scales as $\chi_{\rm f,crit}\propto (r_{\rm s,f}/\ell_{\rm shatter})^{1/4}$ (\fig{ShatteringCriterion}). Alternatively, this can be written in terms of the cloud radius and the ratio of its density to the equilibrium density at the moment when it loses sonic contact, $r^*$ and $\eta^*$. For spheres $r_{\rm s,f}=(\eta^*)^{-1/3}r^*$ while for streams $r_{\rm s,f}=(\eta^*)^{-1/2}r^*$. This is contrary to previous models for shattering \citep[e.g.][]{Gronke.Oh.20b}, where $\chi_{\rm f,crit}$ depends explicitly on $r^*$ rather than $r_{\rm s,f}$. We find the normalization of $\chi_{\rm f,crit}$ for streams to be larger than for spheres by a factor of $\sim 1.5$ (\fig{ShatteringCriterion}). We propose a model for `slow' versus `fast' coagulation, based on the competition between fragmentation of the intial cloud and deceleration and entrainment of the clumps, which agrees well with the empirical shattering criterion. Sheets are always in the fast coagulation regime (Section~\ref{sec:fastvsslow}). 

\item Many of these conclusions may change when additional physics are included, such as externally driven turbulence, magnetic fields, thermal conduction, self-gravity, or shear between the initial cloud and the background. These effects should be explored in future work.

\end{itemize}

The application of our model to cold streams penetrating the hot CGM of massive halos at high-$z$ from the cosmic web (\se{application}) rests on the realization that the confining pressure of these streams in intergalactic filaments is lower than their confining pressure in the CGM \citep{Lu.etal.24}. This could cause streams to shatter as they enter the virial shock, in much the same way as clouds which are underpressurized due to cooling shatter in our simulations.  
We evaluate this pressure contrast ($\eta$), the final density contrast between the cold streams and hot CGM ($\chif$), and the critical overdensity for shattering ($\chi_{\rm f,crit}$) as a function of halo mass and redshift, based on previous models for the properties of cold streams in the CGM \citep{Mandelker.etal.20b}. We find that shattering can be important for halos with $M_{\rm v}> 10^{12}\msun$ at $z>2$. By comparing the shattering timescale to the inflow time of the streams, we find that shattering is more likely to manifest towards the lower end of this redshift range. This offers a possible explanation for the large clumping factors and covering fractions of cold gas in the CGM around such galaxies, and may be related to galaxy quenching by providing a mechanism to prevent cold streams from reaching the central galaxy. These issues will be explored using direct simulations in upcoming work (Yao et al., in preparation).

\section*{Acknowledgements}
We thank Romain Teyssier for technical support and help with the \texttt{Ramses} clump finder. We thank Andrei Antipov, Yuval Birnboim, Frank Van den Bosch, Orly Gnat, Elisha Modelevsky, and Daisuke Nagai for helpful comments and interesting discussions. ZY and NM acknowledge support from BSF grants 2020302 and 2022281 and NSF-BSF grant 2022736. SPO thanks NASA grant 19-ATP19-0205 and NSF grant 240752 for support. AD has been partly supported by the grants ISF 861/20, NSF-BSF 2023730 and NSF-BSF 2023723. The simulations were performed on the \texttt{Moriah} cluster at the Hebrew University.

\texttt{Software:} matplotlib \citep{Hunter.2007}, numpy \citep{Harris.2020}, astropy \citep{astropy.22}, yT \citep{Turk.etal.11}, scikit-image \citep{scikit-image}, powerlaw \citep{Alstott.etal.2014}

\section*{Data Availability}

Simulation data will be shared upon reasonable request. The data supporting the plots within this article are available on reasonable request to the corresponding author.



\bibliographystyle{mnras}
\bibliography{stream_shattering} 




\appendix
\section{The implosion and explosion processes in sheets}
\label{sec:imandex}

Complementary to the qualitative discussion regarding the implosion and explosion processes in all three geometries presented in \se{explosion}, we here offer a rigorous mathematical description of these processes in sheets. 

\subsection{Adiabatic Under-pressurized Sheets}
\label{sec:ad}

\begin{figure*}
    \centering	\includegraphics[width=\textwidth]{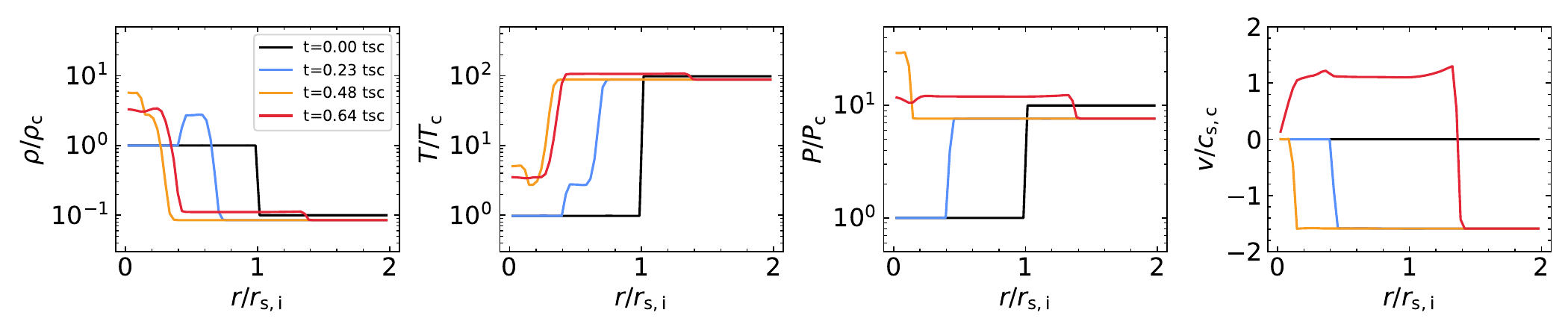}
   \caption{Profiles of density, temperature, pressure, and radial velocity (perpendicular to the surface of the sheet) at four snapshots for an adiabatic under-pressured sheet with $\eta=\chi_{\rm i}=10$, $\chi_\mathrm{f}=100$. The black, blue, orange, and red lines represent four phases of evolution: the initial condition, the implosion, the peak collision of the shock in the centre, and the explosion of the sheet, respectively. All quantities are shown normalized by their initial values in the cold sheet, $r_\mathrm{s,i}$, $\rho_\mathrm{c}$, $T_\mathrm{c}$, $P_\mathrm{c}$, and $c_\mathrm{s,c}$ are the properties of the initial cold sheet. The sound crossing time is defined as $t_{\mathrm{sc}}\equiv r_\mathrm{s,i}/c_\mathrm{s,c}$.}
   \label{fig:EarlyProfile_ad}
\end{figure*}

\begin{figure*}
    \centering
    \includegraphics[width=\textwidth]{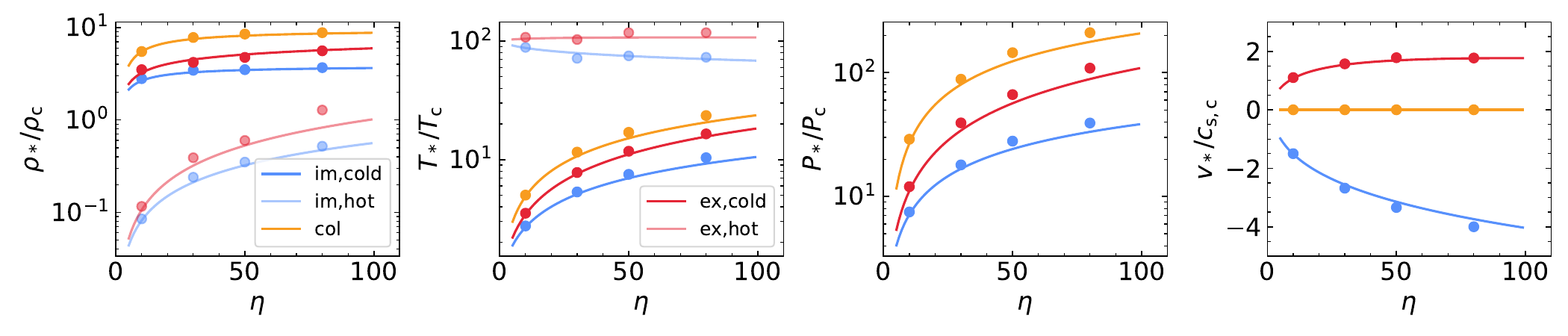}
   \caption{Physical quantities (density, temperature, pressure, and radial velocity from left to right, respectively) in the star region in an adiabatic under-pressured sheets as functions of the initial pressure contrast, $\eta$, for a fixed final overdensity, $\chi_\mathrm{f}\equiv \eta \chi_{\rm i}=100$. Lines represent the model predictions, which agree well with the simulation results shown by the dots. Colours are consistent with \fig{EarlyProfile_ad}, with the faint lines and dots represent the density and temperature of hot gas, and the darker lines and dots representing the cold gas.}
   \label{fig:EarlyVarwithEta_ad}
\end{figure*}

Consider a cold, under-pressurized sheet of density $\rho_{\rm c}$ and pressure $P_{\rm c}$, surrounded by hot gas with density $\rho_{\rm h}=\rho_{\rm c}/\chi_{\rm i}$ and pressure $P_{\rm h}=\eta P_{\rm c}$. This setup, shown by the black lines in \fig{EarlyProfile_ad} for the case $\chi_{\rm i}=\eta=10$, is unstable, as the negative pressure gradient drives the contraction of the sheet towards its centre. The evolution during the contraction, shown in \fig{EarlyProfile_ad} by the blue lines, is similar to the Sod shock tube. 

\smallskip
Inside the sheet (the left-hand side of the profiles in \fig{EarlyProfile_ad}), a shock wave develops according to the Rankine-Hugoniot conditions, describing the conservation of mass, momentum, and energy across the shock:
\begin{align}
    \label{eq:shock_mass}
    &\left(v_\mathrm{L}-v_\mathrm{sh}\right)\rho_\mathrm{L}=\left(v_*-v_\mathrm{sh}\right)\rho_\mathrm{*,L},\\
    \label{eq:shock_mom}
    &\left(v_\mathrm{L}-v_\mathrm{sh}\right)^2\rho_\mathrm{L}+P_\mathrm{L}=\left(v_*-v_\mathrm{sh}\right)^2\rho_\mathrm{*,L}+P_*,\\
    \label{eq:shock_energy}
    &\frac{1}{2}\left(v_\mathrm{L}-v_\mathrm{sh}\right)^2+\frac{\gamma}{\gamma-1}\frac{P_\mathrm{L}}{\rho_\mathrm{L}}=\frac{1}{2}\left(v_*-v_\mathrm{sh}\right)^2+\frac{\gamma}{\gamma-1}\frac{P_*}{\rho_\mathrm{*,L}}.
\end{align}
Here $v_\mathrm{L}=0$, $\rho_\mathrm{L}=\rho_\mathrm{c}$, and $P_\mathrm{L}=P_\mathrm{c}$ are the physical quantities in the sheet (on the left side). The velocities in \equs{shock_mass}-\equm{shock_energy} are in the lab frame, so we subtract the shock velocity, $v_{\rm sh}$, to describe the equations in the co-moving shock frame. The star symbol, $*$, represents the region that develops between the initial cold and hot media (between the left- and right-hand sides of the initial condition). The density and temperature in the star region have sudden transitions across the contact discontinuity (CD), while the velocity and pressure are constant. Therefore, $\rho_\mathrm{*,L}$ denotes the density on the left-hand side of the star region, while $v_*$ and $P_*$ represent the velocity and pressure throughout the whole star region. These three equations can be combined to eliminate $v_{\rm sh}$ and obtain expressions for $v_{\rm *}$ and $\rho_{\rm *,L}$, 
\begin{align}
    \label{eq:vstar_left}
    &v_*=v_\mathrm{L}-c_\mathrm{s,L}\left(\frac{2}{\gamma(\gamma+1)}\right)^{1/2}\left(\frac{P_*}{P_\mathrm{L}}-1\right)\left(\frac{P_*}{P_\mathrm{L}}+\frac{\gamma-1}{\gamma+1}\right)^{-1/2},\\
    \label{eq:dstar_left}
    &\rho_\mathrm{*,L}=\rho_\mathrm{L}\frac{\left(\gamma+1\right)(P_*/P_\mathrm{L})+\gamma-1}{\left(\gamma-1\right)(P_*/P_\mathrm{L})+\gamma+1}.
\end{align}
We find that if $P_*/P_\mathrm{L}\gg 1$ and $v_{\rm L} \ll c_{\rm s,L}$, then $v_*\propto c_\mathrm{s,L}\sqrt{P_*/P_\mathrm{L}}$ and $\rho_\mathrm{*,L}\sim\rho_\mathrm{L}(\gamma+1)/(\gamma-1)$.

\smallskip
In the background (the right-hand side of the profiles in \fig{EarlyProfile_ad}), adiabatic expansion produces a rarefaction wave that conserves entropy and the Generalised Riemann Invariant:
\begin{align}
    \label{eq:rarefaction_entropy}
    &\frac{P_*}{\rho_\mathrm{*,R}^{\gamma}} = \frac{P_\mathrm{R}}{\rho_{R}^{\gamma}},\\
    \label{eq:rarefaction_riemann}
    &v_*-\frac{2c_\mathrm{s*,R}}{\gamma-1} = v_\mathrm{R}-\frac{2c_\mathrm{s,R}}{\gamma-1}.
\end{align}
Here $v_\mathrm{R}=0$, $\rho_\mathrm{R}=\rho_\mathrm{h}$, and $P_\mathrm{R}=P_\mathrm{h}$ are the physical quantities in the background (on the right side). $\rho_\mathrm{*,R}$ and $c_\mathrm{s,*}$ are the density and the sound speed on the right-hand side of the star region, respectively. These equations yield 
\begin{align}
    \label{eq:vstar_right}
    &v_*=v_\mathrm{R}+\frac{2c_\mathrm{s,R}}{\gamma-1}\left[\left(\frac{P_*}{P_\mathrm{R}}\right)^{(\gamma-1)/(2\gamma)}-1\right],\\
    \label{eq:dstar_right}
    &\rho_\mathrm{*,R}=\rho_\mathrm{R}\left(\frac{P_*}{P_\mathrm{R}}\right)^{1/\gamma}.
\end{align}

\smallskip
Equations (\ref{eq:vstar_left}, \ref{eq:dstar_left}, \ref{eq:vstar_right}, \ref{eq:dstar_right}) offer solutions for the implosion quantities, $v_*=v_\mathrm{im}<0$, $P_*=P_\mathrm{im}$, $\rho_\mathrm{*,L}=\rho_\mathrm{im,c}$, and $\rho_\mathrm{*,R}=\rho_\mathrm{im,h}$. Furthermore, from \equ{rarefaction_riemann} we get 
\begin{equation}
    c_\mathrm{s*,R}=c_\mathrm{s,R}+\frac{\gamma-1}{2}\left(v_\mathrm{im}-v_\mathrm{R}\right)=c_\mathrm{s,im,h},
\end{equation}
and subsequently
\begin{equation}
     c_\mathrm{s*,L}= c_\mathrm{s,im,h}\left(\frac{\rho_\mathrm{im,R}}{\rho_\mathrm{im,c}}\right)^{1/2}=c_\mathrm{s,im,c}.
\end{equation}
Finally, the shock velocity $v_\mathrm{sh}$ can be derived from \equ{shock_mass}
\begin{equation}
    v_\mathrm{sh}=\frac{v_\mathrm{L}\rho_\mathrm{L}-v_\mathrm{im}\rho_\mathrm{im,c}}{\rho_\mathrm{L}-\rho_\mathrm{im,c}}=v_\mathrm{im}\frac{\rho_\mathrm{im,c}}{\rho_\mathrm{im,c}-\rho_\mathrm{L}}.
\end{equation}
These solutions for the implosion quantities are shown by the blue lines in \fig{EarlyVarwithEta_ad}. We show these as functions of $\eta$, while fixing $\chi_\mathrm{f}\equiv \eta\chi_\mathrm{i}=100$. The blue dots show results from 3D sheet simulations for comparison. We find that they agree with each other very well.

\begin{figure*}
    \centering	\includegraphics[width=\textwidth]{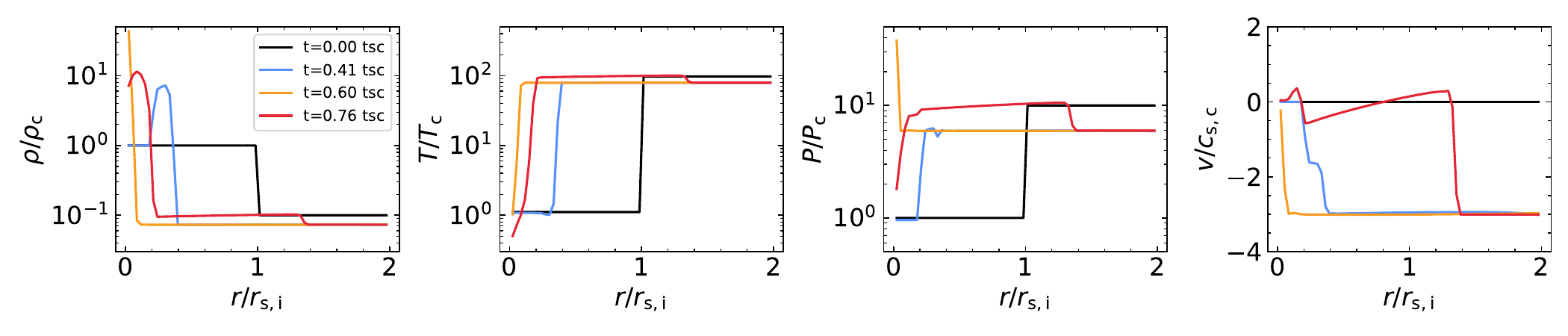}
   \caption{Similar to \fig{EarlyProfile_ad}, but for a simulation with radiative cooling. Gas is assumed to have solar metallicity, with no UV background. We set a cooling floor at $T=T_{\rm c}\sim 10^4\,{\rm K}$ and also shut off cooling at $T>0.8T_{\rm h}$.}
   \label{fig:EarlyProfile_coolsharp}
\end{figure*}

\smallskip
The implosion shock propagates towards the centre of the sheet where it collides with the shock from the other side at $t_{\mathrm{col}}=r_{\mathrm{s,i}}/v_{\mathrm{sh}}$. This head-on shock collision, shown in \fig{EarlyProfile_ad} by the orange lines, is similar to the planer Noh problem. Applying the shock solutions from \equs{vstar_left}- \equm{dstar_left} to both sides, and inserting $v_*=0$, $v_\mathrm{R}=-v_\mathrm{L}=v_\mathrm{im}<0$, $\rho_\mathrm{L}=\rho_\mathrm{R}=\rho_\mathrm{im,c}$, $P_\mathrm{L}=P_\mathrm{R}=P_\mathrm{im}$, $c_\mathrm{s,L}=c_\mathrm{s,R}=c_\mathrm{s,im,c}$ from symmetry arguments, we can derive expressions for $P_*=P_\mathrm{col}$, $\rho_*=\rho_\mathrm{col}$, and $c_\mathrm{s,*}=c_\mathrm{s,col}$. We obtain 
\begin{align}
     \label{eq:Pcol}
    &\frac{P_\mathrm{col}}{P_\mathrm{im}}=1 + \mathcal{M}^2_{\rm im,c} \frac{\gamma (\gamma+1)}{4}\left[1+\left(1+\frac{16}{(\gamma+1)^2\mathcal{M}^2_{\rm im,c}}\right)^{1/2}\right],\\
	&\frac{\rho_\mathrm{col}}{\rho_\mathrm{im,c}}=\left(\frac{P_\mathrm{col}}{P_\mathrm{im}}+\frac{\gamma-1}{\gamma+1}\right)\left(\frac{\gamma-1}{\gamma+1}\frac{P_\mathrm{col}}{P_\mathrm{im}}+1\right)^{-1},
\end{align}
where $\mathcal{M}_{\rm im,c}\equiv v_\mathrm{im}/c_\mathrm{s,im,c}$. For $\mathcal{M_{\rm im,c}}\gg 1$ and $\gamma=5/3$ we have $P_\mathrm{col}/P_\mathrm{im} \sim \mathcal{M}_{\rm im,c}^2$ and $\rho_\mathrm{col}/\rho_\mathrm{im,c} \sim 4$ as expected for strong shocks. Together these imply that $c_\mathrm{s,col}\sim v_\mathrm{im}/2$.
The orange lines and points in \fig{EarlyVarwithEta_ad} compare these solutions with simulation results and find excellent agreement. 

\smallskip
Once all the cold gas has been shock-heated, the inflowing hot gas is unable to prevent the expansion of shocked cold gas. At this stage, a shock wave develops on the right-hand side, propagating into the background, while a rarefaction wave forms on the left-hand side, propagating into the cloud. This stage is shown by the red lines in \fig{EarlyProfile_ad}. Following similar steps to those applied above when deriving the implosion quantities, but switching the left and right sides and using the conditions of $v_\mathrm{L}=0$, $v_\mathrm{R}=v_\mathrm{im}$, $c_\mathrm{s,L}=c_\mathrm{s,col}$, $c_\mathrm{s,R}=c_\mathrm{s,im,h}$, $P_\mathrm{L}=P_\mathrm{col}$, and $P_\mathrm{R}=P_\mathrm{im}$, we can obtain expressions for $P_*=P_\mathrm{ex}$, $v_*=v_\mathrm{ex}$, $\rho_\mathrm{*,L}=\rho_\mathrm{ex,c}$, and $\rho_\mathrm{*,R}=\rho_\mathrm{ex,h}$. During this phase, the reversal of the direction of the rarefaction wave modifies \equ{vstar_right} so that we have 
\begin{equation}
    \label{eq:vexRiemann}
    v_\mathrm{ex}=\frac{2c_\mathrm{s,col}}{\gamma-1}\left[1-\left(\frac{P_*}{P_\mathrm{col}}\right)^{(\gamma-1)/(2\gamma)}\right].
\end{equation}
This equation provides a maximum value for the explosion velocity of $v_{\rm ex,max}=2c_\mathrm{s,col}/(\gamma-1)$. In practice, this is always an overestimate because $P_*$ is bounded from below by $P_{\rm im}$, and furthermore the exponent $(\gamma-1)/(2\gamma)= 0.2$ is weak. The red lines and points in \fig{EarlyVarwithEta_ad} compare our model predictions for the explosion quantities with simulation results, again finding ecellent agreement.

\smallskip
Overall, we find that our model for the implosion and explosion properties predicts the simulation results quite well. Both the implosion and explosion velocities increase with $\eta$, while the latter velocity is smaller than the former because part of the kinetic energy is converted to internal energy when the shock heats the cold gas during the central collision, increasing its temperature.

\subsection{Under-pressurized Sheets with Radiative Cooling}
\label{sec:rad}

We now expand upon the simple adiabatic problem explored above by including radiative cooling, which clearly plays an important role in the shattering process we study in the main text. We consider identical initial conditions to those explored in Appendix \ref{sec:ad}, and assume that both the cold and the hot gas have solar metallicity. We set a temperature floor at the temperature of initial cold gas, $T_\mathrm{c}\sim10^4~\mathrm{K}$, and prohibit radiative cooling above $0.8~T_\mathrm{h}$ as in the main text. 
Similar to the adiabatic case, the initial implosion has a shock solution on the left and a rarefaction wave on the right. However, since radiative cooling is efficient in the cold and dense in the sheet, the post-shock gas cools down immediately after being heated up, effectively forming an isothermal shock, as shown by the blue lines in \fig{EarlyProfile_coolsharp}. 

\subsubsection{Isothermal shocks}
In isothermal shocks, energy conservation (\equnp{shock_energy}) is replaced by 
\begin{equation}
    \frac{P_\mathrm{L}}{\rho_\mathrm{L}}=\frac{P_*}{\rho_\mathrm{*,L}}=\frac{c_\mathrm{s,L}^2}{\gamma},
\end{equation}
where $c_\mathrm{s,L}$ is the adiabatic sound speed in the left-hand fluid. Consequently, the velocity, density, and sound speed in the left-hand star region are given by
\begin{align}
    \label{eq:vstar_left_iso}
    &v_*=v_\mathrm{L}-\gamma^{-1/2}c_\mathrm{s,L}\left[\left(\frac{P_*}{P_\mathrm{L}}\right)^{1/2}-\left(\frac{P_*}{P_\mathrm{L}}\right)^{-1/2}\right], \\
    \label{eq:dstar_left_iso}
    &\rho_\mathrm{*,L}=\rho_\mathrm{L}\frac{P_*}{P_\mathrm{L}},\\
    &c_\mathrm{s*,L}=c_\mathrm{s,L}.
\end{align}
Since radiative cooling is prohibited in the hot gas, the rarefaction wave on the right side has a similar solution to the adiabatic case in \equs{vstar_right}-\equm{dstar_right}. The four equations (\ref{eq:vstar_right}, \ref{eq:dstar_right}, \ref{eq:vstar_left_iso}, \ref{eq:dstar_left_iso}),thus provide the solutions for the implosion quantities in the case of an isothermal shock $v_*=v_\mathrm{im,iso}$, $P_*=P_\mathrm{im,iso}$, $\rho_\mathrm{*,L}=\rho_\mathrm{im,iso,c}$, $\rho_\mathrm{*,R}=\rho_\mathrm{im,iso,h}$, and $c_\mathrm{s*,L}=c_\mathrm{s,floor}$. Comparing $v_\mathrm{im,iso}$ in \equ{vstar_left_iso} to $v_\mathrm{im}$ in \equ{vstar_left}, we find that for $P_*/P_\mathrm{L}\gg 1$, the isothermal implosion velocity is roughly a factor of $\sqrt{(\gamma+1)/2}\sim 1.15$ larger than the adiabatic one.

\smallskip
The head-on shock collision is also isothermal, shown by the orange lines in \fig{EarlyProfile_coolsharp}. By applying \equs{vstar_left_iso}-\equm{dstar_left_iso} on both sides, we have $P_\mathrm{col,iso}/P_\mathrm{im,iso}\sim\rho_\mathrm{col,iso}/\rho_\mathrm{im,iso,c}\sim \gamma (v_\mathrm{im,iso}/c_\mathrm{s,floor})^2$ and $c_\mathrm{s,col}=c_\mathrm{s,floor}$. 

\smallskip
However, the expansion phase is adiabatic for both the shock propagating into the background on the right and the rarefaction wave propagating into the cloud on the left, shown by the red lines in \fig{EarlyProfile_coolsharp}. This is because the shock heats up the hot gas in the background (on the right) to the regime where radiative cooling is prohibited, while the cold gas in the cloud (on the left) is already at the temperature floor and cannot cool, though its temperature can decrease through adiabatic expansion. Therefore, we expect similar solutions to the adiabatic case, and in particular the maximum explosion velocity is predicted to be $v_{\rm ex,max}\sim 2c_\mathrm{s,floor}/(\gamma-1)\sim 3c_{\rm s,floor}$ from \equ{vexRiemann}, larger by a factor of $\sim 2$ than our estimate of $v_{\rm ex}\sim 1.34 c_{\rm s,c}$ from \equ{vex}. However, as noted above, the maximum limit provided by \equ{vexRiemann} is necessarily an overestimate, while the limit provided by \equ{vex} is not.

\subsubsection{Radiative cooling in mixing layers}
\label{subsec:mixing}
In addition to isothermal shocks, radiative cooling also takes place at the interfaces between cold and hot phases, where gas is at intermediate temperatures and consequently has efficient cooling. In the context of our shock model, we assume this occurs at the CD. Radiative cooling causes enthalpy loss and pulls both cold and hot gas into this transition region, inducing turbulent mixing that smooths the CD. Therefore, we hereafter refer to this region as the `mixing layer' instead of the CD. The effects of radiative cooling in the mixing layer can be seen in the blues lines in the pressure and velocity panels of \fig{EarlyProfile_coolsharp}. There is a small dip of pressure at the mixing layer due to cooling. This pressure difference pulls both hot and cold gas into the mixing layer, causing the hot gas to accelerate and the cold gas to decelerate with respect to their implosion velocities in the lab frame. During the explosion phase, the cold gas is accelerated and the hot gas is decelerated near the mixing layer, as can be seen by the red line in the velocity panel.

\smallskip
To study this process quantitatively, we begin with the momentum and energy equations 
\begin{equation}
\label{eq:mom}
    \frac{\partial \rho \vec{v}}{\partial t}=-\nabla\cdot (\rho \vec{v}\otimes \vec{v} +P\vec{\vec{I}}), 
\end{equation}
\begin{equation}
\label{eq:energy}
    \rho\frac{\partial \epsilon}{\partial t}=-\rho \vec{v}\cdot \nabla \epsilon -P\nabla\cdot \vec{v}-n^2\Lambda+E_\mathrm{vis},
\end{equation}
where $\epsilon$ is the specific internal energy. The terms on the right hand side of \equ{energy} are (from left to right) the advection of thermal energy, adiabatic expansion/compression, radiative cooling, and viscous dissipation (caused by turbulence, viscosity, etc.). In steady state, the thermal pressure is roughly the same on either side of the mixing layer so the ram pressure exerted by cold and hot gas inflowing into the mixing layer should roughly balance, 
\begin{equation} 
\label{eq:mix_momentum}
\rho_\mathrm{c}v_\mathrm{ent,c}^2=\rho_\mathrm{h}v_\mathrm{ent,h}^2.
\end{equation}
This can also be deduced from \equ{mom}. Here $v_\mathrm{ent,c}$ and $v_\mathrm{ent,h}$ are the entrainment velocities of cold and hot gas into the mixing layer, respectively. Assuming that the advection of thermal energy into the mixing layer from the hot phase is balanced by the advection of thermal energy out of the mixing layer into the cold phase and that viscous dissipation is negligible, then radiative cooling is only balanced by adiabatic compression (\equnp{energy}), 
\begin{equation}
\label{eq:mix_energy}
    P\frac{v_\mathrm{ent,h}-v_\mathrm{ent,c}}{H}=-\frac{P}{(\gamma-1)t_\mathrm{cool}},
\end{equation}
where $H$ is the width of the mixing layer. Combining \equs{mix_momentum} and \equm{mix_energy}, we can derive the entrainment velocities of cold and hot gas with respect to the mixing layer 
\begin{align}
    &v_\mathrm{ent,c}=\frac{1}{1+\sqrt{\chi}}\frac{H}{(\gamma-1)t_\mathrm{cool}}>0,\\
    &v_\mathrm{ent,h}=-\frac{\sqrt{\chi}}{1+\sqrt{\chi}}\frac{H}{(\gamma-1)t_\mathrm{cool}}<0,
\end{align}
where $\chi=\rho_\mathrm{c}/\rho_\mathrm{h}$ is the density ratio between the cold and hot phases. 

\begin{figure}
    \centering	
    \includegraphics[width=\columnwidth]{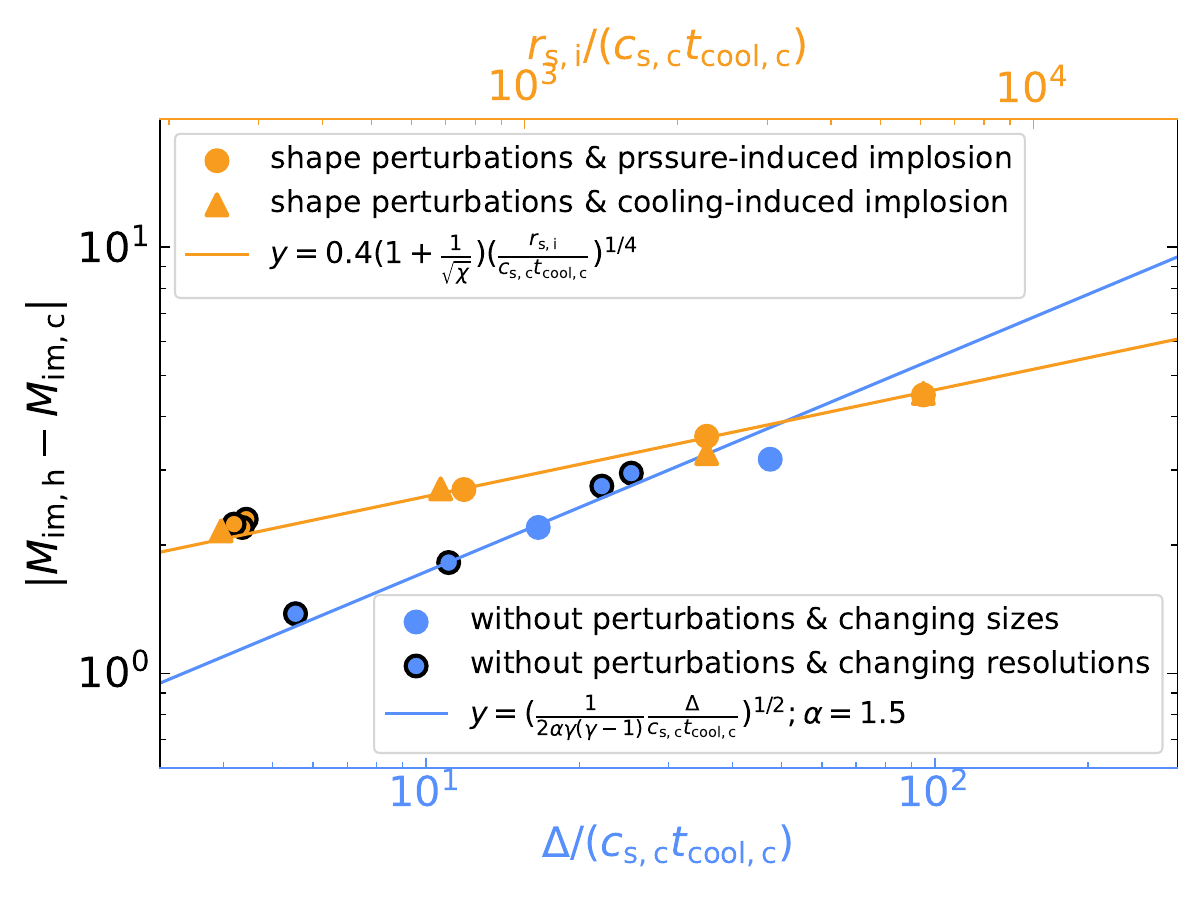}
   \caption{Implosion velocity differences between the hot and cold phases in radiatively cooling sheets with $\eta=10$ and $\chi_\mathrm{f}=100$. Blue dots represent simulations without any initial perturbations, where the velocity difference depends on resolution (shown on the bottom x-axis) and is well fit by \equ{vel_diff} shown by the blue line. Dots with black edges fix the size of sheets and change the number of cells across the sheet, while those without edges fix the number of cells per sheet but change the sheet thickness. Orange dots represent simulations with initial density and interface perturbations, where the velocity difference does not depend on resolution, as shown by those dots with black edges. Instead, it depends only on the sheet thickness (shown on the top x-axis) and is well fit by \equ{vel_diff_2} shown by the orange line. This is true whether the pressure jump was present in the initial conditions (circular points) or was generated due to strong cooling (triangles).}
   \label{fig:EarlyVmix}
\end{figure}

\smallskip
We can now evaluate the velocity of cold and hot gas in the lab frame. Both phases share the same implosion velocity $v_\mathrm{im,iso}$, and we assume the mixing layer has a velocity $v_{\rm ml}$ relative to $v_\mathrm{im,iso}$. If radiative cooling is efficient, the mixing layer moves outwards into the hot phase by accumulating cold gas due to the cooling of intermediate-temperature gas in the mixing layer, $v_{\rm ml}>0$. On the other hand, if radiative cooling is inefficient, the mixing layer moves inwards into the cold phase due to the heating by mixing, $v_{\rm ml}<0$. Thus, in the lab frame, the cold gas velocity during implosion is $v_\mathrm{im,c}=v_\mathrm{im,iso}+v_\mathrm{ml}+v_\mathrm{ent,c}$, while the hot gas velocity is $v_\mathrm{im,h}=v_\mathrm{im,iso}+v_\mathrm{ml}+v_\mathrm{ent,h}$. Therefore, the velocity difference is 
\begin{equation}
    |v_\mathrm{im,h}-v_\mathrm{im,c}|=|v_\mathrm{ent,h}-v_\mathrm{ent,c}|=\frac{H}{(\gamma-1)t_\mathrm{cool}},
\end{equation}
which only depends on the properties of the mixing layer. 

\smallskip
In the simulation shown in \fig{EarlyProfile_coolsharp}, there is an initial pressure jump at the boundary of cold and hot phases, limited by the grid scale, $\Delta$. The width of the mixing layer and the velocity difference thus initially satisfy $|v_{\rm im,c}-v_{\rm im,h}|\propto H\sim\Delta$. However, we find that the velocity difference in this case is proportional to $\Delta^{1/2}$ rather than $\Delta$ (blue dots in \fig{EarlyVmix}). This is consistent with \citet{Tan.etal.21}, who found that the surface brightness, which is proportional to the entrainment velocity, scales as $\Delta^{1/2}$ in the absence of thermal conduction. 

\smallskip
The result $v_\mathrm{ent}\propto H$ was first proposed by \citet{Begelmam.Fabian.90}, but was later found to not hold by recent 3D turbulent mixing layer simulations \citep{Ji.etal.19,Tan.etal.21}. The fundamental reason is that viscous dissipation cannot be neglected. In the absence of strong turbulence, the dissipation on the grid scale (the scale of the initial phase transition) is mainly numerical. According to the incompressible Navier-Stokes equations, the viscous dissipation rate can be described by
\begin{equation}
    E_\mathrm{vis}=2\rho \nu S_{ij}S_{ij},
\end{equation}
where $S_{ij}=\frac{1}{2}(\frac{\partial v_i}{\partial x_j}+\frac{\partial v_j}{\partial x_i})$ is the strain-rate tensor, which has a value of approximately $|v_\mathrm{im,h}-v_\mathrm{im,c}|/\Delta$ across the mixing layer. The numerical kinetic viscosity $\nu$ is estimated by $\nu=\alpha c_\mathrm{s,c}\Delta$. Equating the viscous dissipation term to the cooling term, we have 
\begin{equation}
    2\rho\alpha c_\mathrm{s,c}\Delta \frac{(v_\mathrm{h}-v_\mathrm{c})^2}{\Delta^2}=n^2\Lambda.
\end{equation}
Using $n^2\Lambda=P/[(\gamma-1)t_{\rm cool}]$ and considering the cold phase yields
\begin{equation}
\label{eq:vel_diff}
    |\mathcal{M}_\mathrm{im,h}-\mathcal{M}_\mathrm{im,c}|=\left(\frac{1}{2\alpha \gamma (\gamma-1)}\frac{\Delta}{c_\mathrm{s,c}t_\mathrm{cool,c}}\right)^{1/2},
\end{equation}
where $\mathcal{M}_\mathrm{im,h}$ and $\mathcal{M}_\mathrm{im,c}$ are the Mach numbers of hot and cold gas with respect to $c_\mathrm{s,c}$, respectively. As shown by the blue line in \fig{EarlyVmix}, \equ{vel_diff} with $\alpha=1.5$ is an excellent fit to simulation results, which use \texttt{Ramses} with a MUSCL scheme and HLLC Riemann solver. This is true whether we vary $\Delta$ by keeping the sheet thickness fixed and varying the number of cells across the sheet (points with a black boundary), or whether we keep the number of cells across the sheet fixed and vary the sheet thickness (points without a black boundary). 

\subsubsection{The Impact of Turbulence}
\label{subsec:mixing_2}

\smallskip
In more realistic cases, turbulence is ubiquitous and shapes the properties of mixing layers along with radiative cooling. In this case, the width of the mixing layer is comparable to the largest eddy, which has a comparable size of the width of the sheet. According to \citet{Gronke.Oh.20}, the entrainment velocity of the hot phase is 
\be 
\label{eq:vent_h_2}
v_\mathrm{ent,h}\approx\beta c_\mathrm{s,c}(t_\mathrm{sc}/t_\mathrm{cool,c})^{1/4},
\ee 
when cooling dominates over turbulent mixing, where $\beta=0.2-0.5$ \citep{Gronke.Oh.20,Gronke.Oh.23}. Combining this with \equ{mix_momentum}, the velocity difference is 
\begin{equation}
\label{eq:vel_diff_2}
    |\mathcal{M}_\mathrm{im,h}-\mathcal{M}_\mathrm{im,c}|\approx \beta\left(1+\frac{1}{\sqrt{\chi}}\right)\left(\frac{\rsi}{c_\mathrm{s,c}t_\mathrm{cool,c}}\right)^{1/4} .
\end{equation} 

\begin{figure*}
    \centering
    \includegraphics[width=\textwidth]{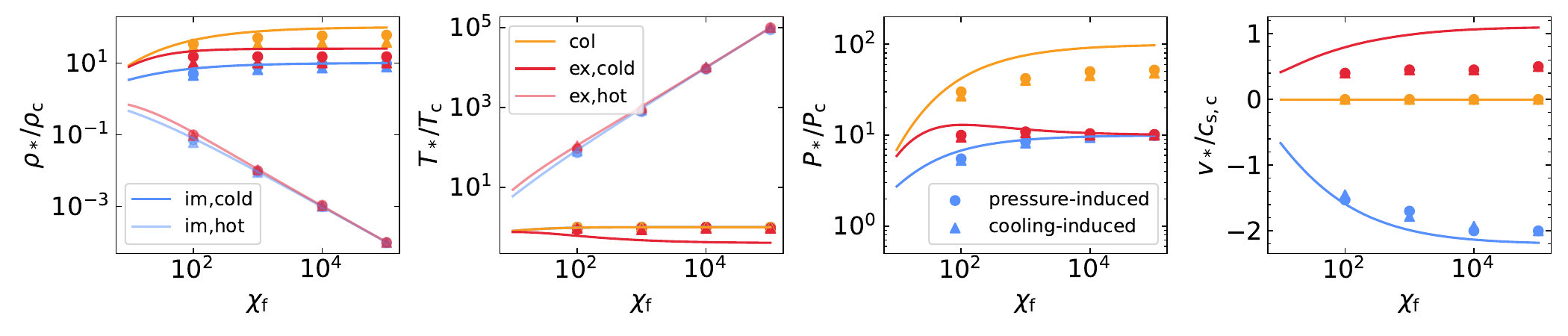}
   \caption{Physical quantities (density, temperature, pressure, and radial velocity from left to right, respectively) in the star region in simulations with radiative cooling, shown as functions of the final overdensity $\chi_\mathrm{f}$ at a fixed $\eta=10$ and $\rsi=3\,$kpc. All simulations have initial surface and density perturbations in the sheet. Circular points represent simulations with initial pressure discontinuities while triangles represent simulations where the pressure jump was induced by strong cooling in a sheet initially in pressure equilibrium with the background, as in the main text. Lines represent the model predictions, which agree reasonably well with the simulation results for the density and temperature. For the pressure and velocity, the model agrees with the simulations during the implosion phase, but overpredicts the peak pressure during the shock collision and the explosion velocity, especially at large $\chif$.}
   \label{fig:EarlyVarwithChi_coolsmooth}
\end{figure*}

\smallskip
In order to develop turbulence during the implosion, we add interface perturbations to the surfaces of sheets and density perturbations inside the sheets, as described in Section~\ref{sec:methods}. As shown by the orange circular dots and orange line in \fig{EarlyVmix}, we find that the velocity differences in simulations are well fit by \equ{vel_diff_2} with $\beta=0.4$. This result is independent on resolution, measured by the number cells across the sheet thickness, as shown by the orange points with a black boundary, where we vary the number of cells by factors of 2, 4, and 8 while keeping the sheet size fixed. 

\smallskip
Until now we have considered a case with an initial pressure discontinuity between the cold and hot gas. We now consider a case as in the main text where the sheet is initially in pressure equilibrium with the hot background, but its temperature is a factor of $\eta$ higher than the temperature floor. Assuming $t_\mathrm{cool}\ll t_\mathrm{sc}$, which is true in our case, the sheet rapidly cools isochorically and induces the pressure contrast. As shown by the orange triangles in \fig{EarlyVmix}, the properties of the mixing layer in this case are similar to the case with an initial pressure contrast. 

\begin{figure*}
    \centering
    \includegraphics[width=\textwidth]{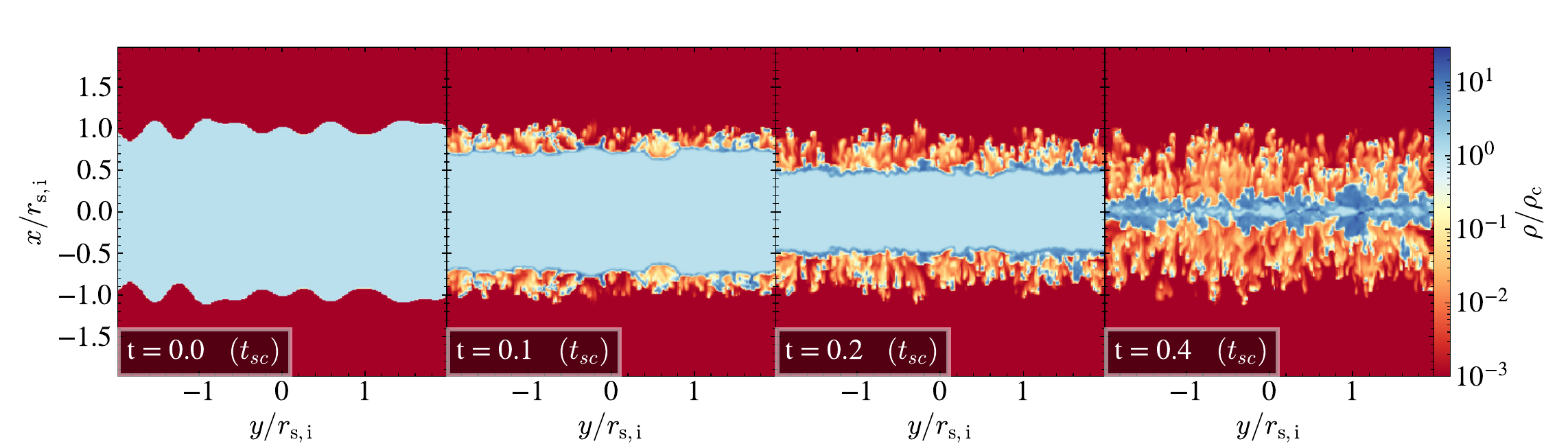}
   \caption{Normalized 2D density slices at four snapshots in the case of $\eta=10$, $\chi_\mathrm{f}=10^{4}$. Fragmentation occurs during implosion. }
   \label{fig:EarlySlice}
\end{figure*}

\smallskip
If the movement of mixing layer with respect to the implosion shock can be neglected, then the total implosion velocities for the cold and hot gas are given by $v_\mathrm{im,c}=v_\mathrm{im,iso}+v_\mathrm{ent,c}$ and $v_\mathrm{im,h}=v_\mathrm{im,iso}+v_\mathrm{ent,h}$, respectively. Furthermore, the thermal pressure of imploding gas decreases near the mixing layer, which drives the inflow into the mixing layer. In other words, the pressure and density are modified due to the entrainment flow
\begin{align}
&P_\mathrm{im}=P_\mathrm{im,iso}-\rho_\mathrm{im,c}v_\mathrm{ent,c}^{2}=P_\mathrm{im,iso}-\rho_\mathrm{im,h}v_\mathrm{ent,h}^{2},\\
&\rho_\mathrm{im,c}=\rho_\mathrm{im,iso}\left(\frac{P_\mathrm{im}}{P_\mathrm{im,iso}}\right)^{1/\gamma},\\
&\rho_\mathrm{im,h}=\rho_\mathrm{im,iso}\left(\frac{P_\mathrm{im}}{P_\mathrm{im,iso}}\right)^{1/\gamma},
\end{align}
where we have assumed the density varies adiabatically because the cold gas is at the cooling floor and the hot gas cannot cool (recall that it is primarily intermediate gas that cools in the mixing layer). We have thus obtained the properties of the implosion, accounting for the effects of the radiatively cooling mixing layer. Similar modifications are applicable to the explosion phase as well, where the entrainment accelerates the cold gas and decelerates the hot gas. 

\smallskip
We compare our model predictions with simulations in \fig{EarlyVarwithChi_coolsmooth}. Here we vary $\chi_\mathrm{f}$ while fixing $\eta=10$ and $\rsi=3\,$kpc. Circles represent simulations with an initial pressure contrast while triangles represent simulations where the pressure contrast was induced by rapid cooling. Note that the triangles at $\chif=100$ correspond to the sheet case shown in \fig{profiles}. For all values of $\chif$, the model is a good fit to the density and temperature values at all stages - implosion, peak collision, and explosion - and a good fit to the pressure and radial veloity during the implosion. However, we overpredict the peak collision pressure and the explosion velocity, especially at large $\chif$. We suspect that this discrepancy arises from the fragmentation of the sheet during the implosion, as illustrated in \fig{EarlySlice}. 
This fragmentation seems to be due to the surface perturbations and a combination of RMI and vorticity induced at the curved surface of the perturbations together with strong cooling. Clumps begin to form and detach from the sheet very early on in the evolution, before the outer cold gas surface has gained much velocity. At high overdensity, these detached clumps are unable to catch up to the implosion front since the timescale for clumps to become entrained in the flow increases linearly with $\chi$ (\se{theory}). These fragmented clumps increase the cold-gas surface area, thereby driving stronger inflows of hot background gas into the mixing layers. Furthermore, because of this fragmentation the imploding gas that reaches the centre is not fully collisional, with small-scale clumps `missing' each other. This leads to a lower collisional density and pressure than predicted, ultimately resulting in a reduced explosion velocity. 

\smallskip
Following the explosion, the cold gas escapes outwards at a velocity of $\sim c_\mathrm{s,c}$, while a mild shock propagates outwards into the hot gas. In the adiabatic case (\fig{EarlyProfile_ad}, red lines), the velocity of the post-shock hot gas is also $\sim c_\mathrm{s,c}$. However, radiative cooling at the surface of the central cold gas sheet induces an entrainment flow through a turbulent mixing layer, as described in \se{theory} of the main text. This causes the hot gas to flow towards the centre in an entrainment flow, while the cold clumps escape outwards. 

\section{Convergence tests}

\begin{figure*}
    \centering
    \includegraphics[width=\textwidth]{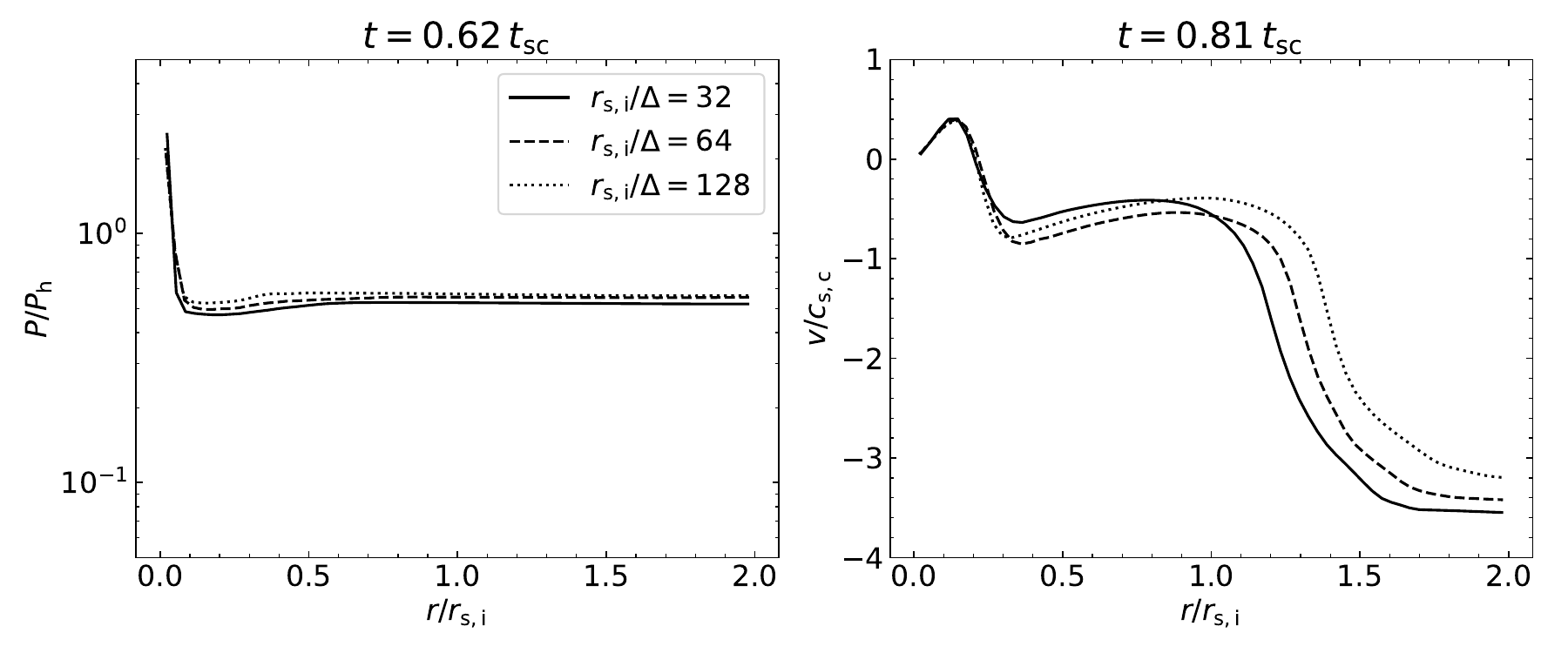}
   \caption{Convergence test of the effect of resolution at the centre of the sheet on the peak pressure during the implosion shock collision (left) and the explosion velocity (right). We find that both of these quantities are converged at our fiducial resolution.}
   \label{fig:EarlyProfile_coolsmooth_res}
\end{figure*}

\begin{figure}
    \centering
    \includegraphics[width=\columnwidth]{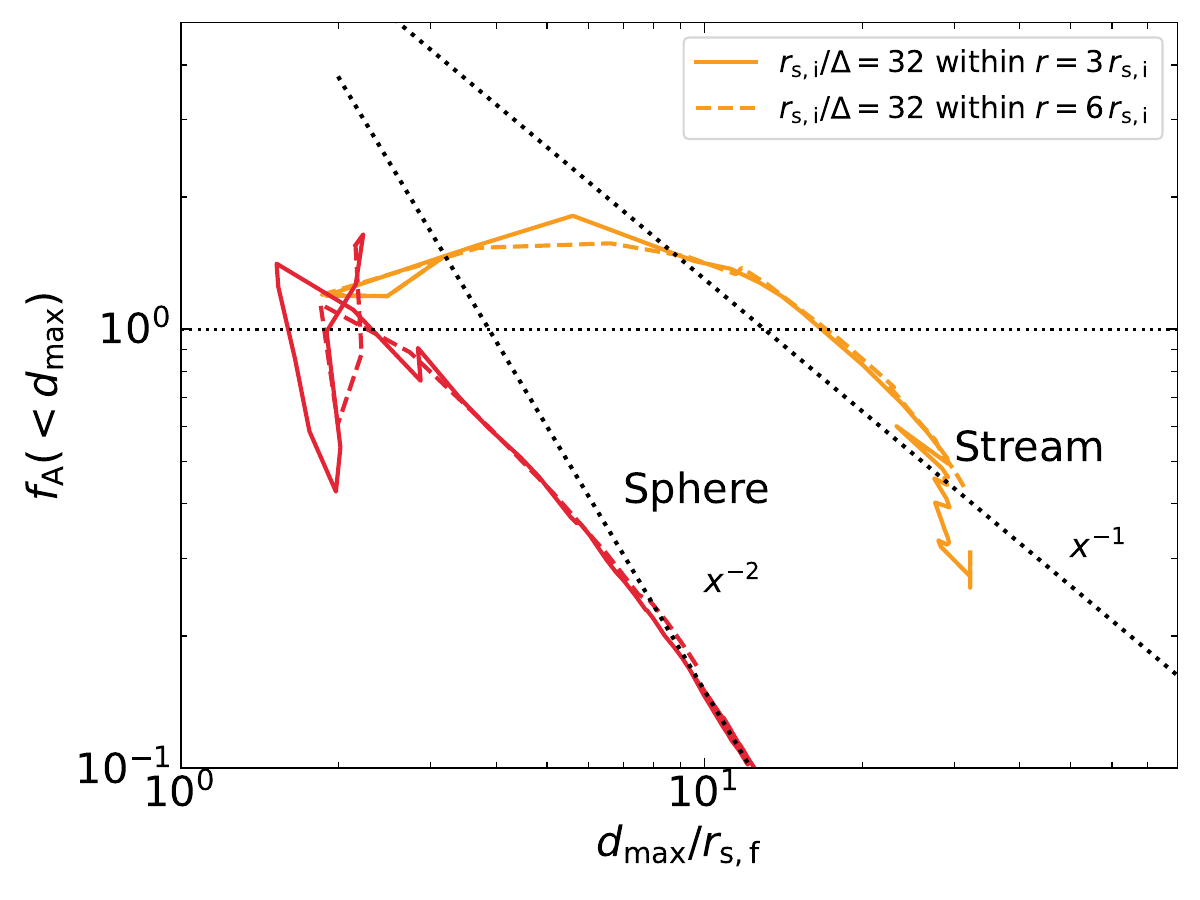}
   \caption{The convergence test of the effect of the size of the high-resolution region on the area modulation factor, $f_{\rm A}$, and thus on $r_\mathrm{fA}$, the radius where $f_{\rm A}$ first decreases and a critical element of our shattering criterion. Solid lines represent simulations with our fiducial grid structure, with the first resolution drop at $r=3\,r_\mathrm{s,i}$, while dashed lines represent simulations with a larger high resolution region where the first resolution drop is at $r=6\,r_\mathrm{s,i}$. Orange lines represent streams while red lines represent spheres, each with $\chif=1000$, $\eta=10$, $r_{\rm s,i}=3\kpc$ and high-$Z$. In all cases, we find that the size of the high resolution region has no impact on $f_{\rm A}$ or $r_{\rm fA}$.}
   \label{fig:Clump_area_res}
\end{figure}

\begin{figure}
    \centering
    \includegraphics[width=\columnwidth]{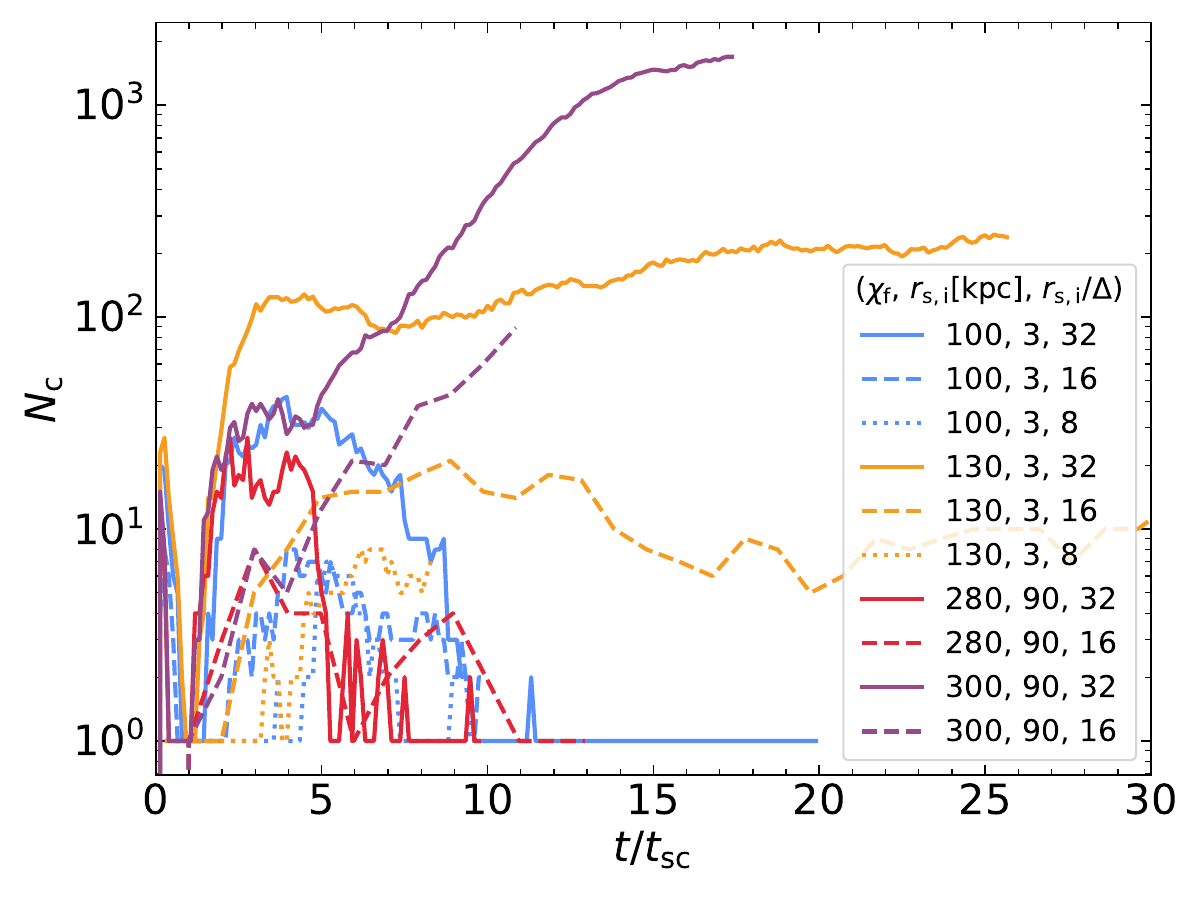}
   \caption{Convergence test of the effect of resolution on the distinction between shattering and coagulation in spheres. We choose four combinations of $\chif$ and $r_{\rm s,i}$ as shown in the legend (with $\eta=10$ and high-$Z$), and vary the number of cells per initial cloud radius. While the number of clumps depends on resolution, the distinction between shattering and coagulation does not.}
   \label{fig:Clump_numberevo_res}
\end{figure}

\subsection{Peak pressure and explosion velocity of sheets}
\label{subsec:peakpres_sheet}
In \fig{profiles}, we found that the peak pressure and the explosion velocity in sheets are both much smaller than in streams and spheres. Since the density and pressure in collapsing isothermal clouds increases fastest for spheres and slowest for sheets, we wished to verify that we would not see a more dramatic increase in the central pressure (and therefore in explosion velocity) in sheets if the resolution were increased allowing further collapse. To this end, we performed sheet simulations with the same parameters as in \fig{profiles}, with three different resolutions within $r<3\,r_\mathrm{s,i}$. The results are shown in \fig{EarlyProfile_coolsmooth_res}. We found that the peak pressure and explosion velocity do not change with resolution. 

\subsection{$f_\mathrm{A}$ and $r_\mathrm{fA}$}
In \se{fastvsslow}, we derive a shattering criterion that depends on $r_{\rm fA}$, the radius where $f_{\rm A}$ began to decrease. In order to verify that this is not sensitive to the artificial disruption of cold gas induced by our statically refined mesh (\se{methods}), we performed two tests with a larger high resolution region, such that the resolution first decreases at $r=6\,r_\mathrm{s,i}$ rather than $3\,r_{\rm s,i}$. We do this for both streams and spheres with $\chi_\mathrm{f}=1000$, as shown in \fig{Clump_area_res}. We find that the size of the high-resolution region does not affect $f_\mathrm{A}$ significantly. 

\subsection{The shattering criterion}
\label{subsec:shatteringcriterion}
Our definition of shattering or coagulation rests primarily on the evolution of the number of clumps. To verify that this is not sensitive to resolution, we chose four simulations that are close to the borderline and lowered their resolutions by factors of 2 and 4, as shown in \fig{Clump_numberevo_res}. While the number of clumps obviously depends on resolution, the distinction between shattering and coagulation, namely whether $N_{\rm c}$ decreases to order unity or remains large until the end of the simulation, does not. 

\section{Volume-weighted average of density projection maps}
\begin{figure*}
    \centering	\includegraphics[width=\textwidth]{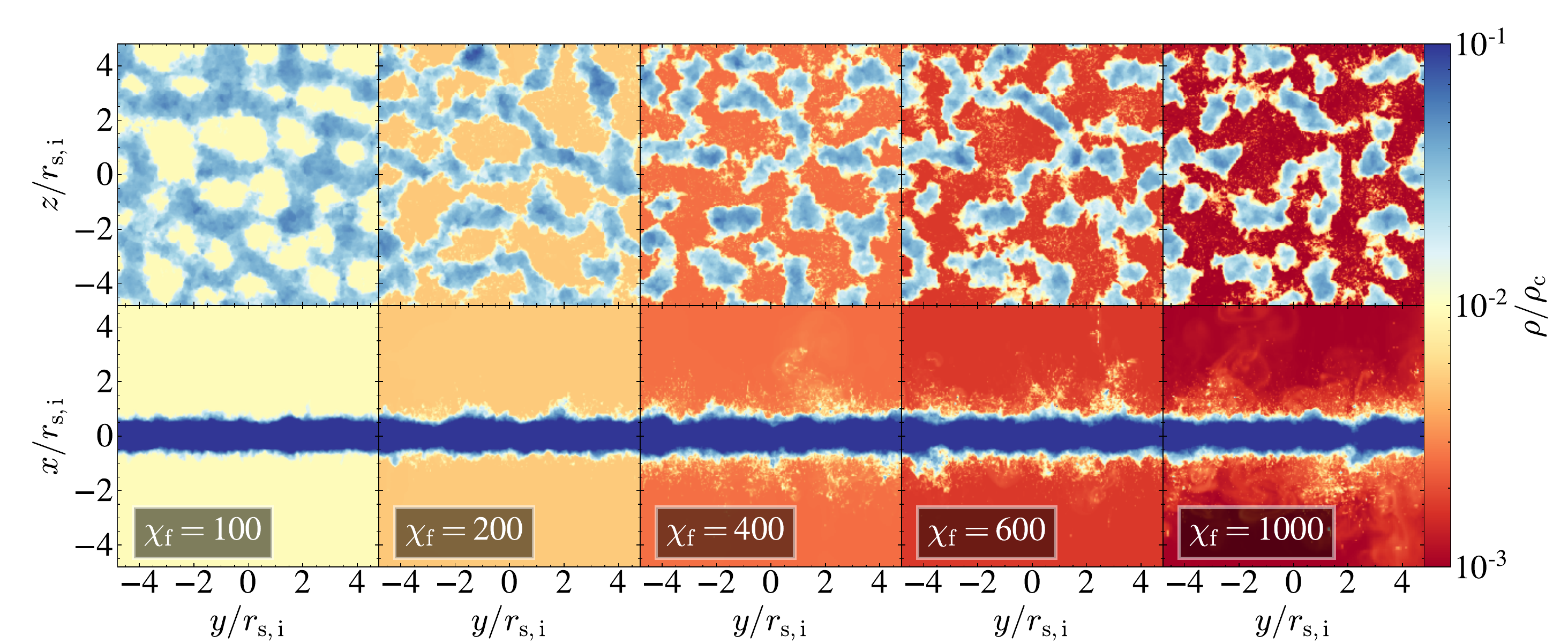}
   \caption{Same as \fig{Sheet_map}, but showing the average density along the line-of-sight rather than the maximal density, to highlight coagulation and the geometry of large clouds rather than small clouds which result from shattering. 
   }
   \label{fig:Sheet_map_avg}
\end{figure*}

\begin{figure*}
    \centering	\includegraphics[width=\textwidth]{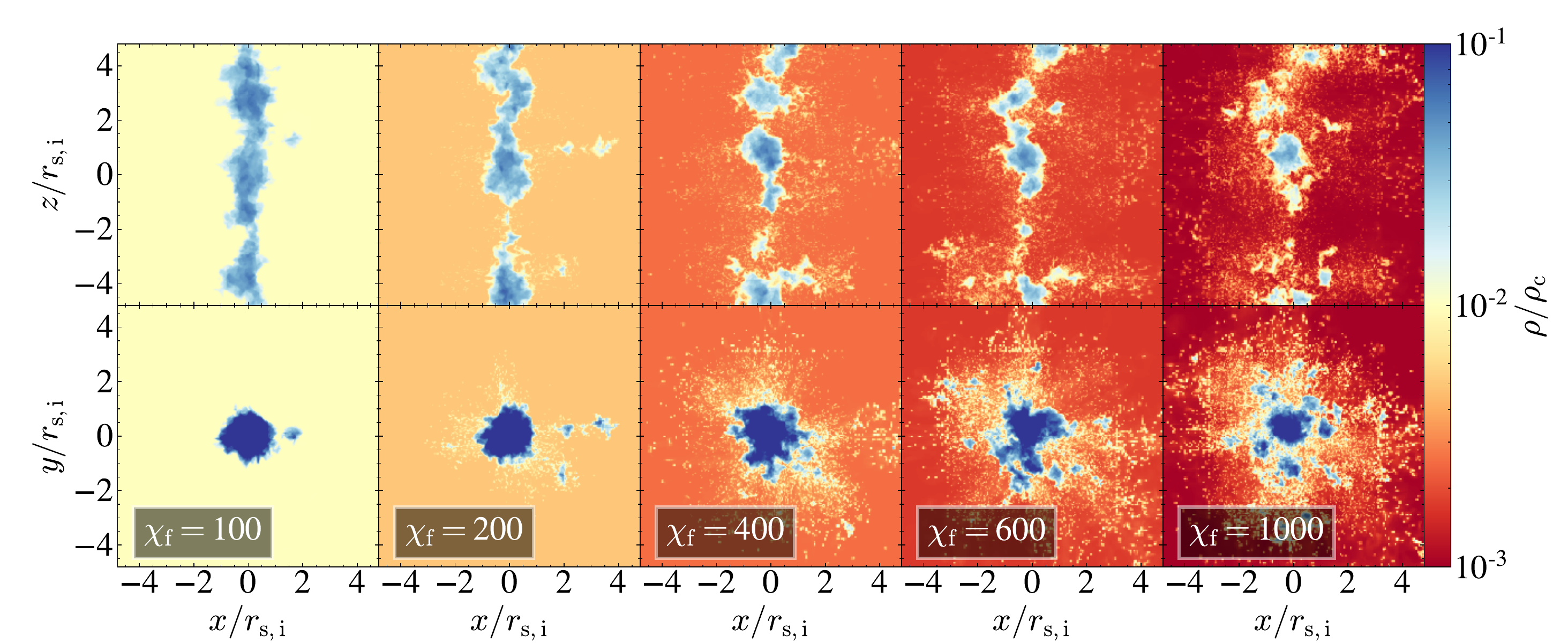}
   \caption{
   Same as \fig{Stream_map}, but showing the average density along the line-of-sight rather than the maximal density, to highlight coagulation and the geometry of large clouds rather than small clouds which result from shattering. 
   }
   \label{fig:Stream_map_avg}
\end{figure*}

\begin{figure*}
    \centering	\includegraphics[width=\textwidth]{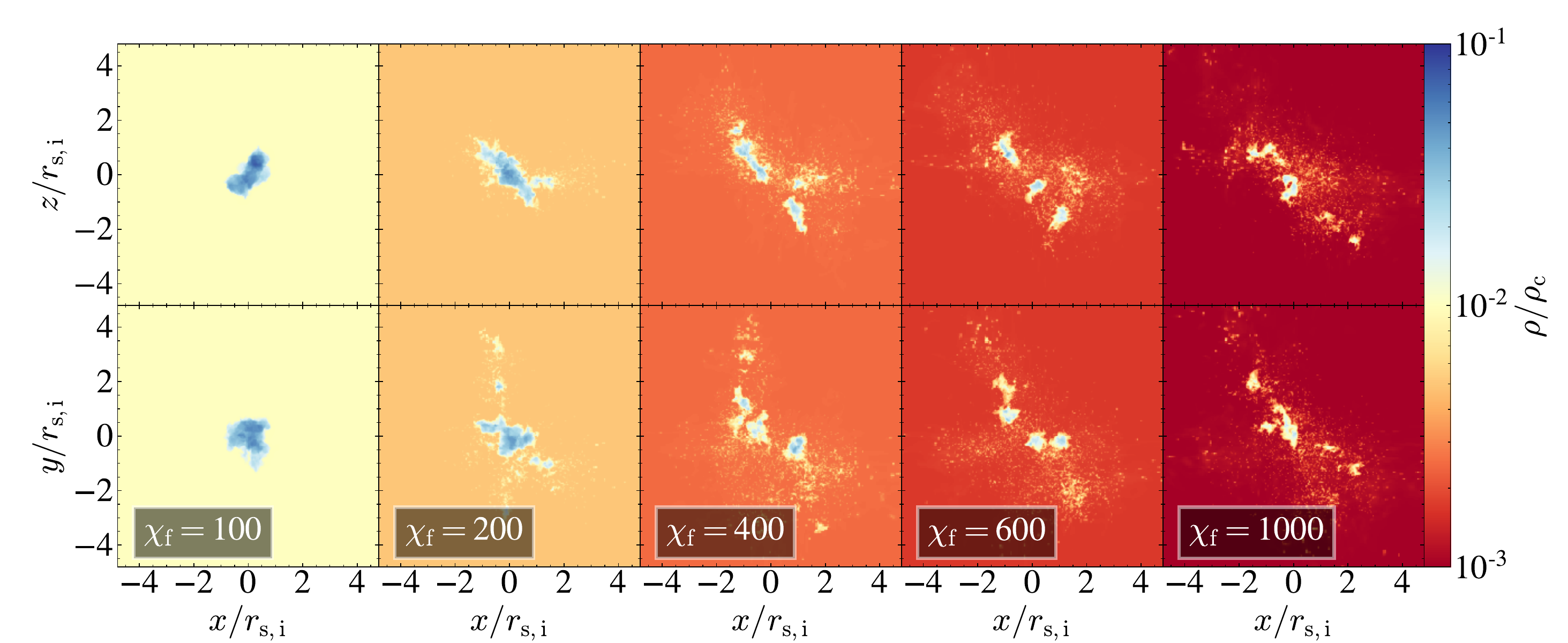}
   \caption{
   Same as \fig{Sphere_map}, but showing the average density along the line-of-sight rather than the maximal density, to highlight coagulation and the geometry of large clouds rather than small clouds which result from shattering. 
   }
   \label{fig:Sphere_map_avg}
\end{figure*}

\begin{figure*}
    \centering	\includegraphics[width=\textwidth]{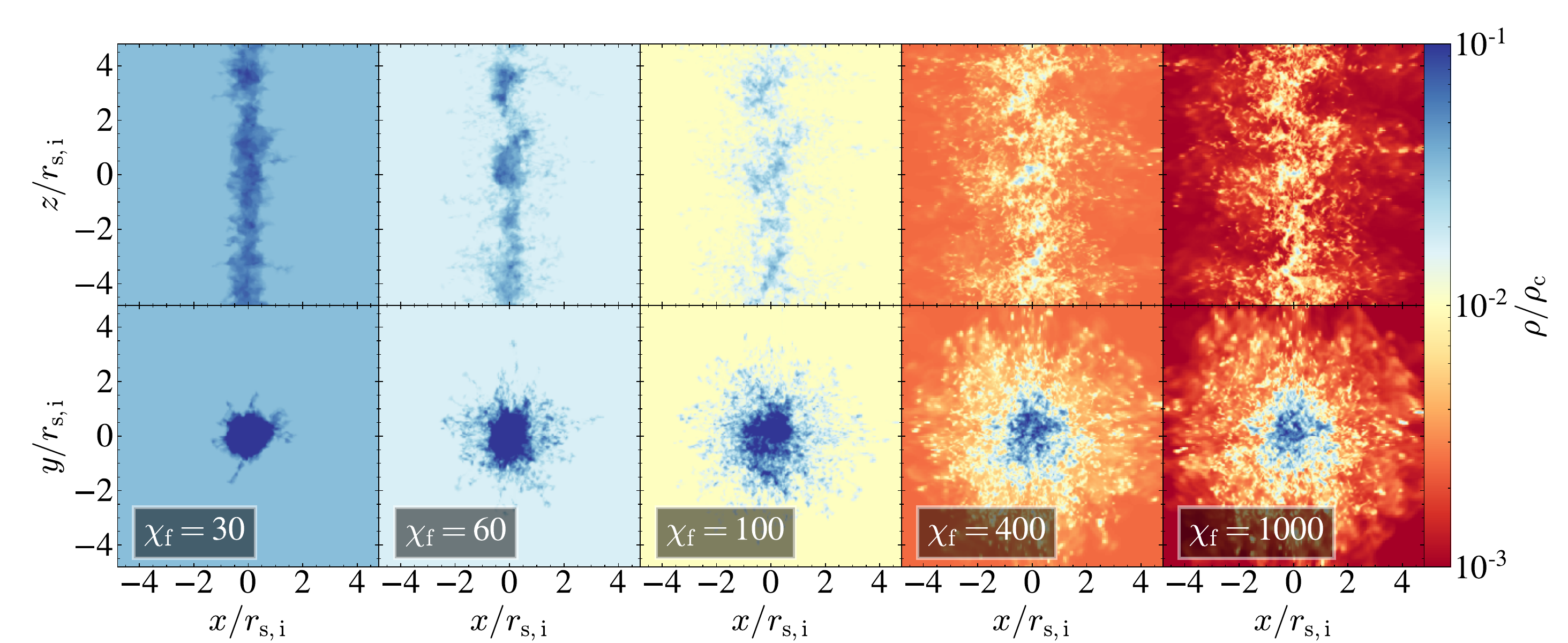}
   \caption{
   Same as \fig{Stream_met_map}, but showing the average density along the line-of-sight rather than the maximal density, to highlight coagulation and the geometry of large clouds rather than small clouds which result from shattering. 
   }
   \label{fig:Stream_met_map_avg}
\end{figure*}

In \figs{Sheet_map_avg}-\figss{Stream_met_map_avg} we present maps of the volume-weighted average density in sheets, streams, and spheres, with different $\chif$ and metallicity values. These are meant to complement, respectively, \figs{Sheet_map}-\figss{Sphere_map} and \fig{Stream_met_map} from the main text, which show the maximal density along the line of sight. As described in the text, the maximal density better highlights shattering and small clumps, while the average density better highlights coagulation and large clumps. This better shows how coagulation is suppressed along the stream axis and in the plane of the sheet, and also how in general radial coagulation is stronger in sheets compared to streams compared to spheres.
 
\section{Lognormal fits to the clump mass distributions}
\label{sec:lognormal}

In \se{sizes} and \fig{Clump_mass_dist} we discussed the distribution of clump sizes, suggesting these are well described by a power law, $N(>m)\propto m^{-1}$. However, many power laws are actually log-normal distributions in disguise \citep{Mitzenmacher.04}, and pure fragmentation, in particular, is expected to produce a log-normal due to the central limit theorem \citep{kolmogorov.41}. We compare these distributions using the python package \textit{powerlaw} \citep{Alstott.etal.2014}, which fits both power laws and log-normals while estimating the p-values of the significance via log-likelihood ratios. When we use all clumps down to the minimal clump mass, a log-normal distribution is significantly favoured over a power, law with a p-value less than $10^{-20}$. However, when allowing the minimal clump mass to vary as part of the fitting procedure, the p-value can rise to as high as $0.3$ when fitting over the range $m_{\rm cl}/m_0>10^{-4}$. This suggests that while the log-normal distribution is still favoured, a power-law is also a good description of the data over this mass range. In either case, the mode of the log-normal distribution is sensitive to resolution even when $\ell_\mathrm{shatter}$ is resolved, consistent with our main conclusion that $\ell_\mathrm{shatter}$ does not set a characteristic value for clump size in our simulations.

\begin{figure}
    \centering	\includegraphics[width=\columnwidth]{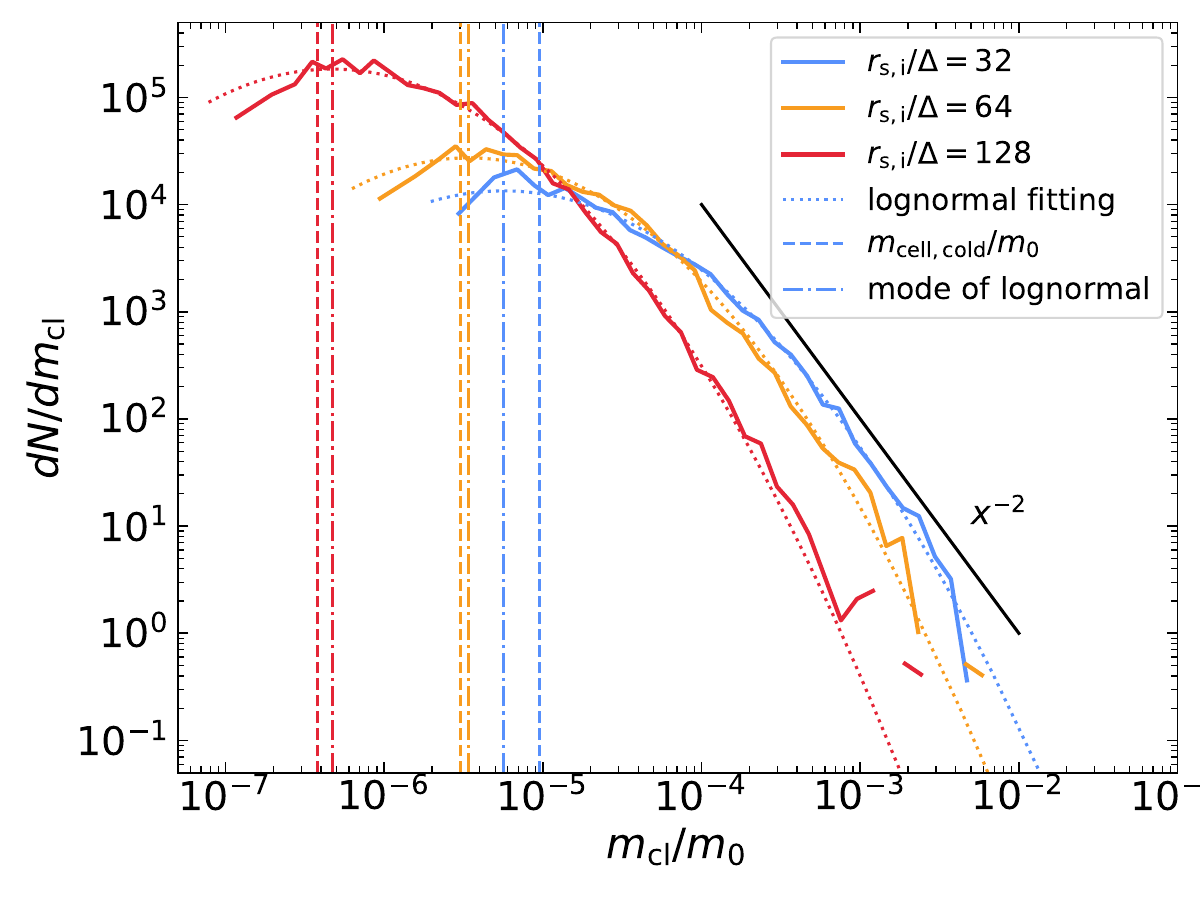}
   \caption{The PDF of clump masses for the low-metallicity stream case with $\rsi=3\,$kpc, $\chif=100$ at different resolutions. The clump mass is normalized by the initial mass of the stream, with colours representing different resolutions. Higher-resolution simulations capture smaller clump masses, extending to lower values of $m_\mathrm{cl}/m_0$. Dotted lines indicate lognormal fits for each resolution, with the modes (dash-dotted lines) aligning with the normalized cold masses in a single cell (dashed lines). Clump masses can fall below this value due to lower densities and higher temperatures. For $m_\mathrm{cl}/m_0 \gtrsim 10^{-4}$, where resolution effects are minimal, the distributions exhibit a power-law feature with an index of $-2$, consistent with the cumulative power-law slope of $-1$ discussed in \se{sizes}.
   }
   \label{fig:Clump_mass_pdf_one}
\end{figure}

\begin{figure}
    \centering	\includegraphics[width=\columnwidth]{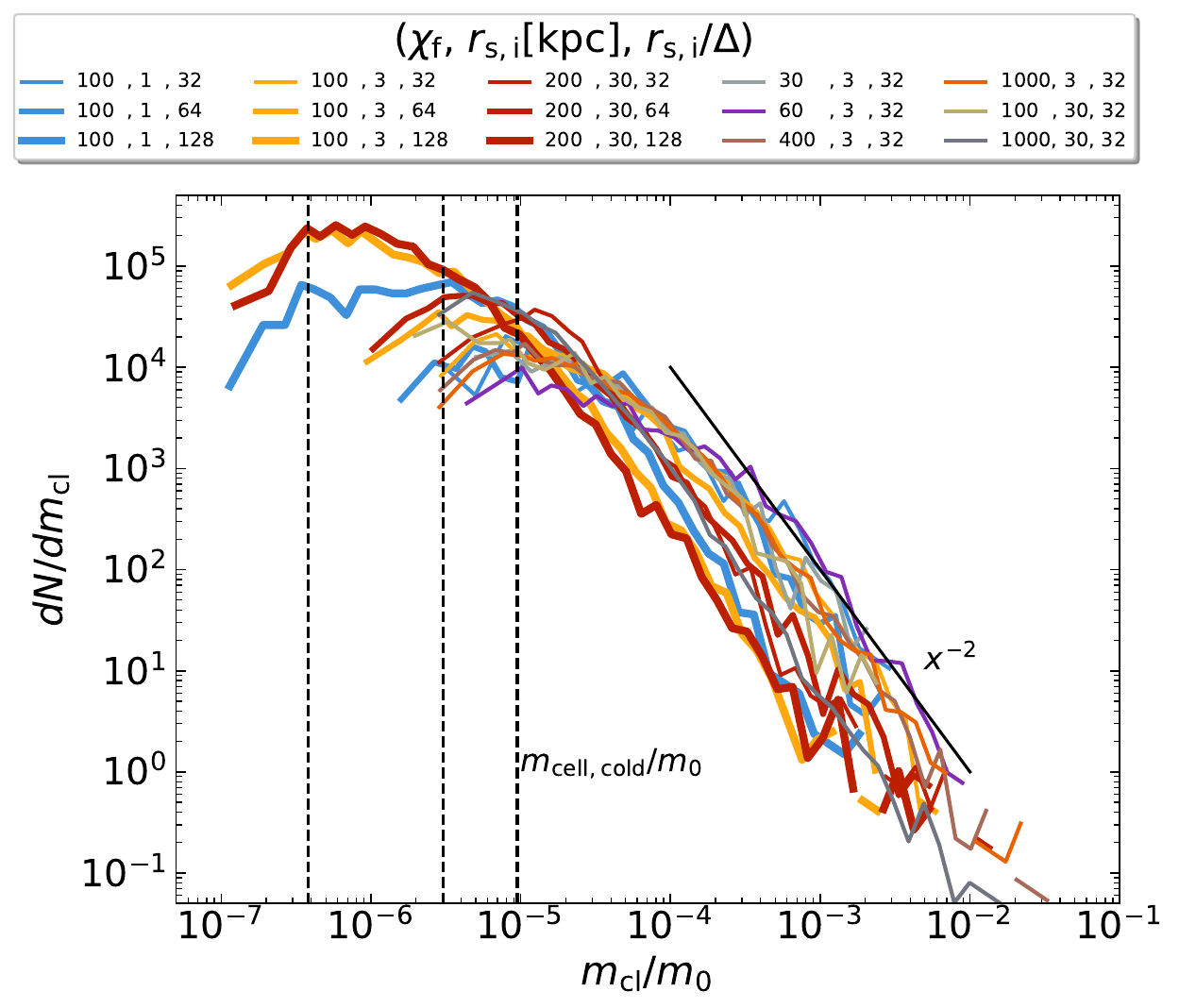}
   \caption{The PDF of clump masses for all low-metallicity stream cases. Line styles and colours are as in \fig{Clump_mass_dist}. The dashed lines represent the normalized cold masses in a single cell, which appear to align with the peaks of the mass distributions.
   }
   \label{fig:Clump_mass_pdf_all}
\end{figure}

\begin{figure}
   \centering	\includegraphics[width=\columnwidth]{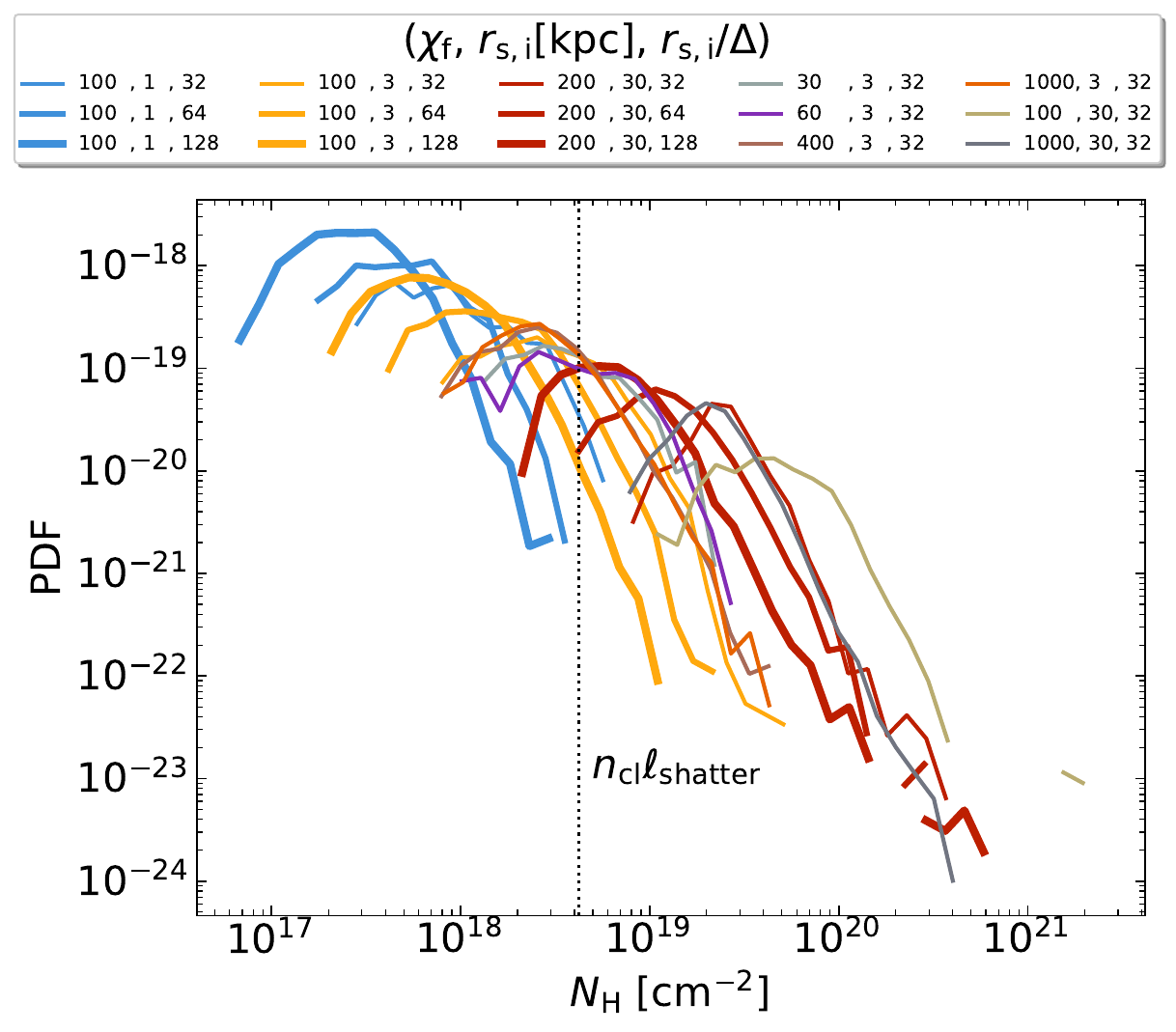}
  \caption{The probability density function (PDF) of column densities in simulations of low-metallicity streams. Line styles and colours are as in \fig{Clump_mass_dist}. The column densities are calculated by $N_\mathrm{H}=2\rho_\mathrm{cl}r_\mathrm{cl}$. The cloud sizes are derived assuming spherical geometry, $r_\mathrm{cl}=[3m_\mathrm{cl}/(4\pi \rho_\mathrm{cl})]^{1/3}$, where $m_\mathrm{cl}$ and $\rho_\mathrm{cl}$ are obtained from the clump finder. The black dashed line represents the column density of cold clumps at the size of $\ell_\mathrm{shatter}$. This column density is higher than $10^{17}\,$cm$^{-2}$ predicted by \citet{McCourt.etal.18} due to our much larger $\ell_\mathrm{shatter}$ caused by lower metallicity and the presence of a UVB. The peaks of the column densities are determined by the resolution rather than by the column density at $\ell_\mathrm{shatter}$.}
  \label{fig:Clump_column_density}
\end{figure}

\smallskip
\fig{Clump_mass_pdf_one} shows the lognormal fits to the case of a low-$Z$ stream with $\rsi=3\,$kpc, $\eta=10$, and $\chif=100$ at different resolutions. \fig{Clump_mass_pdf_all} shows the PDF of all low-$Z$ cases, which can be fit by lognormal distributions similar to \fig{Clump_mass_pdf_one}. Note that the peak of the distribution tracks the numerical resolution. \fig{Clump_column_density} shows the PDF of column densities of all low-metallicity streams. The peaks of the column densities are determined by the
resolution rather than by the column density at $\ell_\mathrm{shatter}$.

\bsp	
\label{lastpage}
\end{document}